\PassOptionsToPackage{left=0.25in,right=1.5in}{geometry}

\documentclass{article}
\usepackage[left=0.5in,right=0.7in,top=0.85in,bottom=0.85in]{geometry}%

\usepackage{endnotes}

\usepackage[final, 
colorlinks=true, 
            linkcolor=teal, 
            citecolor=teal, 
            urlcolor=teal]{hyperref}

\usepackage{graphicx}
\usepackage{amssymb}
\usepackage{amsmath}
\usepackage{mathabx}
\usepackage{ifthen}
\numberwithin{equation}{section}
\usepackage{tipa}

\usepackage{amsthm}
\usepackage{mathrsfs}
\usepackage{dutchcal}
\usepackage{tcolorbox}
\tcbuselibrary{breakable}
\usepackage{empheq}
\usepackage{graphbox}
\usepackage{subfigure}
\usepackage{enumitem} %
\usepackage{soul}
\usepackage{empheq}
\usepackage{sidecap}
\usepackage{xcolor}
\usepackage{ulem}

\usepackage{multirow}
\usepackage{graphbox}
\usepackage{booktabs}
\usepackage{longtable}
\usepackage{verbatimbox}
\usepackage{array}
\usepackage{fancyhdr}
\usepackage{float}

\usepackage{glossaries} %
\usepackage[font=small,labelfont=bf]{caption}

\makeatletter

\newcommand{\coloruline}[2]{%
        \newcommand\temp@reduline{\bgroup\markoverwith
            {\textcolor{#1}{\rule[-1ex]{2pt}{0.8pt}}}\ULon}%
        \temp@reduline{#2}%
    }

 \newcommand{\coloruuline}[2]{%
        \UL@protected\def\temp@uuline{\leavevmode \bgroup
            \UL@setULdepth
            \ifx\UL@on\UL@onin \advance\ULdepth2.8\p@\fi
            \markoverwith{\textcolor{#1}{\lower\dimexpr\ULdepth+2pt\relax\hbox
                {\kern-.03em\vbox{\hrule width.2em\kern1\p@\hrule}\kern-.03em}}}%
        \ULon}%
        \temp@uuline{#2}%
    }
\newcommand{\coloruwave}[2]{%
        \UL@protected\def\temp@uwave{\leavevmode \bgroup 
        \ifdim \ULdepth=\maxdimen \ULdepth 3.5\p@
        \else \advance\ULdepth2\p@ 
        \fi \markoverwith{\textcolor{#1}{\lower\ULdepth\hbox{\sixly \char58}}}\ULon}
        \font\sixly=lasy6 %
        \temp@uwave{#2}%
    }
\makeatother

\definecolor{babyblue}{rgb}{0.54, 0.81, 0.94}
\definecolor{darkpastelblue}{rgb}{0.47, 0.62, 0.8}
\definecolor{glaucous}{rgb}{0.38, 0.51, 0.71}
  \definecolor{bittersweet}{rgb}{1.0, 0.44, 0.37}
  \definecolor{periwinkle}{rgb}{0.8, 0.8, 1.0}

\usepackage{setspace}
\usepackage{tikz}
\usetikzlibrary{decorations.pathmorphing}

\usepackage{cancel}

\newcommand{\W}{\gls{pi}}     %
\newcommand{\ag}[1]{\left[ #1\right]}
\newcommand{\pr}[1]{\left\{ #1\right\}}
\newcommand{\en}[1]{\left( #1\right)}

\newcommand{\myblue}[1]{%
  {\color[RGB]{4, 51, 255}#1}%
}

\renewcommand{\myblue}[1]{}

\newtheorem*{remark}{Remark}

\usepackage{tikz}
\usepackage{cutwin}
\usepackage[framemethod=TikZ]{mdframed}
\usetikzlibrary{arrows.meta,
                positioning,
                quotes,shapes,arrows,fit,calc,automata
                }
\usetikzlibrary{decorations.markings}

\preto\longtable{\setcounter{magicrownumbers}{0}}
\newcounter{magicrownumbers}

\usepackage{mathtools}

\newcommand{\bfF}{{\bf F}}%
\newcommand{\bfn}{{\bf n}}%
\newcommand{\bfT}{{\bf T}}%
\newcommand{\bfx}{{\bf x}}\newcommand{\bfX}{{\bf X}}%
\newcommand{\bfchi}{\boldsymbol{\chi}}%
\newfont{\tenbfit}{cmmib10}%
\newfont{\svnbfit}{cmmib8}%
\newfont{\tenbfsl}{cmbxti10}%
\newfont{\mmit}{cmmi10}%
\newfont{\smit}{cmmi9}%
\newfont{\bfMit}{cmmi5}%
\newfont{\tenbbb}{msbm10}%
\newfont{\svnbbb}{msbm8}%
\newfont{\tenssit}{cmssqi8 at 10pt}%
\newfont{\svnssit}{cmssqi8 at 7pt}%
\newfont{\gothic}{eufm10}%
\newfont{\sgothic}{eufm7}%
\newcommand{\Det}{\hbox{\rm det}\mskip2mu}
\newcommand{\cof}{\hbox{\rm cof}\mskip4mu}

\newcommand{\Blj}{\mbox{$\Big[\kern-0.275em\Big[$}}
\newcommand{\Brj}{\mbox{$\Big]\kern-0.275em\Big]$}}

\newcommand{\X}{\bfX}
\newcommand{\x}{\bfx}
\newcommand{\y}{\bfchi}
\newcommand{\F}{\bfF}

\newcommand{\n}{\bfn}
\renewcommand{\cof}{\scriptscriptstyle\sf{C}}

\newcommand{\grad}{\text{grad}\,}

\newcommand{\T}{\bfT}

\newcommand{\mR}{\bfm_{\mat}}

\renewcommand{\u}[1]{\boldsymbol{#1}}

\newcommand{\norm}[1]{\left\lVert #1\right\rVert}

\newcommand{\abs}[1]{\left\lvert #1\right\rvert}

\newcommand{\ShowComments}{false}
\newcommand{\forus}[1]{\ifthenelse{\equal{\ShowComments}{true}}{{\footnotesize \color{blue!90!black}~[[#1]]}}{}}

\usepackage[T3,T1]{fontenc}
\DeclareSymbolFont{tipa}{T3}{cmr}{m}{n}
\DeclareMathAccent{\invbreve}{\mathalpha}{tipa}{16}

\usepackage[eulergreek]{sansmath}

\newcommand{\tp}[1]{\mathsf{#1}}

\NewDocumentCommand{\cavityRO}{o}{
\IfNoValueTF{#1}
  {\mathcal{C}^{\circ}_{\rm{R}}}
  {{}^{#1}\mathcal{C}^{\circ}_{\rm{R}}}
}
\NewDocumentCommand{\cavityROc}{o}{
\IfNoValueTF{#1}
  {\widebar{\mathcal{C}^{\circ}_{\rm{R}}}}
  {{}^{#1}\widebar{\mathcal{C}^{\circ}_{\rm{R}}}}
}

\newcommand{\matRn}{\mathcal{M}^{\eta}_{\rm{R}}}

\NewDocumentCommand{\cavityDO}{o}{
\IfNoValueTF{#1}
  {\mathcal{C}^{\circ}_{\rm{D}}}
  {{}^{#1}\mathcal{C}^{\circ}_{\rm{D}}}
}
\NewDocumentCommand{\cavityDOc}{o}{
\IfNoValueTF{#1}
  {\widebar{\mathcal{C}^{\circ}_{\rm{D}}}}
  {{}^{#1}\widebar{\mathcal{C}^{\circ}_{\rm{D}}}}
}

\NewDocumentCommand{\bodyRn}{o}{
\IfNoValueTF{#1}
  {\mathcal{B}^\eta_{\rm{R}}}
  {\mathcal{B}^{\eta}_{\rm{R}(#1)}}
}

\NewDocumentCommand{\bodyRO}{o}{
\IfNoValueTF{#1}
  {\mathcal{B}^\circ_{\rm{R}}}
  {\mathcal{B}^{\circ}_{\rm{R}(#1)}}
}

\NewDocumentCommand{\bodyR}{o}{
\IfNoValueTF{#1}
  {\mathcal{B}_{\rm{R}}}
  {\mathcal{B}^{#1}_{\rm{R}}}
}

\NewDocumentCommand{\cavityRn}{o}{
\IfNoValueTF{#1}
  {\mathcal{C}^{\eta}_{\rm{R}}}
  {{}^{#1}\mathcal{C}^{\eta}_{\rm{R}}}
}
\NewDocumentCommand{\cavityRnc}{o}{
\IfNoValueTF{#1}
  {\widebar{\mathcal{C}^{\eta}_{\rm{R}}}}
  {{}^{#1}\widebar{\mathcal{C}^{\eta}_{\rm{R}}}}
}

\NewDocumentCommand{\cavityDn}{o}{
\IfNoValueTF{#1}
  {\mathcal{C}^{\eta}_{\rm{D}}}
  {{}^{#1}\mathcal{C}^{\eta}_{\rm{D}}}
}
\NewDocumentCommand{\cavityDnc}{o}{
\IfNoValueTF{#1}
  {\widebar{\mathcal{C}^{\eta}_{\rm{D}}}}
  {{}^{#1}\widebar{\mathcal{C}^{\eta}_{\rm{D}}}}
}

\RenewDocumentCommand{\n}{o}{
\IfNoValueTF{#1}
  {\tp{n}}
  {\tp{n}_{#1}}
}

\NewDocumentCommand{\nO}{o}{
\IfNoValueTF{#1}
  {\tp{n}^\circ}
  {\tp{n}^{\circ}_{#1}}
}
\NewDocumentCommand{\nn}{o}{
\IfNoValueTF{#1}
  {\tp{n}^\eta}
  {\tp{n}^{\eta}_{#1}}
}

\newcommand{\nna}{\nn[\alpha]}
\newcommand{\nOa}{\nO[\alpha]}

\newcommand{\UO}{\mathcal{U}^\circ}

\NewDocumentCommand{\nnc}{o}{
\IfNoValueTF{#1}
  {\check{\tp{n}}^\eta}
  {\check{\tp{n}}^{\eta}_{#1}}
}

\newcommand{\nnca}{\nnc[\alpha]}

\NewDocumentCommand{\Nh}{o}{
\IfNoValueTF{#1}
  {\widehat{\tp{N}}^\circ}
  {\widehat{\tp{N}}^\circ_{\cdot #1}}
}

\NewDocumentCommand{\Nnh}{o}{
\IfNoValueTF{#1}
  {\widehat{\tp{N}}^\eta}
  {\widehat{\tp{N}}_{\cdot #1}^\eta}
}

\NewDocumentCommand{\NnHh}{o}{
\IfNoValueTF{#1}
  {\widehat{\tp{N}}_{\mathcal{H}}^\eta}
  {\widehat{\tp{N}}^\eta_{\mathcal{H}\cdot #1}}
}

\NewDocumentCommand{\NnMh}{o}{
\IfNoValueTF{#1}
  {\widehat{\tp{N}}_{\mathcal{M}}^\eta}
  {\widehat{\tp{N}}^\eta_{\mathcal{M}\cdot #1}}
}

\NewDocumentCommand{\NnPh}{o}{
\IfNoValueTF{#1}
  {\widehat{\tp{N}}_{(p)}^\eta}
  {\widehat{\tp{N}}^\eta_{(p)\cdot #1}}
}

\NewDocumentCommand{\Nnrh}{o}{
\IfNoValueTF{#1}
  {\widehat{\tp{N}}_{(r)}^\eta}
  {\widehat{\tp{N}}^\eta_{(r)\cdot #1}}
}

\NewDocumentCommand{\NUh}{o}{
\IfNoValueTF{#1}
  {\widehat{\tp{N}}^\circ_{\mathcal{U}}}
  {\widehat{\tp{N}}^\circ_{\mathcal{U}\cdot #1}}
}

\NewDocumentCommand{\NVh}{o}{
\IfNoValueTF{#1}
  {\widehat{\tp{N}}^\circ_{\mathcal{V}}}
  {\widehat{\tp{N}}^\circ_{\mathcal{V}\cdot #1}}
}

\NewDocumentCommand{\NCh}{o}{
\IfNoValueTF{#1}
  {\widehat{\tp{N}}^\circ_{\mathcal{C}}}
  {\widehat{\tp{N}}^\circ_{\mathcal{C}\cdot #1}}
}

\NewDocumentCommand{\bd}{o}{%
\IfNoValueTF{#1}
  {\partial}
  {\partial #1}
}

\NewDocumentCommand{\clos}{o}{%
\IfNoValueTF{#1}
  {Error}
  {\overline{#1}}
}

\NewDocumentCommand{\GammagRO}{o}{%
\IfNoValueTF{#1}
  {\Gamma^{g, \circ}_{\textrm{R}}}
  {\Gamma^{g, \circ}_{\textrm{R}, #1}}
}
\newcommand{\GammagROa}{\Gamma^{g,\circ}_{\textrm{R}, \alpha}}

\NewDocumentCommand{\GammagROsn}{o}{%
\IfNoValueTF{#1}
  {\GammagRO[\n[i]]}
  {\GammagRO[\n[#1]]}
}

\NewDocumentCommand{\GammahRO}{o}{%
\IfNoValueTF{#1}
  {\Gamma^{h, \circ}_{\textrm{R}}}
  {\Gamma^{h, \circ}_{\textrm{R}, #1}}
}
\newcommand{\GammahROa}{\Gamma^{h,\circ}_{\textrm{R}, \alpha}}

\NewDocumentCommand{\GammahROsn}{o}{%
\IfNoValueTF{#1}
  {\GammahRO[\n[i]]}
  {\GammahRO[\n[#1]]}
}

\NewDocumentCommand{\GammagRn}{o}{%
\IfNoValueTF{#1}
  {\Gamma^{g, \eta}_{\textrm{R}}}
  {\Gamma^{g, \eta}_{\textrm{R}, #1}}
}

\NewDocumentCommand{\GammagR}{oo}{%
\IfNoValueTF{#1}
  {\Gamma^{g, \eta}_{\textrm{R}}}
  {\Gamma^{g, #1}_{\textrm{R}, #2}}
}

\newcommand{\GammaIn}{\Gamma^{\eta}_{\textrm{I}}}
\newcommand{\GammaIO}{\Gamma^{\circ}_{\textrm{I}}}

\newcommand{\GammaErn}{\Gamma^{\eta}_{(r),\textrm{E}}}

\NewDocumentCommand{\GammaECn}{o}{%
\IfNoValueTF{#1}
  {\Gamma^{\eta}_{(p),\textrm{E}}}
  {\Gamma^{\eta}_{(#1),\textrm{E}}}
}

\newcommand{\GammaErO}{\Gamma^{\circ}_{(r),\textrm{E}}}

\newcommand{\GammagRna}{\Gamma^{g,\eta}_{\textrm{R}, \alpha}}

\NewDocumentCommand{\GammagRnsn}{o}{%
\IfNoValueTF{#1}
  {\GammagRn[\n[i]]}
  {\GammagRn[\n[#1]]}
}

\NewDocumentCommand{\GammahRn}{o}{%
\IfNoValueTF{#1}
  {\Gamma^{h, \eta}_{\textrm{R}}}
  {\Gamma^{h, \eta}_{\textrm{R}, #1}}
}
\newcommand{\GammahRna}{\Gamma^{h,\eta}_{\textrm{R}, \alpha}}

\NewDocumentCommand{\GammahRnsn}{o}{%
\IfNoValueTF{#1}
  {\GammahRn[\n[i]]}
  {\GammahRn[\n[#1]]}
}

\RenewDocumentCommand{\u}{o}{%
\IfNoValueTF{#1}
  {\tp{u}}
  {\tp{u}_{\cdot #1}}
}

\NewDocumentCommand{\usn}{o}{%
\IfNoValueTF{#1}
  {\u[\n[i]]}
  {\u[\n[#1]]}
}

\NewDocumentCommand{\ubOroman}{o}{%
\IfNoValueTF{#1}
  {\overline{u}^{\hspace{0.1em}\circ}}
  {\overline{u}^\circ_{\cdot #1}}
}

\NewDocumentCommand{\ubnroman}{o}{%
\IfNoValueTF{#1}
  {\overline{u}^{\hspace{0.1em}\eta}}
  {\overline{u}^\eta_{\cdot #1}}
}

\NewDocumentCommand{\ubO}{o}{%
\IfNoValueTF{#1}
  {\overline{\tp{u}}^{\hspace{0.1em}\circ}}
  {\overline{\tp{u}}^\circ_{\cdot #1}}
}

\NewDocumentCommand{\ubOsn}{o}{%
\IfNoValueTF{#1}
  {\ubO[\n[i]]}
  {\ubO[\n[#1]]}
}

\NewDocumentCommand{\ub}{o}{%
\IfNoValueTF{#1}
  {\overline{\tp{u}}}
  {\overline{\tp{u}}_{\cdot #1}}
}

\NewDocumentCommand{\ubsn}{o}{%
\IfNoValueTF{#1}
  {\ub[\n[i]]}
  {\ub[\n[#1]]}
}

\NewDocumentCommand{\uh}{o}{%
\IfNoValueTF{#1}
  {\widehat{\tp{u}}}
  {\widehat{\tp{u}}_{\cdot #1}}
}

\NewDocumentCommand{\ut}{o}{%
\IfNoValueTF{#1}
  {\tilde{\tp{u}}}
  {\tilde{\tp{u}}_{\cdot#1}}
}

\NewDocumentCommand{\un}{o}{%
\IfNoValueTF{#1}
  {\tp{u}^\eta}
  {\tp{u}^{\eta}_{\cdot #1}}
}

\NewDocumentCommand{\ubn}{o}{%
\IfNoValueTF{#1}
  {\overline{\tp{u}}^\eta}
  {\overline{\tp{u}}^{\eta}_{\cdot #1}}
}

\newcommand{\OmegaO}{\Omega^\circ}

\NewDocumentCommand{\uhn}{o}{%
\IfNoValueTF{#1}
  {\hat{\tp{u}}^\eta}
  {\hat{u}^{\eta}_{#1}}
}

\NewDocumentCommand{\ubnc}{o}{%
\IfNoValueTF{#1}
  {\check{\overline{\tp{u}}}^\eta}
  {\check{\overline{\tp{u}}}^\eta_{\cdot #1}}
}

\NewDocumentCommand{\ubc}{o}{%
\IfNoValueTF{#1}
  {\check{\overline{\tp{u}}}}
  {\check{\overline{\tp{u}}}^{#1}}
}

\RenewDocumentCommand{\grad}{o}{%
\IfNoValueTF{#1}
{\nabla}
{\nabla #1}
}

\RenewDocumentCommand{\T}{o}{%
\IfNoValueTF{#1}
  {\tp{T}}
  {\tp{T}_{\cdot #1}}
}

\NewDocumentCommand{\TbO}{o}{%
\IfNoValueTF{#1}
  {\overline{\rule{0pt}{2.7mm}\tp{T}}^\circ}
  {\overline{\rule{0pt}{2.7mm}\tp{T}}^{\circ}_{\cdot #1}}
}

\NewDocumentCommand{\Th}{o}{%
\IfNoValueTF{#1}
  {\widehat{\tp{T}}}
  {\widehat{\tp{T}}_{\cdot #1}}
}

\NewDocumentCommand{\TtO}{o}{%
\IfNoValueTF{#1}
  {\widetilde{\tp{T}}^\circ}
  {\widetilde{\tp{T}}^{\circ}_{\cdot #1}}
}

\NewDocumentCommand{\Tte}{o}{%
\IfNoValueTF{#1}
  {\widetilde{\tp{T}}^\epsilon}
  {\widetilde{\tp{T}}^{\epsilon}_{\cdot #1}}
}

\NewDocumentCommand{\Tt}{o}{%
\IfNoValueTF{#1}
  {\widetilde{\tp{T}}}
  {\widetilde{\tp{T}}_{\cdot #1}}
}

 \NewDocumentCommand{\TbOH}{o}{%
\IfNoValueTF{#1}
  {\overline{\tp{T}}^\circ_{\mathcal{H}}}
  {\overline{\tp{T}}^{\circ}_{\mathcal{H}\cdot #1}}
}

 \NewDocumentCommand{\TbOP}{o}{%
\IfNoValueTF{#1}
  {\overline{\tp{T}}^\circ_{(p)}}
  {\overline{\tp{T}}^{\circ}_{(p)\cdot #1}}
}

 \NewDocumentCommand{\TbOr}{o}{%
\IfNoValueTF{#1}
  {\overline{\tp{T}}^\circ_{(r)}}
  {\overline{\tp{T}}^{\circ}_{(r)\cdot #1}}
}

 \NewDocumentCommand{\TbOR}{o}{%
\IfNoValueTF{#1}
  {\overline{\tp{T}}^\circ_{\mathcal{R}}}
  {\overline{\tp{T}}^{\circ}_{\mathcal{R}\cdot #1}}
}

\NewDocumentCommand{\TbOM}{o}{%
\IfNoValueTF{#1}
  {\overline{\tp{T}}^\circ_{\mathcal{M}}}
  {\overline{\tp{T}}^{\circ}_{\mathcal{M}\cdot #1}}
}

\NewDocumentCommand{\TbOB}{o}{%
\IfNoValueTF{#1}
  {\overline{\tp{T}}^\circ_{\mathcal{B}}}
  {\overline{\tp{T}}^{\circ}_{\mathcal{B}\cdot #1}}
}

\NewDocumentCommand{\TbnH}{o}{%
\IfNoValueTF{#1}
  {\overline{\tp{T}}^\eta_{\mathcal{H}}}
  {\overline{\tp{T}}^{\eta}_{\mathcal{H}\cdot #1}}
}

\NewDocumentCommand{\TbnM}{o}{%
\IfNoValueTF{#1}
  {\overline{\tp{T}}^\eta_{\mathcal{M}}}
  {\overline{\tp{T}}^{\eta}_{\mathcal{M}\cdot #1}}
}

\NewDocumentCommand{\TbnP}{o}{%
\IfNoValueTF{#1}
  {\overline{\tp{T}}^\eta_{(p)}}
  {\overline{\tp{T}}^{\eta}_{(p)\cdot #1}}
}

\NewDocumentCommand{\Tbnr}{o}{%
\IfNoValueTF{#1}
  {\overline{\tp{T}}^\eta_{(r)}}
  {\overline{\tp{T}}^{\eta}_{(r)\cdot #1}}
}

\NewDocumentCommand{\TbnC}{o o}{%
\IfNoValueTF{#2}
  {\overline{\tp{T}}^\eta_{(#1)}}
  {\overline{\tp{T}}^{\eta}_{(#1)\cdot #2}}
}

\NewDocumentCommand{\TbOC}{o o}{%
\IfNoValueTF{#2}
  {\overline{\tp{T}}^\circ_{(#1)}}
  {\overline{\tp{T}}^{\circ}_{(#1)\cdot #2}}
}

\NewDocumentCommand{\TbnB}{o}{%
\IfNoValueTF{#1}
  {\overline{\tp{T}}^\eta_{\mathcal{B}}}
  {\overline{\tp{T}}^{\eta}_{\mathcal{B}\cdot #1}}
}

\NewDocumentCommand{\Tbn}{o}{%
\IfNoValueTF{#1}
  {\overline{\tp{T}}^\eta}
  {\overline{\tp{T}}^{\eta}_{\cdot #1}}
}

\NewDocumentCommand{\Tbnc}{o}{%
\IfNoValueTF{#1}
  {\check{\overline{\tp{T}}}^\eta}
  {\check{\overline{\tp{T}}}^\eta_{\cdot #1}}
}

\NewDocumentCommand{\Thn}{o}{%
\IfNoValueTF{#1}
  {\widehat{\tp{T}}^\eta}
  {\widehat{\tp{T}}^{\eta}_{\cdot #1}}
}

\NewDocumentCommand{\Ttn}{o}{%
\IfNoValueTF{#1}
  {\widetilde{\tp{T}}^\eta}
  {\widetilde{\tp{T}}^{\eta}_{\cdot #1}}
}

\NewDocumentCommand{\Thnc}{o}{%
\IfNoValueTF{#1}
  {\check{\widehat{\tp{T}}}^\eta}
  {\check{\widehat{\tp{T}}}^\eta_{\cdot #1}}
}

\NewDocumentCommand{\Ttnc}{o}{%
\IfNoValueTF{#1}
  {\check{\widetilde{\tp{T}}}^\eta}
  {\check{\widetilde{\tp{T}}}^\eta_{\cdot #1}}
}

\NewDocumentCommand{\Ttc}{o}{%
\IfNoValueTF{#1}
  {\check{\widetilde{\tp{T}}}}
  {\check{\widetilde{\tp{T}}}^{#1}}
}

\NewDocumentCommand{\Ttncsn}{o}{%
\IfNoValueTF{#1}
{\Ttnc[\n[i]]}
{\Ttnc[\n[#1]]}
}

\NewDocumentCommand{\Ttncsnn}{o}{%
\IfNoValueTF{#1}
{\Ttnc[\nn[i]]}
{\Ttnc[\nn[#1]]}
}

\RenewDocumentCommand{\x}{o}{%
\IfNoValueTF{#1}
  {\tp{x}}
  {\tp{x}_{\cdot #1}}
}

\NewDocumentCommand{\z}{o}{%
\IfNoValueTF{#1}
{\tp{z}}
{\tp{z}_{\cdot #1}}
}

\RenewDocumentCommand{\y}{o}{%
\IfNoValueTF{#1}
  {\tp{y}}
  {\tp{y}_{\cdot #1}}
}

\NewDocumentCommand{\yn}{o}{%
\IfNoValueTF{#1}
  {\tp{y}^\eta}
  {\tp{y}^{\eta}_{\cdot #1}}
}

\NewDocumentCommand{\varphiO}{o}{%
\IfNoValueTF{#1}
  {\hspace{0.1em}{\sansmath{\varphi}}^{\circ}}
  {\hspace{0.1em}{\sansmath{\varphi}}^{\circ}_{\cdot #1}}
}

\NewDocumentCommand{\varphin}{o}{%
\IfNoValueTF{#1}
  {\hspace{0.1em}{\sansmath{\varphi}}^{\eta}}
  {\hspace{0.1em}{\sansmath{\varphi}}^{\eta}_{\cdot #1}}
}

\RenewDocumentCommand{\F}{o}{%
\IfNoValueTF{#1}
  {\tp{F}}
  {\tp{F}_{\cdot #1}}
}

\NewDocumentCommand{\xnb}{o}{%
\IfNoValueTF{#1}
  {\breve{\tp{x}}^\eta}
  {\breve{\tp{x}}^\eta_{\cdot #1}}
}

\NewDocumentCommand{\dOn}{o}{%
\IfNoValueTF{#1}
  {\tp{d}^{\circ\to\eta}}
  {\tp{d}^{\circ\to\eta}_{\cdot #1}}
}

\NewDocumentCommand{\dOO}{o}{%
\IfNoValueTF{#1}
  {\tp{d}^{\circ\to\circ}}
  {\tp{d}^{\circ\to\circ}_{\cdot #1}}
}

\NewDocumentCommand{\dOnc}{o}{%
\IfNoValueTF{#1}
  {\check{\tp{d}}^{\circ\to\eta}}
  {\check{\tp{d}}^{\circ\to\eta}_{\cdot #1}}
}

\NewDocumentCommand{\G}{o}{%
\IfNoValueTF{#1}
  {G^\eta}
  {G^\eta_{#1}}
}

\NewDocumentCommand{\Gn}{o}{%
\IfNoValueTF{#1}
  {G^\eta}
  {G^\eta_{#1}}
}

\NewDocumentCommand{\Gs}{o}{%
\IfNoValueTF{#1}
{G^\eta_{\rm{s}}}
  {G^\eta_{\rm{s}}}
}

\NewDocumentCommand{\Gns}{o}{%
\IfNoValueTF{#1}
{G^\eta_{\rm{s}}}
  {G^\eta_{\rm{s}}}
}

\NewDocumentCommand{\SigmaC}{o o}{%
\IfNoValueTF{#1}
  {\hspace{0.1em}{\mathsf{\Sigma}}}
  {\hspace{0.1em}{\mathsf{\Sigma}}_{\cdot #1\cdot #2}}
}

\NewDocumentCommand{\SigmaO}{o o}{%
\IfNoValueTF{#1}
  {\hspace{0.1em}{\mathsf{\Sigma}^{\circ}}}
  {\hspace{0.1em}{\mathsf{\Sigma}^{\circ}}_{\cdot #1 \cdot #2}}
}

\NewDocumentCommand{\SigmaOH}{o o}{%
\IfNoValueTF{#1}
  {\hspace{0.1em}{\mathsf{\Sigma}_{\mathcal{H}}^{\circ}}}
  {\hspace{0.1em}{\mathsf{\Sigma}_{\mathcal{H}\cdot #1 \cdot #2}^{\circ}}}
}

\NewDocumentCommand{\SigmaOP}{o o}{%
\IfNoValueTF{#1}
  {\hspace{0.1em}{\mathsf{\Sigma}_{(p)}^{\circ}}}
  {\hspace{0.1em}{\mathsf{\Sigma}_{(p)\cdot #1 \cdot #2}^{\circ}}}
}

\NewDocumentCommand{\SigmaOr}{o o}{%
\IfNoValueTF{#1}
  {\hspace{0.1em}{\mathsf{\Sigma}_{(r)}^{\circ}}}
  {\hspace{0.1em}{\mathsf{\Sigma}_{(r)\cdot #1 \cdot #2}^{\circ}}}
}

\NewDocumentCommand{\SigmaOR}{o o}{%
\IfNoValueTF{#1}
  {\hspace{0.1em}{\mathsf{\Sigma}_{\mathcal{R}}^{\circ}}}
  {\hspace{0.1em}{\mathsf{\Sigma}_{\mathcal{R}\cdot #1 \cdot #2}^{\circ}}}
}

\NewDocumentCommand{\SigmaOM}{o o}{%
\IfNoValueTF{#1}
  {\hspace{0.1em}{\mathsf{\Sigma}_{\mathcal{M}}^{\circ}}}
  {\hspace{0.1em}{\mathsf{\Sigma}_{\mathcal{M}\cdot #1 \cdot #2}^{\circ}}}
}

\newcommand{\zetabO}{\overline{\zeta}^\circ}

\newcommand{\zetabOr}{\overline{\zeta}_{(r)}^\circ}

\newcommand{\zerovec}{\textsf{0}}
\newcommand{\zeroMat}{\tp{O}}

\newcommand{\zetaO}{\hspace{0.1em}{\sansmath{\zeta}}^\circ}

\newcommand{\zetaOr}{\hspace{0.1em}{\sansmath{\zeta}}^\circ_{(r)}}

\NewDocumentCommand{\zetaOC}{o}{%
\IfNoValueTF{#1}
  {\hspace{0.1em}{\sansmath{\zeta}}^\circ}
  {\hspace{0.1em}{\sansmath{\zeta}}^\circ_{(#1)}}
}

\newcommand{\AC}{\mathcal{A}}
\newcommand{\An}{\mathcal{A}^\eta}
\newcommand{\Anc}{\check{\mathcal{A}}^\eta}
\newcommand{\AO}{\mathcal{A}^\circ}

\NewDocumentCommand{\phiC}{o}{%
\IfNoValueTF{#1}
  {\hspace{0.1em}{\sansmath{\phi}}}
  {\hspace{0.1em}{\sansmath{\phi}}_{\cdot #1}}
}

\NewDocumentCommand{\phiO}{o}{%
\IfNoValueTF{#1}
  {\hspace{0.1em}{\sansmath{\phi}}^\circ}
  {\hspace{0.1em}{\sansmath{\phi}}^{\circ}_{\cdot #1}}
}

\NewDocumentCommand{\phiOt}{o}{%
\IfNoValueTF{#1}
  {\hspace{0.1em}{\widetilde{\sansmath{\phi}}}^\circ}
  {\hspace{0.1em}{\widetilde{\sansmath{\phi}}}^{\circ}_{\cdot #1}}
}

\newcommand{\baseval}{\eta^\circ}

\NewDocumentCommand{\vu}{o}{%
\IfNoValueTF{#1}
  {\hspace{0.1em}\delta \tp{u}}
  {\hspace{0.1em}\delta \tp{u}_{\cdot #1}}
}

\NewDocumentCommand{\psiC}{o}{%
\IfNoValueTF{#1}
  {\hspace{0.1em}{\sansmath{\psi}}}
  {\hspace{0.1em}{\sansmath{\psi}}_{\cdot #1}}
}

\NewDocumentCommand{\psiO}{o}{%
\IfNoValueTF{#1}
  {\hspace{0.1em}{\sansmath{\psi}^\circ}}
  {\hspace{0.1em}{\sansmath{\psi}^\circ}_{\cdot #1}}
}

\NewDocumentCommand{\psie}{o}{%
\IfNoValueTF{#1}
  {\hspace{0.1em}{\sansmath{\psi}^\epsilon}}
  {\hspace{0.1em}{\sansmath{\psi}^\epsilon}_{\cdot #1}}
}

\NewDocumentCommand{\chiC}{o}{%
\IfNoValueTF{#1}
  {\hspace{0.1em}{\sansmath{\chi}}}
  {\hspace{0.1em}{\sansmath{\chi}}_{\cdot #1}}
}

\NewDocumentCommand{\chiO}{o}{%
\IfNoValueTF{#1}
  {\hspace{0.1em}{\sansmath{\chi}^\circ}}
  {\hspace{0.1em}{\sansmath{\chi}^\circ}_{\cdot #1}}
}

\NewDocumentCommand{\chie}{o}{%
\IfNoValueTF{#1}
  {\hspace{0.1em}{\sansmath{\chi}^\epsilon}}
  {\hspace{0.1em}{\sansmath{\chi}^\epsilon}_{\cdot #1}}
}

\NewDocumentCommand{\f}{o}{%
\IfNoValueTF{#1}
  {f}
  {f_{#1}}
}

\NewDocumentCommand{\fn}{o}{%
\IfNoValueTF{#1}
  {f^\eta}
  {f^\eta_{#1}}
}
\NewDocumentCommand{\fnC}{o o}{%
\IfNoValueTF{#2}
  {f^\eta_{(#1)}}
  {f^\eta_{(#1),#2}}
}

\NewDocumentCommand{\fnH}{o}{%
\IfNoValueTF{#1}
  {f^\eta_{\mathcal{H}}}
  {f^\eta_{\mathcal{H} #1}}
}

\NewDocumentCommand{\fnM}{o}{%
\IfNoValueTF{#1}
  {f^\eta_{\mathcal{M}}}
  {f^\eta_{\mathcal{M} #1}}
}

\NewDocumentCommand{\fnP}{o}{%
\IfNoValueTF{#1}
  {f^\eta_{(p)}}
  {f^\eta_{(p) #1}}
}

\NewDocumentCommand{\fnr}{o}{%
\IfNoValueTF{#1}
  {f^\eta_{(r)}}
  {f^\eta_{(r) #1}}
}

\NewDocumentCommand{\fnB}{o}{%
\IfNoValueTF{#1}
  {f^\eta_{\mathcal{B}}}
  {f^\eta_{\mathcal{B} #1}}
}

\NewDocumentCommand{\PiC}{o}{%
\IfNoValueTF{#1}
  {\Pi}
  {\Pi^{\hspace{0.1em}#1}}
}

\NewDocumentCommand{\Pib}{o}{%
\IfNoValueTF{#1}
  {\overline{\Pi}}
  {\overline{\Pi}^{#1}}
}

\NewDocumentCommand{\vPi}{o}{%
\IfNoValueTF{#1}
  {\delta\Pi}
  {\delta\Pi^{\hspace{0.1em}#1}}
}

\newcommand{\SEDensity}{\mathscr{W}}
\newcommand{\bodyforcepotential}{\Phi}

\renewcommand{\W}{\SEDensity}
\newcommand{\PhiC}{\bodyforcepotential}

\NewDocumentCommand{\FP}{o}{%
\IfNoValueTF{#1}
  {\overline{\tp{P}}}
  {\overline{\tp{P}}_{\cdot #1}}
}

\NewDocumentCommand{\FPO}{o o}{%
\IfNoValueTF{#1}
  {\overline{\tp{P}}^\circ}
  {\overline{\tp{P}}^\circ_{\cdot #1 \cdot #2}}
}

\NewDocumentCommand{\FPn}{o o}{%
\IfNoValueTF{#1}
  {\overline{\tp{P}}^\eta}
  {\overline{\tp{P}}^\eta_{\cdot #1 \cdot #2}}
}

\newcommand{\an}{\alpha^\eta}
\newcommand{\aO}{\alpha^\circ}

\NewDocumentCommand{\FPnH}{o o}{%
\IfNoValueTF{#1}
  {\overline{\tp{P}}_{\mathcal{H}}^\eta}
  {\overline{\tp{P}}^\eta_{\mathcal{H}\cdot #1 \cdot #2}}
}

\NewDocumentCommand{\FPnM}{o o}{%
\IfNoValueTF{#1}
  {\overline{\tp{P}}_{\mathcal{M}}^\eta}
  {\overline{\tp{P}}^\eta_{\mathcal{M}\cdot #1 \cdot #2}}
}

\NewDocumentCommand{\FPnP}{o o}{%
\IfNoValueTF{#1}
  {\overline{\tp{P}}_{(p)}^\eta}
  {\overline{\tp{P}}^\eta_{(p)\cdot #1 \cdot #2}}
}

\NewDocumentCommand{\FPnr}{o o}{%
\IfNoValueTF{#1}
  {\overline{\tp{P}}_{(r)}^\eta}
  {\overline{\tp{P}}^\eta_{(r)\cdot #1 \cdot #2}}
}

\NewDocumentCommand{\AS}{o}{%
\IfNoValueTF{#1}
  {\mathcal{A}}
  {\mathcal{A}^{#1}}
}

\NewDocumentCommand{\e}{o}{%
\IfNoValueTF{#1}
  {\tp{e}}
  {\tp{e}_{#1}}
}

\NewDocumentCommand{\E}{o}{%
\IfNoValueTF{#1}
  {\tp{E}}
  {\tp{E}_{#1}}
}

\NewDocumentCommand{\Reals}{o}{%
\IfNoValueTF{#1}
  {\mathbb{R}}
  {\mathbb{R}^{#1}}
}

\NewDocumentCommand{\Matrix}{o}{%
\IfNoValueTF{#1}
  {M}
  {M^{#1}}
}
\newcommand{\dV}{\,dV}
\newcommand{\dS}{\,dS}
\NewDocumentCommand{\Bv}{o}{%
\IfNoValueTF{#1}
  {\Big\vert}
  {\Big\vert_{#1}}
}

\RenewDocumentCommand{\H}{o o}{%
\IfNoValueTF{#1}
  {\tp{H}}
  {\tp{H}_{\cdot #1\cdot #2}}
}
\newcommand{\Id}{\tp{I}}

\RenewDocumentCommand{\cof}{o}{%
\IfNoValueTF{#1}
  {\mathrm{Cof}}
  {\mathrm{Cof}\ag{#1}}
}

\RenewDocumentCommand{\det}{o}{%
\IfNoValueTF{#1}
  {\mathrm{Det}}
  {\mathrm{Det}\ag{#1}}
}

\NewDocumentCommand{\Divs}{o}{%
\IfNoValueTF{#1}
  {\mathrm{Div}_{\mathrm{s}}}
  {\mathrm{Div}_{\mathrm{s}}\ag{#1}}
}

\NewDocumentCommand{\DivC}{o}{%
\IfNoValueTF{#1}
  {\mathrm{Div}}
  {\mathrm{Div}\ag{#1}}
}

\NewDocumentCommand{\comp}{o}{%
\IfNoValueTF{#1}
  {\circ}
  {\circ #1}
}

\NewDocumentCommand{\oC}{o}{%
\IfNoValueTF{#1}
  {\mathcal{o}}
  {\mathcal{o}\en{#1}}
}

\NewDocumentCommand{\xnbi}{o}{%
\IfNoValueTF{#1}
  {\invbreve{\tp{x}}^\eta}
  {\invbreve{\tp{x}}^\eta_{\cdot #1}}
}

\NewDocumentCommand{\psib}{o}{%
\IfNoValueTF{#1}
   {\hspace{0.1em}\overline{{\sansmath{\psi}}}}
   {\hspace{0.1em}\overline{{\sansmath{\psi}}}_{\cdot #1}}
}

\NewDocumentCommand{\tauC}{o}{%
\IfNoValueTF{#1}
   {\hspace{0.1em}{\sansmath{\tau}}}
   {\hspace{0.1em}{\sansmath{\tau}}_{\cdot #1}}
}

\NewDocumentCommand{\bC}{o}{%
\IfNoValueTF{#1}
  {\overline{\tp{b}}}
  {\overline{\tp{b}}_{\cdot #1}}
}
\NewDocumentCommand{\bO}{o}{%
\IfNoValueTF{#1}
  {\overline{\tp{b}}^\circ}
  {\overline{\tp{b}}^\circ_{\cdot #1}}
}
\NewDocumentCommand{\bn}{o}{%
\IfNoValueTF{#1}
  {\overline{\tp{b}}^\eta}
  {\overline{\tp{b}}^\eta_{\cdot #1}}
}

\NewDocumentCommand{\bCC}{o o}{%
\IfNoValueTF{#2}
  {\overline{\tp{b}}_{(#1)}}
  {\overline{\tp{b}}_{(#1)\cdot #2}}
}

\NewDocumentCommand{\bH}{o}{%
\IfNoValueTF{#1}
  {\tp{b}_{\mathcal{H}}}
  {\tp{b}_{\mathcal{H}\cdot #1}}
}

\NewDocumentCommand{\bP}{o}{%
\IfNoValueTF{#1}
  {\tp{b}_{(p)}}
  {\tp{b}_{(p)\cdot #1}}
}

\NewDocumentCommand{\br}{o}{%
\IfNoValueTF{#1}
  {\overline{\tp{b}}^\eta_{(r)}}
  {\overline{\tp{b}}^\eta_{(r)\cdot #1}}
}

\NewDocumentCommand{\bM}{o}{%
\IfNoValueTF{#1}
  {\tp{b}_{\mathcal{M}}}
  {\tp{b}_{\mathcal{M}\cdot #1}}
}

\newcommand{\Wn}{\W^{\eta}}

\newcommand{\Phin}{\PhiC^{\eta}}
\newcommand{\PiO}{\PiC^{\hspace{0.1em}\circ}}
\newcommand{\PibO}{\Pib^{\hspace{0.1em}\circ}}
\newcommand{\Pin}{\PiC^{\eta}}
\newcommand{\Pibn}{\Pib^{\eta}}

\newcommand{\vPin}{\vPi^{\hspace{0.1em}\eta}}

\NewDocumentCommand{\uhnc}{o}{%
\IfNoValueTF{#1}
  {\widecheck{\widehat{\tp{u}}}^\eta}
  {\widecheck{\widehat{u}}^\eta_{#1}}
}

\NewDocumentCommand{\psihO}{o}{%
\IfNoValueTF{#1}
  {\widehat{\psi}^{\circ}}
  {\widehat{\psi}^{\circ}_{#1}}
}

\newcommand{\psihOa}{\widehat{\psi}^{\circ}_{\alpha}}

\NewDocumentCommand{\uhO}{o}{%
\IfNoValueTF{#1}
  {\widehat{\tp{u}}^\circ}
  {\widehat{u}^\circ_{ #1}}
}

\NewDocumentCommand{\uhc}{o}{%
\IfNoValueTF{#1}
  {\widecheck{\widehat{\tp{u}}}}
  {\widecheck{\widehat{\tp{u}}}_{\cdot #1}}
}

\NewDocumentCommand{\uc}{o}{%
\IfNoValueTF{#1}
  {\widecheck{\tp{u}}}
  {\widecheck{\tp{u}}_{\cdot #1}}
}

\NewDocumentCommand{\gammaC}{o}{%
\IfNoValueTF{#1}
  {\hspace{0.1em}{\sansmath{\gamma}}}
  {\hspace{0.1em}\gamma_{#1}}
}

\NewDocumentCommand{\fO}{o}{%
\IfNoValueTF{#1}
  {f^\circ}
  {f^\circ_{#1}}
}
\NewDocumentCommand{\fOC}{o o}{%
\IfNoValueTF{#2}
  {f^\circ_{(#1)}}
  {f^\circ_{(#1),#2}}
}

\NewDocumentCommand{\fOH}{o}{%
\IfNoValueTF{#1}
  {f^\circ_{\mathcal{H}}}
  {f^\circ_{\mathcal{H}#1}}
}

\NewDocumentCommand{\fOB}{o}{%
\IfNoValueTF{#1}
  {f^\circ_{\mathcal{B}}}
  {f^\circ_{\mathcal{B}#1}}
}

\NewDocumentCommand{\fOM}{o}{%
\IfNoValueTF{#1}
  {f^\circ_{\mathcal{M}}}
  {f^\circ_{\mathcal{M}#1}}
}

\NewDocumentCommand{\fOP}{o}{%
\IfNoValueTF{#1}
  {f^\circ_{(p)}}
  {f^\circ_{(p)#1}}
}

\NewDocumentCommand{\fOr}{o}{%
\IfNoValueTF{#1}
  {f^\circ_{(r)}}
  {f^\circ_{(r)#1}}
}

\newcommand{\Xn}{\mathsf{X}^\eta}
\newcommand{\XO}{\mathsf{X}^\circ}

\RenewDocumentCommand{\X}{o}{%
\IfNoValueTF{#1}
  {\mathsf{X}}
  {\mathsf{X}_{\cdot #1}}
}

\NewDocumentCommand{\fbO}{o}{%
\IfNoValueTF{#1}
  {\bar{f}^\circ}
  {\bar{f}^\circ_{#1}}
}

\newcommand{\prr}{potential energy release rate~}

\NewDocumentCommand{\argsO}{m}{\x,\ubO\ag{\x}, \grad \ubO}

\newcommand{\mL}{m_{\rm{L}}}
\renewcommand{\mR}{m_{\rm{R}}}
\newcommand{\hL}{h_{\rm{L}}}
\newcommand{\hR}{h_{\rm{R}}}

\begin{document}

\makeatother
\setstretch{1.3}

\newpage
\title{Potential energy release rate for solids with bounded domains during evolutions involving changes to boundaries and boundary data}
\author{%
Mrityunjay Kothari\textsuperscript{1}, Sayaka Kochiyama\textsuperscript{2}, and Haneesh Kesari\textsuperscript{2,*}\\[0.75em]
\small\textsuperscript{1}Department of Mechanical Engineering and Integrated Applied Mathematics,\\
\small University of New Hampshire, Durham, New Hampshire 03824, USA\\
\small\textsuperscript{2}School of Engineering, Brown University, Providence, Rhode Island 02912, USA\\
\small\textsuperscript{*}Corresponding author: \href{mailto:Haneesh_Kesari@brown.edu}{\texttt{Haneesh\_Kesari@brown.edu}}
}
\date{}
\maketitle

\begin{abstract}
We investigate the rate of change of equilibrium potential energy for nonlinear elastic solids undergoing evolutions that may involve changes in body geometry, material properties, applied tractions, and prescribed displacements. We represent such an evolution by a one-parameter family of variational problems and identify a class of these families for which the potential-energy release rate is well defined. Particular attention is given to the formulation of the potential energy at each evolution state: each constituent variational problem must yield the appropriate equilibrium state, while its equilibrium value must also correctly represent the physical change in potential energy between states. Within this framework, we derive a closed-form expression for the potential-energy release rate for solids with bounded domains, allowing for spatially nonuniform material evolution, sharp material interfaces, and evolving boundary geometry and prescribed boundary data. The expression is determined entirely by the equilibrium solution at the state under consideration and by the quantities prescribing the local evolution, so neighboring equilibrium problems need not be solved. Finally, we apply the result to the extension of a notch in a two-dimensional elastic body and show that the resulting potential-energy release rate reduces to the classical $J$-integral.
\end{abstract}

\section{Introduction}
\label{sec:overview}

In this article, we investigate the rate of change of potential energy as a conservative mechanical system evolves. The mechanical system we focus on consists of a nonlinear elastic body. When we say that a mechanical system evolves, we consider that the geometry of the body, its material properties, applied traction, and applied displacement fields can all evolve. For concreteness, the reader may picture, as a simple example of geometric evolution, a configurational change such as the migration of an inclusion or cavity. We restrict attention to those evolutions in which the body remains in an equilibrium state as the evolution takes place. 

We formulate such an evolution within the following variational framework. Specifically, we model the evolution by a one-parameter family of nonlinear variational problems, indexed by $\eta$.  This is to say that at each $\eta$, the mechanical system is represented by a nonlinear variational problem, whose equilibrium state is characterized by a local minimizer\footnote{Conditions guaranteeing the existence of such local minimizers are beyond the scope of this work. We therefore assume their existence without imposing additional conditions on the potential energy functional or the admissible set for the purpose of establishing existence.} $\ubn\ag{\cdot}$ of the potential energy functional $\Pin\ag{\cdot}$ over the set of admissible displacement fields $\An$. The rate of change of equilibrium potential energy, or more precisely its negative, that we wish to investigate is then the quantity
\begin{equation}
F_c^\eta=-\lim_{\epsilon\to0}\frac{\Pi^{\eta+\epsilon}\ag{\overline{\tp{u}}^{\eta+\epsilon}}-\Pin[\ubn]}{\epsilon}.
\end{equation}
Henceforth, we will also refer to $F_c^\eta$ as the \textit{\prr}. In what follows, we consider this rate at an arbitrary state $\eta=\baseval$ and denote its value there by $F_c^\circ$.

The different ways in which the mechanical system can evolve within our framework are made explicit through the $\eta$-indexed quantities entering $\Pin\ag{\cdot}$ and $\An$. 
Briefly, the potential energy functional is given by

\begin{subequations}
\label{eq:VE}
\begin{align}
\Pin\ag{\u}&=
\int_{\bodyRn}
\fn\ag{\z,\u\ag{\z},\grad\u\ag{\z}}\, dV
-
\int_{\bd\bodyRn}\Ttn\ag{\z}\cdot\en{\dOn\ag{\z}+\u\ag{\z}}\,d S,
\label{eq:PibnDef}
\end{align}
where $\bodyRn$ is the domain of the elastic body, $\fn\ag{\cdot}$ is the generalized potential energy density function, $\bd\bodyRn$ is the boundary of the body, and $\Ttn\ag{\cdot}$ is the applied traction field. The unfamiliar looking $\dOn\ag{\cdot}$ is the \textit{geometric displacement field} relative to the $\baseval$-state;  its precise definition is given in \S\ref{subsec:domainevol}. For a fixed $p>m$, the admissible set is taken to be 
\begin{equation}
\An=\pr{\u\ag{\cdot}\in W^{1,p}(\bodyRn;\Reals[m])~~\vert~~\u\ag{\z}\cdot\nn[\alpha]\ag{\z}=\uhn[\alpha]\ag{\z} 
\quad\text{for a.e. }\z\in\GammagRna,\quad
\alpha=1,\ldots,m}\footnote{Here and throughout, $W^{1,p}(\Omega;\Reals[m])$ denotes the usual Sobolev space of $\Reals[m]$-valued fields whose components and first weak derivatives belong to $L^p(\Omega)$. Boundary values of Sobolev functions are understood in the trace sense.}.
\label{eq:dispcondition}
\end{equation}
\end{subequations}
In \eqref{eq:dispcondition}, $\GammagRna$ is the portion of the body's boundary on which the displacement is prescribed in the local direction $\nna\ag{\cdot}$, and $\uhn[\alpha]\ag{\cdot}$ is the corresponding prescribed displacement field.

We refer to the one-parameter family of variational problems defined by \eqref{eq:VE} as a variational evolution. Within this framework, our main contributions are threefold: (i) to identify a class of variational evolutions for which $F_c^\circ$ is well defined; (ii) to formulate the variational evolution so that each constituent problem yields the appropriate equilibrium state, while the resulting equilibrium values correctly represent the physical change in potential energy across evolution states; and (iii) to derive a closed-form expression for $F_c^\eta$. We develop the first contribution in \S\ref{sec:VE}, explain the rationale for the second in \S\ref{sec:evolutionarilyconsistent}, and derive the third in \S\ref{sec:PERR}. Section~\ref{sec:preliminaries} provides the preliminary results used in the derivation, while in \S\ref{sec:Demos} we demonstrate the utility of the derived expression by applying it to a notch-extension problem and recovering the classical $J$-integral.

\section{Variational evolution}
\label{sec:VE}

In this section, we identify a class of variational evolutions for which the \prr $F_c^\circ$ is well defined. As briefly stated in \S\ref{sec:overview}, we allow for the geometry of the body, its material properties, and the applied traction and displacement fields to all evolve.  We begin by giving a physical picture of the range of evolutions accommodated by this class. 
In \S\ref{subsec:domainevol}--\S\ref{subsec:dispevol} to follow, we then make this picture mathematically precise by specifying the conditions placed on the $\eta$-indexed quantities defining the variational problem \eqref{eq:VE}, and how these quantities may vary with $\eta$. We further describe the evolution of the resulting equilibrium displacement field in \S\ref{subsec:equilDispevol}.

\begin{figure}[htbp]
\centering
\includegraphics[draft=false,width=\textwidth]{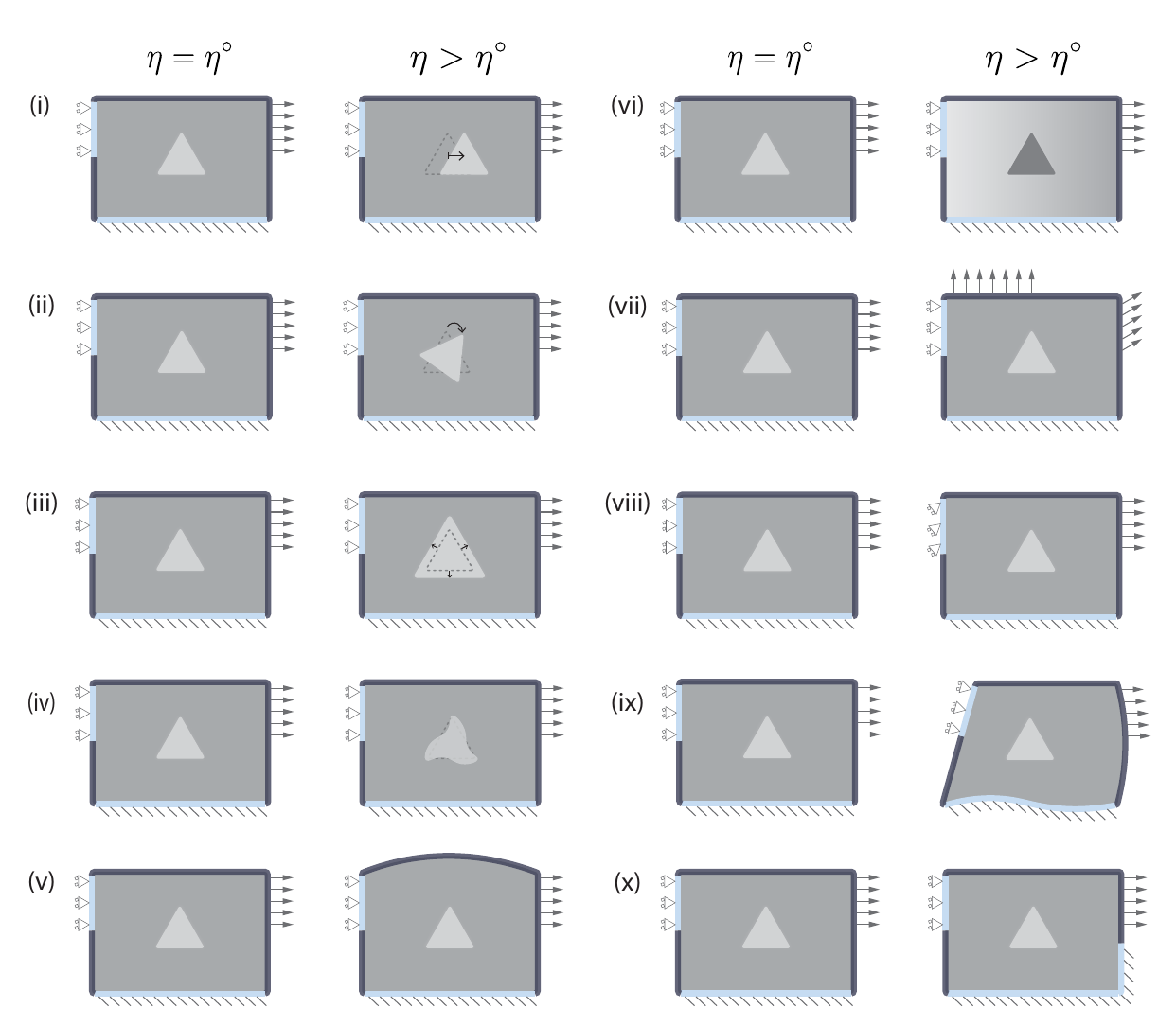}
\caption{Representative variational evolutions accommodated by the framework. The gray region represents the body, and the light-gray triangle represents an internal material region, such as an inclusion. The dark-blue curve delineates the external boundary, while light-blue highlighting identifies portions on which displacement is prescribed. Hatching and roller symbols indicate fixed and roller conditions, respectively, and arrows indicate applied traction. In each panel, the state at $\eta=\baseval$ (left) is compared with an evolved state at $\eta>\baseval$ (right). Panels (i)--(x) illustrate the following possibilities. (i)--(iv) An internal material region may translate, rotate, expand, or undergo a more general change in shape. (v) The external boundary geometry may evolve. (vi) Material properties may evolve spatially nonuniformly within individual material regions and discontinuously across internal material interfaces, as represented schematically by the changes in shading. (vii) Applied traction may evolve in magnitude, direction, and the portion of the boundary on which it is nonzero; in the schematic, traction is applied to a previously traction-free portion of the boundary, while the direction of the existing traction also changes. (viii) Prescribed displacement conditions may evolve in magnitude and direction; in the schematic, the prescribed direction associated with an initially horizontal roller condition becomes oblique. (ix) These different modes of evolution need not occur separately and may arise simultaneously. As one illustrative combination, the boundary geometry and roller direction evolve together, while the fixed displacement condition and applied traction are transported to the evolved boundary without otherwise changing. Other combinations, including simultaneous evolution of the material properties, geometry, applied traction, and prescribed displacement conditions, are likewise accommodated. (x) Even when the external boundary geometry remains fixed, the portion of the boundary carrying a prescribed displacement condition may evolve. In the schematic, a portion of the boundary that is initially traction-free becomes subject to a fixed-displacement condition.}
\label{fig:configurationalChanges}
\end{figure}

We begin by making concrete the physical picture associated with the geometric evolution. 
By an evolution of the geometry of the body, we mean that one or both of the following may occur: (i) regions within the body, such as an inclusion, may translate, rotate, expand, or undergo more general changes in shape and (ii) the external boundary of the body itself may evolve. Representative geometric evolutions are illustrated schematically in Fig.~\ref{fig:configurationalChanges}(i)--(v).

Turning to material properties, our framework accommodates, for example, the modeling of stiffening associated with a phase change. The change in material properties need not be spatially uniform; for example, the degree of stiffening may vary continuously across the body. In the presence of material heterogeneities, such as inclusions, both the material properties and their evolution may be discontinuous across material interfaces. This type of evolution is illustrated in Fig.~\ref{fig:configurationalChanges}(vi).

We now turn to the applied traction and prescribed displacement conditions. The applied traction may change in both magnitude and direction. For example, a surface that is traction-free at one state may have a nonzero traction applied at another, and a traction initially applied in one direction may subsequently act in another, as illustrated in Fig.~\ref{fig:configurationalChanges}(vii). Prescribed displacement conditions may evolve in analogous ways. A boundary portion that is fixed at zero displacement at one state may have a nonzero displacement prescribed at another. The direction in which displacement is prescribed may also evolve. For instance, the prescribed direction associated with a roller condition may change while the geometry of the boundary remains fixed, as illustrated in Fig.~\ref{fig:configurationalChanges}(viii). Alternatively, consider an initially vertical boundary surface that evolves into a slanted surface, with displacement constrained in the direction normal to the surface at each state, as illustrated in Fig.~\ref{fig:configurationalChanges}(ix). In the latter case, along with the geometric evolution of the surface, the direction in which displacement is prescribed also evolves. These possibilities are illustrated schematically in Fig.~\ref{fig:configurationalChanges}(vii)--(ix).

In addition to the traction and displacement fields themselves, our framework also accommodates evolution of the portions of the boundary on which traction and displacement conditions are prescribed. A simple example is an evolution representing a changing contact region: a portion of the boundary that is subject to a displacement constraint while in contact may become traction-free upon loss of contact. Conversely, a previously traction-free portion of the boundary may enter into contact and become subject to a displacement constraint. Such a change in the boundary regions on which the conditions are imposed is illustrated schematically in Fig.~\ref{fig:configurationalChanges}(x). 

We now make this physical description mathematically precise. Specifically, in \S\ref{subsec:domainevol} we place conditions on the map $\eta\mapsto\bodyRn$ specifying geometric evolution (from which we also define the associated map $\eta\mapsto\dOn\ag{\cdot}$); in \S\ref{subsec:fnevol}, on the map $\eta\mapsto\fn\ag{\cdot}$ accounting for the evolution of material properties; in \S\ref{subsec:appliedtrac}, on the map $\eta\mapsto\Ttn\ag{\cdot}$ specifying the evolution of the applied traction field; 
and in \S\ref{subsec:dispevol} we treat the prescribed displacement conditions collectively, placing conditions on the maps $\eta\mapsto\uhn[\alpha]\ag{\cdot}$, $\eta\mapsto\nna\ag{\cdot}$, and $\eta\mapsto\GammagRna$. Having thereby specified the evolution of the variational problem itself, in \S\ref{subsec:equilDispevol} we place conditions on the corresponding family of equilibrium displacement fields $\eta\mapsto\ubn\ag{\cdot}$. Unless otherwise specified, we assume throughout that the quantities associated with the variational evolution possess sufficient regularity and integrability for the functionals and mathematical operations used in the derivation of $F_c^\circ$ in \S\ref{sec:PERR} to be well defined. Specific regularity requirements are stated where needed.

We formulate the conditions in \S\ref{subsec:domainevol}--\S\ref{subsec:equilDispevol} locally about the arbitrary state $\eta=\baseval$ introduced in \S\ref{sec:overview}. Throughout, a superscript $\circ$ denotes evaluation of the corresponding quantity at this state, and the evolution is characterized for $\eta$ sufficiently close to $\baseval$.

\subsection{Domain and its evolution}
\label{subsec:domainevol}

One of the simplest variational evolutions is one in which only the geometry of the body evolves. Such an evolution is fully specified by the map $\eta\mapsto\bodyRn$.
In this work, we model this map as 
\begin{equation}
\label{eq:DomainEvolution}
\bodyRn=\xnb\ag{\bodyRO}.
\end{equation}
Here and throughout, the subscript $\mathrm{R}$ denotes association with the reference configuration of each $\eta$-variational problem. In \eqref{eq:DomainEvolution}, $\bodyRO\subset\Reals[m]$ denotes the domain at $\eta=\baseval$ and serves as the fixed domain relative to which the local evolution is described.  
We refer to $\xnb\ag{\cdot}:\clos[\bodyRO]\to\Reals^m$ as the \textit{geometric map}, where $\clos[\bodyRO]$ denotes the closure of $\bodyRO$. The geometric map thus describes how the physical geometry of the body would change if the specified geometric evolution were to take place. For each $\x\in\clos[\bodyRO]$, we write
\[
\z:=\xnb\ag{\x}\in\clos[\bodyRn],
\]
for its image under the geometric map. It should be understood, however, that this point-by-point correspondence need not represent the physical motion of individual material points.

Geometrically, $\xnb\ag{\cdot}$ may represent, for example, self-similar expansion of the body, extension of an existing crack, or migration of a cavity. 
Motivated by these physical pictures, we now make precise the class of geometric evolutions accommodated within our framework by stating the assumptions we place on $\xnb\ag{\cdot}$.
These assumptions also provide the regularity that enables the derivation of the closed-form expression for $F_c^\circ$ given in \S\ref{sec:PERR}. 
Specifically, we first require that its restriction to $\bodyRO$ be a diffeomorphism onto $\bodyRn$. We also require that its restriction to the boundary maps $\bd\bodyRO$ onto $\bd\bodyRn$.
We further require that the geometric evolution is smooth to the first order about $\eta=\baseval$, by placing the assumption that the map $\xnb\ag{\cdot}$ admits the first-order asymptotic expansion
\begin{align}
    \xnb\ag{\x}
        &= \x
        + \en{\eta-\baseval}\,\phiO\ag{\x}
        + \oC[\eta-\baseval]
    \label{eq:xnbAsymp}
\end{align}
as $\eta\to\baseval$, uniformly in $\x\in\clos[\bodyRO]$. In \eqref{eq:xnbAsymp}, the first-order term $\phiO\ag{\cdot}$  characterizes the local geometric evolution about the state $\eta=\baseval$ and is a key quantity which appears in the derivation of the closed-form expression of \prr. We take $\phiO\ag{\cdot}$ to be a smooth vector field up to the boundary of $\bodyRO$.  In particular, \eqref{eq:xnbAsymp} implies that the geometric map reduces to the identity at the state about which the evolution is considered.
To allow for the derivation of a closed-form expression for $F_c^\circ$, we additionally assume that the asymptotic expansion in \eqref{eq:xnbAsymp} is preserved under spatial differentiation. Thus,
\begin{align}
    \grad\xnb\ag{\x}
        &= \tp{I}
        + \en{\eta-\baseval}\,\grad\phiO\ag{\x}
        + \oC[\eta-\baseval]
    \label{eq:gradxnbAsymp}
\end{align}
as $\eta\to\baseval$,  uniformly in $\x\in\clos[\bodyRO]$. In \eqref{eq:gradxnbAsymp}, $\grad$ denotes the spatial gradient, as will be formally introduced in \S\ref{subsec:partialderiv}.

We now relate the geometric map to the geometric displacement field $\dOn\ag{\cdot}:\clos[\bodyRn]\to\Reals[m]$, introduced in \S\ref{sec:overview}.  The motivation for this terminology becomes clear from its definition below. We define the field  $\dOn\ag{\cdot}$ through its pullback onto $\clos[\bodyRO]$, denoted by
\[
\dOnc\ag{\cdot}:=\en{\dOn\comp\xnb}\ag{\cdot},
\]
and require that
\begin{equation}
\dOnc\ag{\x}=\xnb\ag{\x}-\x.
\label{eq:dOncDef}
\end{equation}
The quantity on the right-hand side of \eqref{eq:dOncDef} is the geometric-evolution analog of the mechanical displacement: just as the latter is given by the difference between the deformation and identity maps, the former is given by the difference between the geometric and identity maps.
Thus, at the $\eta^\circ$ state, the geometric map reduces to the identity, and hence the corresponding geometric displacement field, $\dOO\ag{\cdot}$, vanishes

We complete the specification of geometric evolution which can be considered within our framework, by specifying the class of domains considered. Throughout, we take each $\bodyRn$ to be a nonempty, open, connected, and bounded domain in $\Reals[m]$ with Lipschitz boundary. Within this setting, the body may contain geometric features such as cavities and crack-like defects, provided that they are represented as full-dimensional features of the geometry in $\Reals[m]$ (rather than as lower-dimensional features). The diffeomorphism requirement on $\xnb\ag{\cdot}$, however, excludes evolutions involving the formation of new geometric singularities, such as crack nucleation or kinking.

\vspace{1em}

\begin{figure}[!t]
\centering
\includegraphics[draft=false,width=0.65\textwidth]{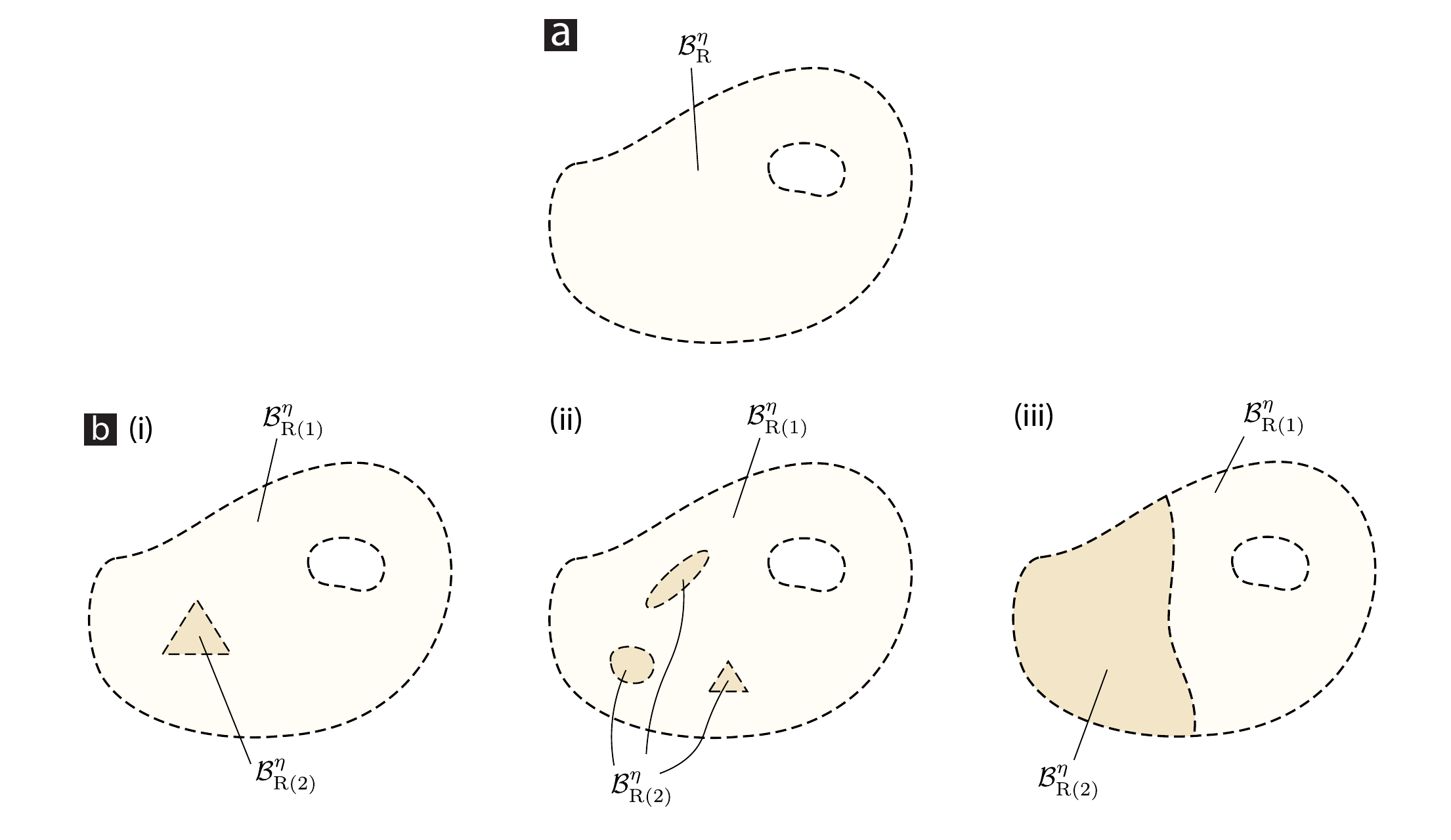}
\caption{Domains considered in the framework. (a) At each $\eta$, the body is represented by the domain $\bodyRn$, which may, for example, contain an internal cavity. (b) Representative decompositions of the body into two material regions $\bodyRn[1]$ and $\bodyRn[2]$: (i) a single inclusion embedded within a surrounding matrix, with the inclusion represented by $\bodyRn[2]$ and the matrix by $\bodyRn[1]$; (ii) multiple disconnected inclusions, collectively represented by $\bodyRn[2]$, embedded within the matrix $\bodyRn[1]$; and (iii) two adjoining material regions, neither of which is required to be contained within the other.}
\label{fig:geometrySetting}
\end{figure}

\textbf{Bodies with sharp material interfaces} We also consider bodies consisting of distinct material regions separated by sharp material interfaces. For simplicity, we restrict the current presentation to two material regions; nevertheless, the framework extends directly to any finite number of such regions. We denote these regions at the $\eta$-state by $\bodyRn[1]$ and $\bodyRn[2]$.  To give a physical picture, $\bodyRn[2]$ may, for example, represent one or more distinct inhomogeneities having the same material properties, while $\bodyRn[1]$ may represent the surrounding matrix (see Fig.~\ref{fig:geometrySetting}(b)(i)--(ii)). As with the body itself, we take the evolution of the material regions to be induced by the geometric map $\xnb\ag{\cdot}$. Thus,
\[\bodyRn[r]=\xnb\ag{\bodyRO[r]}, \quad r=1,2\]
In particular, the geometric evolution may involve changes in the location, orientation, size, or shape of the individual material regions, in addition to changes in the geometry of the body as a whole. 

We now make precise the class of geometries each material region may assume within this framework. For $r=1,2$, we take $\bodyRn[r]$ to be a nonempty, open, and bounded subset of $\Reals[m]$. 
Each $\bodyRn[r]$ may be disconnected. Each of its connected components, however, is assumed to have a sufficiently regular boundary. Specifically, we assume that the divergence theorem applies, that the boundary integrals arising in the subsequent derivations are well defined, and that an outward unit normal exists almost everywhere with respect to surface measure.
Distinct connected components belonging to the same material region are assumed to be positively separated. The two material regions are disjoint and together occupy the body. 
More precisely,
\[\bodyRn[1]\cap\bodyRn[2]=\emptyset\]
and 
\[\clos[\bodyRn[1]]\cup\clos[\bodyRn[2]]=\clos[\bodyRn].\]
To clarify the physical picture, the labels (1) and (2) serve only to distinguish the two material regions; neither region is required to be contained within the other (see Fig.~\ref{fig:geometrySetting}(b)(iii)), and either may meet the boundary of the body\footnote{Requiring the material regions to be open subsets of $\Reals[m]$ ensures that they are $m$-dimensional. Moreover, since $\bodyRn$ is connected and the two nonempty, disjoint material regions partition the body, their common internal boundary must separate the two regions. By standard separation results~\cite{engelking1978dimension}, such a separating set cannot have topological dimension less than $(m-1)$. That is, the construction excludes contact of two material regions only at isolated points in two dimensions or only at points or curves in three dimensions.}. We require that whenever a material region meets the boundary of the body, the intersection of their boundaries contains a nonempty open subset of the boundary of the body, i.e., the contact cannot occur only at isolated points in two dimensions or only at points or curves in three dimensions.

\subsection{Generalized potential energy and its evolution}
\label{subsec:fnevol}
As we outlined previously, evolution can also be one in which the material properties of the body evolve.
Within our framework, such changes, together with evolution of the body-force potential, are specified through the map $\eta\mapsto\fn\ag{\cdot}$; to recall, $\fn\ag{\cdot}$ is the integrand appearing in the volume integral of Eq.~\eqref{eq:PibnDef}. We construct $\fn\ag{\cdot}$ to account for both the constitutive response of the body, or, more precisely, its strain energy, as well as the body-force potential. Accordingly, we refer to $\fn\ag{\cdot}$ as the \textit{generalized potential energy density} and write, at each $\eta$,
\begin{equation}
\fn\ag{\x,\u,\H}=\W^\eta\ag{\x,\H}+\PhiC^\eta\ag{\x,\u},
\label{eq:fnDecomp}
\end{equation}
where $\W^\eta\ag{\cdot}:\Reals[m]\times\Matrix[m]\to\Reals$ is the strain energy density and $\PhiC^\eta\ag{\cdot}:\Reals[m]\times\Reals[m]\to\Reals$ is the body-force potential.
Thus, for example, we may model stiffening through phase transformation  by the map $\eta\mapsto\Wn\ag{\cdot}$. The decomposition in \eqref{eq:fnDecomp} additionally allows for evolution of the body-force potential through the map $\eta\mapsto\Phin\ag{\cdot}$. For example, $\Phin\ag{\cdot}$ may represent a gravitational or electromagnetic potential, whose evolution may arise through changes in either properties of the body, such as its mass or charge density, or the external field acting on it.

In defining $\Wn\ag{\cdot}$ and $\Phin\ag{\cdot}$ above, we have chosen their spatial argument $\x$ to range over the ambient space $\Reals[m]$, rather than restricting it to the particular domain $\bodyRn$ occupied by the body at each $\eta$. Consequently, $\fn\ag{\cdot}$ is defined on $\Reals[m]\times\Reals[m]\times\Matrix[m]$. 
This choice is motivated by the following perturbation assumption we place on the family of $\fn\ag{\cdot}$, which facilitates the derivation of a closed-form expression for $F_c^\circ$ in \S\ref{sec:PERR}. Specifically, we assume that $\fn\ag{\cdot}$ admits the first-order asymptotic expansion about $\eta=\baseval$,
\begin{equation}
\fn\ag{\x,\u,\H}=\fO\ag{\x,\u,\H}+\en{\eta-\baseval}\zetaO\ag{\x,\u,\H}+\oC[\eta-\baseval]
\label{eq:fAsymptoticForm}
\end{equation}
as $\eta\to\baseval$, for each fixed $(\x,\u,\H)$.
In \eqref{eq:fAsymptoticForm}, the zeroth-order term of the expansion, $\fO\ag{\cdot}$, is the generalized potential energy density at $\eta=\baseval$. The first-order term of the expansion, $\zetaO\ag{\cdot}$, is a key quantity that appears in the final closed-form expression for $F_c^\circ$. We take $\zetaO\ag{\cdot}$ to be a smooth function with respect to its arguments. The precise sense in which the remainder in \eqref{eq:fAsymptoticForm} is understood is specified in \S\ref{subsec:fnLimits}. For the pointwise comparison in \eqref{eq:fAsymptoticForm} to be well posed, $\fn\ag{\cdot}$ and $\fO\ag{\cdot}$ must share a common domain. Our choice to define their spatial argument over the ambient space $\Reals[m]$, rather than restricting it to $\bodyRn$, ensures that this condition is satisfied throughout the domain evolution.

For the derivation of the closed-form expression for $F_c^\circ$, we require that $\fn\ag{\x,\u,\H}$ be Fr\'echet differentiable in all its arguments. We enforce this by taking $\Wn\ag{\x,\H}$ and $\Phin\ag{\x,\u}$ to each be Fr\'echet differentiable in their respective arguments. 
Since Fr\'echet differentiability implies continuity, in particular with respect to the spatial argument $\x$, the properties of the body that enter the formulation through these functions may vary spatially, but only in a continuous manner. That is, spatial jumps in such properties are excluded. 
For example, a sharp interface between regions with different constitutive properties would produce a discontinuity in $\Wn\ag{\cdot}$ and thus cannot be accommodated; similarly, under gravitational loading, a jump in mass density would produce a discontinuity in $\Phin\ag{\cdot}$. 
Consequently, the generalized potential energy density $\fn\ag{\cdot}$, and thus the variational formulation given in \eqref{eq:VE}, cannot, in general, accommodate the modeling of a body with sharp material interfaces, such as the interface between an inclusion and the surrounding material.

\vspace{1em}
\phantomsection
\label{sec:bodiessharp}
\textbf{Bodies with sharp material interfaces} Nevertheless, bodies containing sharp material interfaces are of central interest in materials science and configurational mechanics. We therefore accommodate such bodies in the following manner. As described in \S\ref{subsec:domainevol}, 
when the body at the $\eta$-state contains sharp material interfaces,  we model it by the two distinct material regions $\bodyRn[r]$, $r=1,2$. We specify the generalized potential energy density associated with such a body by assigning to each material region a corresponding map $\fnC[r]\ag{\cdot}$. Each map is defined on the ambient space $\Reals[m]\times\Reals[m]\times\Matrix[m]$ and is used to model the material properties of the corresponding region.

We impose on each $\fn_{(r)}\ag{\cdot}$ the same assumptions as those placed on $\fn\ag{\cdot}$ above. In particular, each $\fn_{(r)}\ag{\cdot}$ admits a decomposition of the form given in \eqref{eq:fnDecomp}, with its constituent strain energy density and body-force potential satisfying the differentiability assumptions stated above, and satisfies the first-order perturbation assumption in \eqref{eq:fAsymptoticForm}. 
Specifically,  we write
\begin{equation}
\fnr\ag{\x,\u,\H}=\fOr\ag{\x,\u,\H}+\en{\eta-\baseval}\zetaOr\ag{\x,\u,\H}+\oC[\eta-\baseval].
\end{equation}
Here, the zeroth-order term $\fOr\ag{\cdot}$ is the generalized potential energy density function associated with the material region $\bodyRO[r]$ at $\eta=\baseval$, while $\zetaOr\ag{\cdot}$ is the corresponding first-order term. Analogous to $\zetaO\ag{\cdot}$ above, the terms $\zetaOr\ag{\cdot}$ appear in the final closed-form expression of $F_c^\circ$ when the body contains sharp material interfaces.

Accordingly, for a body with sharp material interfaces, we generalize the potential energy functional in \eqref{eq:PibnDef} to
\begin{equation}
\Pin\ag{\u}:=\sum_{r=1}^2\int_{\bodyRn[r]}\fnr\ag{\z,\u\ag{\z},\grad\u\ag{\z}} \dV -\int_{\bd\bodyRn}\Ttn\ag{\z}\cdot\en{\dOn\ag{\z}+\u\ag{\z}}\dS.
\label{eq:Pibnhet}
\end{equation}

\subsection{Applied traction field and its evolution}
\label{subsec:appliedtrac}

Evolution can also be one in which the applied traction field evolves. At each $\eta$, we represent the applied traction by the map $\Ttn\ag{\cdot}:\bd\bodyRn\to\Reals[m]$, which, to recall, appears in the second integral of \eqref{eq:PibnDef}. We later explain in \S\ref{sec:evolutionarilyconsistent} why $\Ttn\ag{\cdot}$ is defined on the entire boundary $\bd\bodyRn$, rather than only on the Neumann boundary as is the customary choice in standard continuum mechanics framework. 
For now, we focus on characterizing the evolution of the applied traction field itself, by which we mean changes in its magnitude and orientation. Since the domain of $\Ttn\ag{\cdot}$ itself evolves with $\eta$, such changes cannot be characterized independently of the geometric evolution directly through the family $\eta\mapsto\Ttn\ag{\cdot}$.
We therefore pull back $\Ttn\ag{\cdot}$ to the fixed boundary $\bd\bodyRO$ using the geometric map $\xnb\ag{\cdot}$ and define
\begin{equation*}
\Ttnc\ag{\x}:=\en{\Ttn\comp\xnb}\ag{\x},
\qquad \x\in\bd\bodyRO.
\end{equation*}
The map $\eta\mapsto\Ttnc\ag{\cdot}$ thereby allows the applied traction fields at different $\eta$-states to be compared at corresponding boundary points on the fixed boundary $\bd\bodyRO$. Accordingly, throughout this work, when we refer to the evolution of the applied traction field, we mean specifically the evolution of its pulled-back representation, $\eta\mapsto\Ttnc\ag{\cdot}$.

We assume that the applied traction field, when pulled back to the fixed boundary $\bd\bodyRO$, is piecewise continuous\footnote{More precisely, we assume that the regions defining the piecewise-continuous structure form a fixed finite partition of $\bd\bodyRO$, independent of $\eta$, and that $\Ttnc\ag{\cdot}$ extends continuously to the closure of each such region}. To derive the closed-form expression for $F_c^\circ$, we assume that the applied traction field evolves smoothly to first order in $\eta$. Specifically, we assume that $\Ttnc\ag{\cdot}$ admits the asymptotic expansion
\begin{equation}
\Ttnc\ag{\x}=\TtO\ag{\x}+\en{\eta-\baseval}\chiO\ag{\x}+\oC[\eta-\baseval]
\label{eq:TtncAsymp}
\end{equation}
as $\eta\to\baseval$, uniformly in $\x\in\bd\bodyRO$\footnote{More precisely, the expansion holds uniformly on each region of the fixed finite partition of $\bd\bodyRO$ on which the pulled-back traction fields are continuous.}.
In the above expansion, the zeroth-order term $\TtO\ag{\cdot}$ is also the applied traction field of the variational problem at $\eta=\baseval$. The first-order term $\chi^{\circ}\ag{\cdot}$ is a key quantity that appears in the final closed-form expression for $F_c^\circ$. We take $\chiO\ag{\cdot}$ to be smooth with respect to the partition induced by the spatial discontinuities of the field $\Ttnc\ag{\cdot}$.

\subsection{Prescribed displacement condition and its evolution}
\label{subsec:dispevol}

Evolution can also be one in which the prescribed displacement conditions evolve. In our framework we specify prescribed displacement conditions by the quantities $\GammagRna$, $\nna\ag{\cdot}$, and $\uhn[\alpha]\ag{\cdot}$. For convenience, we briefly recall their roles. For each $\alpha=1,\ldots,m$, we denote by $\GammagRna\subset\bd\bodyRn$ the region of the boundary on which displacement is prescribed along the direction $\nna\ag{\cdot}$; we refer to $\GammagRna$ as the \textit{$\alpha$-Dirichlet boundary}. The corresponding prescribed displacement condition is
\[
    \u\ag{\z}\cdot\nna\ag{\z}=\uhn[\alpha]\ag{\z} \quad\text{for a.e. } \z\in\GammagRna, \quad \alpha=1,\ldots, m.
\]
where $\uhn[\alpha]\ag{\cdot}$ specifies the prescribed value of displacement along $\nna\ag{\cdot}$.

Accordingly, we distinguish two aspects of the evolution of prescribed displacement conditions: evolution of the $\alpha$-Dirichlet boundaries, concerning the maps $\eta\mapsto\GammagRna$; and evolution of the prescribed displacement field itself, concerning the maps  $\eta\mapsto\nna\ag{\cdot}$ and $\eta\mapsto\uhn[\alpha]\ag{\cdot}$. We consider these aspects in \S\ref{subsec:alphaDirEvol} and \S\ref{sec:prescribedDispEvol}, respectively. As preliminaries to these discussions, we first provide further details on the construction of the fields and regions used to represent the prescribed displacement conditions. These details make precise the underlying modeling machinery but are not needed to understand the main idea of the evolution described in \S\ref{subsec:alphaDirEvol} and \S\ref{sec:prescribedDispEvol}; they may therefore be skipped on a first reading.

\begin{figure}[htbp]
\centering
\includegraphics[draft=false,width=0.7\textwidth]{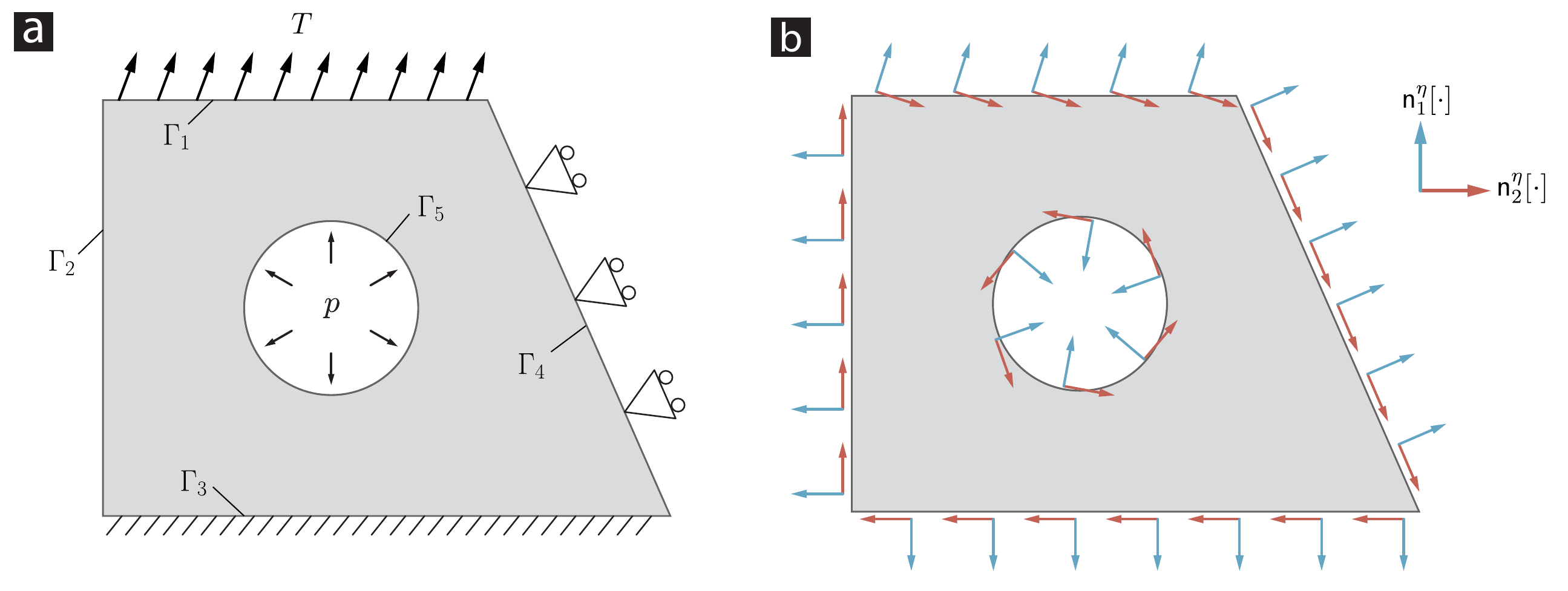}
\caption{Schematic representation of prescribed displacement and applied traction data using an auxiliary orthonormal frame field. (a) The reference configuration of a two-dimensional body at the $\eta$-state, with boundary portions $\Gamma_1,\ldots,\Gamma_5$ subject, respectively, to an applied traction, a traction-free condition, a fixed condition, a roller condition, and pressure loading on the cavity boundary. (b) One corresponding choice of auxiliary orthonormal frame field, with $\nn[1]\ag{\cdot}$ shown in blue and $\nn[2]\ag{\cdot}$ shown in red. The associated $\alpha$-Dirichlet boundaries and the corresponding prescribed displacement and applied traction fields are specified in the \hyperref[ex:fig3BoundaryData]{accompanying example}.}
\label{fig:orthonormalFrame}
\end{figure}

\vspace{1em }
\phantomsection
\label{sec:construction}
\paragraph*{Construction of the prescribed displacement fields} 
We represent prescribed displacement conditions using an auxiliary orthonormal frame field. Figure~\ref{fig:orthonormalFrame} illustrates this construction for a two-dimensional body. The accompanying example makes explicit the associated $\alpha$-Dirichlet boundaries and prescribed displacement fields, together with the applied traction field depicted in the figure. This representation offers considerable flexibility in specifying where displacement is prescribed and how its directions and values vary along the boundary, while retaining a componentwise form that facilitates the subsequent derivation of the boundary equilibrium conditions and the closed-form expression for $F_c^\circ$. We now describe its construction.

For each $\eta$, we choose a frame field
\[
\en{\nn[1]\ag{\cdot},\ldots,\nn[m]\ag{\cdot}}
\]
to form an auxiliary orthonormal frame field defined for almost every point of $\bd\bodyRn$. That is,
\begin{equation}
\nn[i]\ag{\z}\cdot\nn[j]\ag{\z}
=
\delta_{ij},
\qquad i,j=1,\ldots,m, \quad \text{a.e.\,}\z\in\bd\bodyRn,
\end{equation}
where $\delta_{ij}$ is Kronecker's delta. At each boundary point where displacement is prescribed in $k$ mutually orthogonal directions, we choose $k$ of the frame vectors to coincide with those prescribed directions, while the remaining $m-k$ frame vectors are chosen to complete the orthonormal frame.\footnote{Accordingly, the present framework is restricted to problems in which, whenever displacement is prescribed in more than one direction at a boundary point, the prescribed directions are mutually orthogonal.} The assignment of the $k$ prescribed directions to particular frame vectors is not unique and may be chosen freely. The remaining $m-k$ frame vectors may be oriented conveniently; for example, wherever compatible with the prescribed displacement directions, a remaining frame vector may be aligned with the direction of the applied traction. Under pressure loading, a normal--tangential frame is a natural choice. Such alignment is optional and imposes no restriction on the applied traction field. In all cases, these choices are made so that the resulting orthonormal frame field is piecewise smooth along $\bd\bodyRn$. In what follows, we index the members of this frame field by $\alpha=1,\ldots,m$, writing $\nna\ag{\cdot}$ for its $\alpha$-th member. We also use $\en{\nna\ag{\cdot}}_{\alpha=1}^m$ to denote the frame field $\en{\nn[1]\ag{\cdot},\ldots,\nn[m]\ag{\cdot}}$. 

Having chosen the frame field, for each $\alpha=1,\ldots,m$, we denote by $\GammagRna\subset\bd\bodyRn$ the open, possibly disconnected region on which displacement is prescribed along $\nna\ag{\cdot}$. We represent the prescribed value of displacement along this direction by the scalar field
\begin{equation}
\uhn[\alpha]\ag{\cdot}:
\GammagRna\to\Reals.
\end{equation}
We additionally require $\uhn[\alpha]\ag{\cdot}$ to be piecewise continuous. 
Specifically, spatial discontinuities of the chosen frame field partition each $\GammagRna$ into maximal connected regions on which the frame field is smooth, and we require $\uhn[\alpha]\ag{\cdot}$  to be continuous on each such region.

For later use in the derivation of the equilibrium conditions, we also introduce notation for the portion of the boundary complementary to each $\alpha$-Dirichlet boundary. For each $\alpha=1,\ldots,m$, we define
\[
\GammahRna:=\bd\bodyRn\setminus\clos[\GammagRna].
\]
Thus, up to their common boundary, which is understood to have surface measure zero, $\bd\bodyRn=\GammagRna\cup\GammahRna$.

\phantomsection
\label{ex:fig3BoundaryData}
\begin{tcolorbox}[
  breakable,
  title={Example: prescribed displacement and applied traction data in Fig.~\ref*{fig:orthonormalFrame}},
  colback=gray!4,
  colframe=black!55,
  boxrule=0.5pt,
  arc=1.5pt
]
\small
In Fig.~\ref{fig:orthonormalFrame}(b), $\nn[1]\ag{\cdot}$ and
$\nn[2]\ag{\cdot}$ are represented by the blue and red vectors,
respectively. With this choice of frame, the corresponding $1$- and
$2$-Dirichlet boundaries are, up to the endpoints of the illustrated
boundary portions,
\[
\Gamma_{R,1}^{g,\eta}=\Gamma_3\cup\Gamma_4,
\qquad
\Gamma_{R,2}^{g,\eta}=\Gamma_3.
\]
The complementary boundary portions are therefore
\[
\GammahRn[1]=\Gamma_1\cup\Gamma_2\cup\Gamma_5,
\qquad
\GammahRn[2]=\Gamma_1\cup\Gamma_2\cup\Gamma_4\cup\Gamma_5.
\]
The corresponding prescribed displacement fields are
\[
\uhn[1]\ag{\z}=0
\quad\text{for }\z\in\Gamma_3\cup\Gamma_4,
\qquad
\uhn[2]\ag{\z}=0
\quad\text{for }\z\in\Gamma_3.
\]
Thus, both displacement components are prescribed on the fixed boundary
$\Gamma_3$, whereas only the component along $\nn[1]\ag{\cdot}$ is
prescribed on the roller boundary $\Gamma_4$. No displacement component is
prescribed on $\Gamma_1$, $\Gamma_2$, or $\Gamma_5$.

The applied traction field, which is defined on the entire boundary, is
given by
\[
\Ttn\ag{\z}
=
\begin{cases}
T\nn[1]\ag{\z},
  & \z\in\Gamma_1,\\
\mathbf 0,
  & \z\in\Gamma_2\cup\Gamma_3\cup\Gamma_4,\\
-p\nn[1]\ag{\z},
  & \z\in\Gamma_5.
\end{cases}
\]
On $\Gamma_2,\ldots,\Gamma_5$, $\nn[1]\ag{\cdot}$ has been taken to be normal to the boundary. On $\Gamma_1$, by contrast, it has been taken to align with the oblique applied traction and therefore is not normal to the boundary. These choices are illustrative; in general, the auxiliary frame need not be aligned with either the boundary normal or the applied traction.
\end{tcolorbox}

\subsubsection{Evolution of the $\alpha$-Dirichlet boundaries}
\label{subsec:alphaDirEvol}

We model evolution of the $\alpha$-Dirichlet boundary, which is specified by the map $\eta\mapsto\GammagRna$, as that induced by the geometric map introduced in \S\ref{subsec:domainevol}. Specifically, we take
\begin{equation}
    \GammagRna
    =
    \xnb\ag{\GammagROa},
\end{equation}
where $\GammagROa$ denotes the $\alpha$-Dirichlet boundary at
$\eta=\baseval$. 

The most immediate interpretation of the evolution of the $\alpha$-Dirichlet boundary induced by $\xnb\ag{\cdot}$ is that it accompanies geometric evolution of the body: as the shape of the body's boundary evolves according to the geometric map $\xnb\ag{\cdot}$, the geometric map correspondingly transports the $\alpha$-Dirichlet boundary. Importantly, however, $\xnb\ag{\cdot}$ may specify evolution of the $\alpha$-Dirichlet boundary even when the domain occupied by the body remains unchanged. In this case, the boundary remains fixed as $\bd\bodyRO$, but the subset of it comprising $\GammagRna$ varies with $\eta$. We interpret this as evolution of the boundary partition. For example, in the changing-contact problem described previously, the geometry of the body may remain unchanged while the portion of its boundary subject to a prescribed displacement condition evolves as contact is established or lost along different portions of the boundary.

\subsubsection{Evolution of the prescribed displacement field}
\label{sec:prescribedDispEvol}
We now consider evolution of the prescribed displacement field itself. Within our framework, this consists of both the evolution of the directions of prescribed displacement and the evolution of the values of prescribed displacement. As described above, the prescribed directions are represented by the members of the auxiliary orthonormal frame field $\en{\nn[a]\ag{\cdot}}_{\alpha=1}^m$ on their corresponding $\alpha$-Dirichlet boundaries, while the prescribed values are represented by $\uhn[\alpha]\ag{\cdot}$. Accordingly, we characterize these two aspects of the evolution through the maps $\eta\mapsto\nna\ag{\cdot}$ and $\eta\mapsto\uhn[\alpha]\ag{\cdot}$, respectively. We consider them in turn below.

\paragraph*{Evolution of the directions of prescribed displacement.} Consider first the evolution of the directions of prescribed displacement. Since these directions are represented by the auxiliary orthonormal frame field, we specify their evolution through the maps $\eta\mapsto\nna\ag{\cdot}$.  For each $\alpha=1,\ldots,m$, the $\alpha$-th member of the frame field $\nna\ag{\cdot}:\bd\bodyRn\to\Reals[m]$ is defined on the boundary of the body, which itself evolves with $\eta$. Thus, the evolution of the frame field cannot be characterized independently of the evolution of its domain directly through the maps $\eta\mapsto\nna\ag{\cdot}$. We therefore introduce the fields $\nnca\ag{\cdot}:\bd\bodyRO\to\Reals[m]$, $\alpha=1,\ldots,m$, obtained by pulling back the corresponding fields $\nna\ag{\cdot}$ onto the fixed boundary $\bd\bodyRO$ via the geometric map $\xnb\ag{\cdot}$:
\[
\nnca[\x]
:=
(\nna\circ\xnb)[\x].
\]
The maps $\eta\mapsto\nnca\ag{\cdot}$, $\alpha=1,\ldots,m$, thereby allow the orthonormal frame fields at different $\eta$-states to be compared at corresponding boundary points on the fixed boundary $\bd\bodyRO$. Accordingly, throughout this work, when we refer to the evolution of the directions of prescribed displacement, we mean specifically the evolution of the pulled-back frame field, characterized by the maps $\eta\mapsto\nnca\ag{\cdot}$, $\alpha=1,\ldots,m$.

We restrict our attention to evolutions for which the frame field evolves smoothly to first order in $\eta$. Specifically, for each $\alpha=1,\ldots,m$, we assume that $\nnca\ag{\cdot}$ admits the first-order asymptotic expansion about $\eta=\baseval$
\begin{equation}
\nnca[\x]
=\nOa[\x]
+\en{\eta-\baseval}\,\OmegaO[\x]\,\nOa[\x]
+\oC[\eta-\baseval]
\label{eq:nncaAsymp}
\end{equation}
as $\eta\to\baseval$, uniformly in $\x\in\bd\bodyRO$. Here, $\nOa\ag{\cdot}:\bd\bodyRO\to\Reals[m]$ is the $\alpha$-th member of the orthonormal frame field at $\eta=\baseval$, while $\OmegaO\ag{\cdot}:\bd\bodyRO\to\mathfrak{so}(m)$ is an infinitesimal rotation field. By using the same infinitesimal rotation field $\OmegaO\ag{\cdot}$ in the first-order expansions of $\nnca\ag{\cdot}$ for all $\alpha=1,\ldots,m$, we ensure that the orthonormal frame evolves, to first order, by a common infinitesimal rotation. We take $\OmegaO\ag{\cdot}$ to be piecewise smooth with respect to the partition induced by spatial discontinuities in the frame field at $\eta=\baseval$.

\paragraph*{Evolution of the prescribed displacement values.}
Consider next the evolution of the prescribed displacement values. As we have explained, these values are specified by the scalar fields $\uhn[\alpha]\ag{\cdot}$, $\alpha=1,\ldots,m$. Accordingly, we characterize their evolution through the maps $\eta\mapsto\uhn[\alpha]\ag{\cdot}$,  $\alpha=1,\ldots,m$. For each $\alpha=1,\ldots,m$, the field $\uhn[\alpha]\ag{\cdot}:\GammagRna\to\Reals$ is defined on the $\alpha$-Dirichlet boundary. Since $\alpha$-Dirichlet boundary may itself evolve with $\eta$ as described above, the evolution of the prescribed displacement values itself cannot be characterized independently of the evolution of its domain through the map $\eta\mapsto\uhn[\alpha]\ag{\cdot}$.
Proceeding analogously to the treatment of the prescribed directions, we introduce the fields
$\uhnc[\alpha]\ag{\cdot}:\GammagROa\to\Reals$, $\alpha=1,\ldots,m$, obtained by pulling back the corresponding fields $\uhn[\alpha]\ag{\cdot}$ onto the fixed $\alpha$-Dirichlet boundary $\GammagROa$ via the geometric map $\xnb\ag{\cdot}$:
\begin{equation*}
    \uhnc[\alpha]\ag{\x}
    :=
    \en{\uhn[\alpha]\comp\xnb}\ag{\x}.
\end{equation*}
The maps $\eta\mapsto\uhnc[\alpha]\ag{\cdot}$, $\alpha=1,\ldots,m$, thereby allow the prescribed displacement values at different $\eta$-states to be compared at corresponding points on the fixed $\alpha$-Dirichlet boundaries $\GammagROa$. Accordingly, throughout this work, when we refer to the evolution of the values of prescribed displacement, we mean specifically the evolution of their pulled-back representations, characterized by the maps $\eta\mapsto\uhnc[\alpha]\ag{\cdot}$, $\alpha=1,\ldots,m$.

We similarly restrict our attention to evolutions for which the prescribed
displacement values evolve smoothly to first order in $\eta$. Specifically, for each $\alpha=1,\ldots,m$, we assume that $\uhnc[\alpha]\ag{\cdot}$ admits the first-order asymptotic expansion about $\eta=\baseval$
\begin{equation}
    \uhnc[\alpha]\ag{\x}
    =
    \uhO[\alpha]\ag{\x}
    +
    \en{\eta-\baseval}\,\psihOa\ag{\x}
    +
    \oC[\eta-\baseval]
    \label{eq:uhncAsymp}
\end{equation}
as $\eta\to\baseval$, uniformly in $\x\in\GammagROa$\footnote{More precisely, uniformly on each member of the partition of $\GammagROa$}.  Here, the zeroth-order term $\uhO[\alpha]\ag{\cdot}:\GammagROa\to\Reals$ is the field of prescribed displacement value at $\eta=\baseval$, while the first-order term $\psihOa\ag{\cdot}:\GammagROa\to\Reals$ is a key quantity which appears in the closed-form expression of $F_c^\circ$. We take $\psihOa\ag{\cdot}$ to be piecewise smooth with respect to the partition of $\GammagROa$  induced by spatial discontinuities in the frame field at $\eta=\baseval$. 
{}

\subsection{Equilibrium displacement field and its evolution}
\label{subsec:equilDispevol}

The equilibrium displacement field evolves as a result of the generalized potential energy, applied traction field, and prescribed displacement conditions evolving in the manner specified above. At each $\eta$, the equilibrium displacement field $\ubn\ag{\cdot}:\bodyRn\to\Reals[m]$ is a local minimizer of the potential energy given in \eqref{eq:PibnDef} over the admissible set $\An$ specified in \S\ref{sec:overview}. Accordingly, its evolution is specified by the map\footnote{Because the local minimizer need not be unique, this map denotes a selected branch of equilibrium displacement fields for $\eta$ sufficiently close to $\baseval$. The existence of such a branch is assumed here.} $\eta\mapsto\ubn\ag{\cdot}$. As with the fields considered above, however, this evolution cannot be specified independently of the geometric evolution. 
Thus,  we introduce the field $\ubnc\ag{\cdot}:\bodyRO\to\Reals[m]$ obtained as  the pullback of $\ubn\ag{\cdot}$ onto $\bodyRO$ via the geometric map $\xnb\ag{\cdot}$. That is, 
\[
    \ubnc\ag{\x}:=\en{\ubn\comp\xnb}\ag{\x}.
\]
The map $\eta\mapsto\ubnc\ag{\cdot}$ thereby allows the equilibrium displacement fields at different $\eta$-states to be compared at corresponding points on the fixed domain $\bodyRO$.
We then assume that $\ubnc\ag{\cdot}$ admits the first-order asymptotic expansion about $\eta=\baseval$
\begin{equation}
\ubnc\ag{\x}=\ubO\ag{\x}+\en{\eta-\baseval}\psiO\ag{\x}+\oC[\eta-\baseval] \label{eq:ubncAsymp}
\end{equation}
as $\eta\to\baseval$,  where the expansion is assumed to hold uniformly\footnote{More precisely, the expansion in \eqref{eq:ubncAsymp}, together with the expansion below in \eqref{eq:gradubncAsymp}, is understood as the differentiability of the selected pulled-back equilibrium branch at $\eta=\baseval$ in $W^{1,p}(\bodyRO;\Reals[m])$, with derivative $\psiO$. Since $\bodyRO$ is a bounded Lipschitz domain and $p>m$, the continuous Sobolev embedding $W^{1,p}(\bodyRO;\Reals[m])\hookrightarrow C^0(\clos[\bodyRO];\Reals[m])$
implies that convergence in $W^{1,p}$ yields uniform convergence of the corresponding continuous representatives. This gives the uniform displacement expansion in \eqref{eq:ubncAsymp}, while the gradient expansion in \eqref{eq:gradubncAsymp} holds in $L^p$. The differentiability of the selected equilibrium branch is assumed and is not established here.} in $\x\in\clos[\bodyRO]$. In the above expansion, the zeroth order term $\ubO\ag{\cdot}:\bodyRO\to\Reals[m]$ is equivalently the equilibrium displacement field of the variational problem at $\eta=\baseval$. The first order term $\psiO\ag{\cdot}:\bodyRO\to\Reals[m]$ is a key quantity which appears in the derivation of the closed-form expression of $F_c^\circ$. 

For the derivation of $F_c^\circ$, we make the additional assumption that the asymptotic expansion is preserved under spatial differentiation in $L^p$. Specifically, we assume that
\begin{equation}
\grad\ubnc\ag{\cdot}=\grad\ubO\ag{\cdot}+\en{\eta-\baseval}\grad\psiO\ag{\cdot}+\oC[\eta-\baseval],
\label{eq:gradubncAsymp}
\end{equation}
as $\eta\to\baseval$, where the remainder is understood in $L^p(\bodyRO;\Matrix[m\times m])$.

Unlike the first-order fields introduced above to prescribe the variational evolution, $\psiO\ag{\cdot}$ is not prescribed independently. Rather, it is a resulting quantity determined by the selected evolution of the equilibrium displacement field. Its required regularity is imposed through \eqref{eq:ubncAsymp} and \eqref{eq:gradubncAsymp}.

\section{Evolutionarily consistent variational formulation}
\label{sec:evolutionarilyconsistent}
In this section, we explain the rationale for adopting the potential energy functional of the form given in \eqref{eq:PibnDef}. Readers willing to take this form as given may proceed directly to \S\ref{sec:preliminaries}.

In the present work, we seek a potential energy functional that gives rise to an \textit{evolutionarily consistent variational formulation}. By this, we mean a formulation satisfying two requirements. (E1) It must correctly characterize the equilibrium displacement field associated with each individual $\eta$-problem. (E2) The equilibrium values of its potential energy functional must correctly capture the physical change in potential energy of the underlying mechanical system as $\eta$ evolves.

The \textit{standard variational formulation} employed in nonlinear elasticity is constructed to correctly characterize equilibrium at each fixed state and thus satisfies requirement (E1). 
As our evolutionarily consistent formulation may be understood as a modification of this standard formulation, we begin by recalling its form.
At each fixed value of $\eta$, we take the standard variational formulation to characterize equilibrium through local minimization of the \textit{standard potential energy functional} $\Pin_{\rm{S}}\ag{\cdot}$ over the admissible set $\An$, given respectively by
\begin{subequations}
\begin{equation}
\Pin_{\rm{S}}\ag{\u}=
\int_{\bodyRn}
\fn\ag{\z,\u\ag{\z},\grad\u\ag{\z}}\, dV
-\int_{\GammahRn}\Ttn\ag{\z}\cdot\u\ag{\z}\,d S,
\label{eq:PinStandard}
\end{equation}
and
\begin{equation}
\An=\pr{\u\ag{\cdot}\in W^{1,p}(\bodyRn;\Reals[m])~~\vert~~\u\ag{\z}\cdot\nn[\alpha]\ag{\z}=\uhn[\alpha]\ag{\z} 
\quad\text{for a.e. }\z\in\GammagRn,\quad
\alpha=1,\ldots,m}.
\end{equation}
\label{eq:standardformulation}
\end{subequations}
In \eqref{eq:standardformulation}, $\GammagRn$ and $\GammahRn$ denote the Dirichlet and Neumann parts of the boundary, respectively\footnote{For simplicity of presentation, we have taken the $\alpha$-Dirichlet boundaries to coincide, so that $\GammagRn=\GammagRna$, $\alpha=1,\ldots, m$, and $\GammahRn:=\bd\bodyRn\setminus\clos[\GammagRn]$. }
\footnote{When the $\alpha$-Dirichlet boundaries do not coincide, the standard potential energy functional reads, in our framework, \[\Pin_{\rm{S}}\ag{\u}=\int_{\bodyRn}\fn\ag{\z,\u\ag{\z},\grad\u\ag{\z}}\, dV -\sum_{\alpha=1}^m\int_{\GammahRna}\en{\Ttn\ag{\z}\cdot\nna\ag{\z}}\en{\u\ag{\z}\cdot\nna\ag{\z}} \dS.\]}. Minimizing $\Pin_{\rm{S}}\ag{\cdot}$ over $\An$  yields the correct equilibrium displacement field for the corresponding $\eta$-problem, in accordance with (E1). Such standard variational formulation, however, in general does not satisfy (E2): the equilibrium values of $\Pin_{\rm{S}}\ag{\cdot}$ at different $\eta$-states need not correctly reproduce the physical change in potential energy of the underlying mechanical system.

We now consider the variational formulation introduced in \S\ref{sec:overview}.  For direct comparison with \eqref{eq:standardformulation}, we recall its corresponding form\footnote{That is, we make the same simplifying assumption as in \eqref{eq:standardformulation} that the $\alpha$-Dirichlet boundaries coincide.} here:
\begin{subequations}
\begin{equation*}
\Pin\ag{\u}=
\int_{\bodyRn}
\fn\ag{\z,\u\ag{\z},\grad\u\ag{\z}}\, dV
-\int_{\bd\bodyRn}\Ttn\ag{\z}\cdot\en{\dOn\ag{\z}+\u\ag{\z}}\,d S.
\end{equation*}
and
\begin{equation*}
\An=\pr{\u\ag{\cdot}\in W^{1,p}(\bodyRn;\Reals[m])~~\vert~~\u\ag{\z}\cdot\nn[\alpha]\ag{\z}=\uhn[\alpha]\ag{\z} 
\quad\text{for a.e. }\z\in\GammagRn,\quad
\alpha=1,\ldots,m}.
\end{equation*}
\end{subequations}
As we establish in this section, this formulation satisfies both requirements (E1) and (E2), and hence constitutes an evolutionarily consistent variational formulation in the sense defined above. Accordingly, we refer to $\Pin\ag{\cdot}$ as the \textit{evolutionarily consistent potential energy functional}. 
To see how this functional differs from the standard potential energy functional, we compare its form above with \eqref{eq:PinStandard}. This comparison reveals the following two modifications relative to $\Pin_{\rm{S}}\ag{\cdot}$:
(M1) The potential associated with the applied traction field is expressed in terms of both the geometric and mechanical displacement fields; that is, the integrand of the second integral is modified by replacing $\Ttn\ag{\z}\cdot\u\ag{\z}$ by $\Ttn\ag{\z}\cdot\en{\dOn\ag{\z}+\u\ag{\z}}$;  
(M2) The second integral is evaluated over the entire boundary $\bd\bodyRn$ rather than only over the Neumann boundary $\GammahRn$. %

We first verify that the modifications (M1) and (M2) preserve requirement (E1). The difference between the evolutionarily consistent and the standard potential energy functionals is
\[
\Pin\ag{\u}-\Pin_{\rm{S}}\ag{\u}=-\int_{\bd\bodyRn}\Ttn\ag{\z}\cdot\dOn\ag{\z} \dS-\int_{\GammagRn}\Ttn\ag{\z}\cdot\u\ag{\z}\dS .
\]
The first integral is independent of $\u\ag{\cdot}$ and is thereby constant across $\An$. So too is the second integral, since it depends only on the prescribed displacement condition. That is, on its domain of integration $\GammagRn$, every admissible displacement field $\u\ag{\cdot}\in\An$ satisfies $\u\ag{\z}\cdot\nna\ag{\z}=\uhn[\alpha]\ag{\z}$, $\alpha=1,\ldots, m$. Thus, the displacement entering the second integral is completely determined by the prescribed displacement condition. 
Consequently, $\Pin\ag{\cdot}$ and $\Pin_{\rm{S}}\ag{\cdot}$ differ only by a constant over $\An$ and therefore give rise to the same equilibrium displacement field\footnote{Analogous arguments apply to the general case in which the $\alpha$-Dirichlet boundaries do not coincide}. Since the standard variational formulation satisfies (E1), the evolutionarily consistent variational formulation satisfies (E1) as well.

It remains to establish (E2). 
In \S\ref{subsec:onedimexample}, we use a one-dimensional example to show that the standard potential energy functional may fail to reproduce the physical change in potential energy across evolution states.
We then construct an evolutionarily consistent potential energy formulation for this example and verify that it satisfies (E1) and (E2). The form of this formulation subsequently motivates modifications (M1) and (M2) in the general setting.

\subsection{One-dimensional example}
\label{subsec:onedimexample}
We introduce a one-dimensional example to illustrate the distinction between the standard and evolutionarily consistent variational formulations. Specifically, we show that the analogue of the standard formulation in the present setting satisfies (E1) but fails to satisfy (E2), whereas the analogue of the evolutionarily consistent formulation satisfies both requirements.

\vspace{1em}

\paragraph*{Mechanical setup and evolution }

\begin{figure}[htbp]
\centering
\includegraphics[draft=false,width=\textwidth]{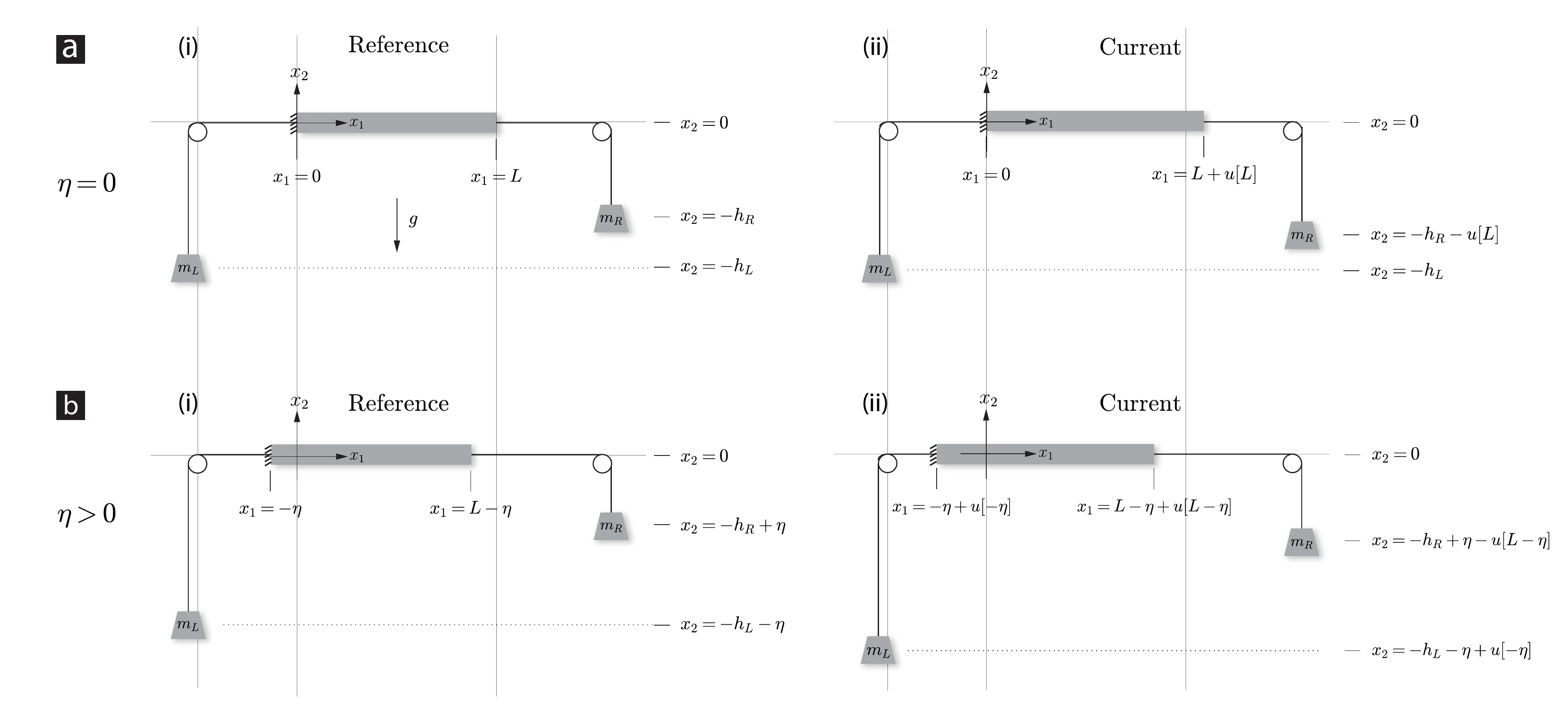}
\caption{Evolution of the one-dimensional elastic bar--mass system. (a) Base state $\eta=0$ and (b) evolved state $\eta>0$. For each state, (i) shows the reference configuration and (ii) the corresponding current configuration determined by the displacement field.}
\label{fig:thoughtExperiment}
\end{figure}

We consider a one-dimensional elastic bar undergoing the evolution shown schematically in Fig.~\ref{fig:thoughtExperiment}. 
We refer to the state $\eta=\baseval$ as the \textit{base} state and, for convenience, take $\baseval=0$ in the present example. We begin by describing the corresponding \textit{base} problem, whose reference configuration is shown in Fig.~\ref{fig:thoughtExperiment}(a)(i). 
The bar has Young's modulus $E$, cross-sectional area $A$, length $L$, and  its reference configuration occupies the region $(0,L)\times \pr{0}\subset \mathbb{R}^2$. Since we are only interested in a one-dimensional treatment of this problem, we have suppressed the bar's transverse dimensions in this representation.
Point masses $\mL$ and $\mR$ hang from the bar's left and right ends, respectively, using inextensible strings of negligible mass that pass over pulleys toward the ground. For the sake of concreteness, we take the point mass $\mL$ to be heavier than $\mR$, i.e., $\mL>\mR$. Let $x_1$ and $x_2$ denote the standard Cartesian coordinates in $\Reals[2]$. The lengths of the inextensible strings and the locations of the pulleys used to hang $\mL$ and $\mR$ are chosen so that the $x_2$-coordinates of $\mL$ and $\mR$ are $-\hL$ and $-\hR$, respectively, where $\hL, \hR>0$. Finally, we assume that the bar is fixed at its left end. 

We compare this base problem with an evolved problem at $\eta>0$, whose reference configuration is shown in Fig.~\ref{fig:thoughtExperiment}(b)(i). In the evolved problem, the bar has undergone geometric evolution and now occupies the region $\en{-\eta, L-\eta}\times\{0\}\subset\Reals[2]$.
The pulleys remain fixed in space, while the same point masses $\mL$ and $\mR$ remain suspended from the corresponding ends of the bar by the same inextensible strings of negligible mass. Consequently, as the reference configuration of the bar effectively translates in space, the positions of the hanging masses also translate. In particular, their $x_2$-coordinates are now $-(\hL+\eta)$ and $-(\hR-\eta)$, respectively. We further assume that the prescribed displacement at the left end of the bar has evolved from $0$ in the base problem to $-\eta$ in the evolved problem.

\vspace{1em}
\paragraph*{Standard potential energy formulation }

We first formulate the base and evolved variational problems using the one-dimensional analogue of the standard variational formulation introduced in \eqref{eq:standardformulation}, and show that it fails to correctly capture the physical change in potential energy.

For the base problem, the standard potential energy functional and a corresponding admissible set are given by
\begin{subequations}
\begin{equation}
    \PiO_{\rm{S}}\ag{u}=\int_{0}^L \frac{1}{2}E A (u'[x_1])^2\, dx_1 -(\mR g)u\ag{L},
\label{eq:typical1DbasePE}
\end{equation}
where $g$ is the gravitational acceleration, and 
\begin{equation}
  \AO=\pr{
  u\ag{\cdot}\in W^{1,2}\en{\en{0,L};\Reals}\,~\vert~\,u\ag{0}=0}.
\label{eq:typical1Dbaseu}
\end{equation}
The variational problem defined by \eqref{eq:typical1DbasePE}--\eqref{eq:typical1Dbaseu} admits the unique solution $\ubOroman\ag{\cdot}:(0,L)\to \Reals$,
\begin{equation}
\ubOroman\ag{x_1}= \frac{\mR g}{E A}x_1
\label{eq:ubO1D}
\end{equation}
\end{subequations}
The equilibrium configuration of the base problem is represented schematically in Fig.~\ref{fig:thoughtExperiment}(a)(ii).

For the evolved problem, the standard potential energy functional and a corresponding admissible set are given by
\begin{subequations}
\begin{equation}
      \Pin_{\rm{S}}\ag{u}=\int_{-\eta}^{L-\eta} \frac{1}{2}E A (u'[x_1])^2 dx_1 - (\mR g)u\ag{L-\eta},
\label{eq:typical1DEvPE}
\end{equation}
and
\begin{equation}
    \An=\pr{u\ag{\cdot}\in W^{1,2}((-\eta,L-\eta);\Reals)\,~\vert~u\ag{-\eta}=-\eta}.
\label{eq:typical1DEvu}
\end{equation}
\label{eq:typical1DEvadmissible}
The variational problem defined by \eqref{eq:typical1DEvPE}--\eqref{eq:typical1DEvu} admits the unique solution $\ubnroman\ag{\cdot}:(-\eta,L-\eta)\to \Reals$,
\begin{equation}
\ubnroman\ag{x_1}= \frac{\mR g}{E A}(x_1+\eta)-\eta.
\label{eq:typical1DEvub}
\end{equation}
\end{subequations}
The equilibrium configuration of the evolved problem is represented schematically in Fig.~\ref{fig:thoughtExperiment}(b)(ii).

We now compute the change in potential energy predicted by the standard variational formulation, and in particular, the associated standard potential energy functional. 
Let $\PibO_{\rm{S}}:=\PiO_{\rm{S}}\ag{\ubOroman}$ and $\Pibn_{\rm{S}}:=\Pin_{\rm{S}}\ag{\ubnroman}$ denote the equilibrium potential energies of the base and evolved problems, respectively. We obtain
\[
    \PibO_{\rm{S}}=-\frac{(\mR g)^2L}{2AE}, \quad \Pibn_{\rm{S}}=-\frac{(\mR g)^2L}{2AE}+(\mR g)\eta.
\]
Accordingly, the standard potential energy functionals predict that the equilibrium potential energy changes by
\begin{equation}
\Delta\overline{\Pi}_{\rm{S}}:=\Pibn_{\rm{S}}-\PibO_{\rm{S}}=(\mR g)\eta,
\label{eq:naivePred}
\end{equation}
as the system evolves from the state $\eta=0$ to $\eta>0$. \textit{Is this change consistent with the actual change in the total potential energy of the underlying physical system?}

To determine the actual change in the total potential energy, consider the limit $L\to0$ of the equilibrium configurations of the base and evolved problems. In this limit, the two point masses are connected directly to each other, and the equilibrium total potential energy reduces to that of a simple particle system. It is therefore sufficient to determine the positions of the two masses in the equilibrium configurations. The equilibrium configurations for arbitrary $L>0$ are shown schematically in Fig.~\ref{fig:thoughtExperiment}(a)(ii) and (b)(ii). Taking the limit $L\to0$ in these configurations, we find that the $x_2$-coordinates of the masses $\mL$ and $\mR$ in the base problem are $-\hL$ and $-\hR$, respectively, whereas in the evolved problem, they are $-(\hL+2\eta)$ and  $-(\hR-2\eta)$, respectively. It follows directly that the change in equilibrium total potential energy is $-2(\mL-\mR)g\eta$. A straightforward calculation shows that the same change in equilibrium total potential energy is obtained for arbitrary $L>0$. Hence, the actual change in equilibrium total potential energy associated with the evolution of the system is 
\begin{equation}
\Delta\overline{\Pi}_{\rm{A}}=-2(\mL-\mR)g\, \eta.
\label{eq:actChange}
\end{equation}
Comparison with \eqref{eq:naivePred} gives $\Delta\Pib_{\rm_{S}}\neq\Delta\Pib_{\rm_{A}}$.
Therefore, although the variational formulation with standard potential energy functional satisfies requirement (E1), it fails to satisfy requirement (E2). That is, while it correctly characterizes equilibrium within each individual problem, its equilibrium potential energy values do not correctly reproduce the physical change in total potential energy across the base and evolved problems.

\paragraph*{Limitation of the standard potential energy formulation}

The source of this discrepancy in the preceding example lies in how the standard potential energy functional is constructed. Since its purpose is to characterize equilibrium, it is customary to omit contributions that do
not affect the attained equilibrium solution. In particular, contributions that are constant over the admissible set may be omitted. While such contributions are immaterial to equilibrium at a fixed $\eta$, a contribution that is constant over the admissible set need not be constant with respect to $\eta$ and may therefore contribute to the change in total potential energy across evolution states. It is precisely this change in total potential energy that the evolutionarily consistent formulation must correctly capture. We therefore examine the base and evolved problems more closely to identify contributions to the total potential energy that have been omitted from their standard potential energy functionals.

Consider first the base problem. The total potential energy of the system consists of the elastic strain energy stored in the bar and the gravitational potential energies of the two hanging masses.\footnote{We neglect the
gravitational potential energy associated with the weight of the bar itself, as it is not relevant to the conceptual point of this example.}  The elastic strain energy is fully retained in the standard potential energy functional \eqref{eq:typical1DbasePE}; we therefore focus on the gravitational contributions. 
In the current configuration of the base problem, the vertical coordinates of $\mL$ and $\mR$ are $-\hL+u\ag{0}$ and $-\hR-u\ag{L}$, respectively (see Fig.~\ref{fig:thoughtExperiment}(a)(ii)). Taking $x_2=0$ as the gravitational potential-energy datum, and since the left end of the bar is prescribed by $u[0]=0$, their gravitational potential energies are 
\[ 
-(\mL g) \hL
\qquad\text{and}\qquad
-(\mR g)(\hR+ u\ag{L}),
\]
respectively.
Consequently, the total potential energy of the base problem, evaluated at $u[\cdot]\in\AO$, may be written as
\begin{equation}
    \int_0^L \frac{1}{2}EA\bigl(u'[x_1]\bigr)^2\,dx_1
    -(\mL g)\hL
    -\mR g(\hR+u\ag{L})
    \label{eq:baseTotalPot}
\end{equation}
Comparison with \eqref{eq:typical1DbasePE} shows that this differs from the standard potential energy functional by the terms
\begin{equation}
    -(\mL g)\hL-(\mR g)\hR.
    \label{eq:omittedBase}
\end{equation}
These terms represent the contributions which are constant over $u\ag{\cdot}\in\AO$ and hence have been omitted in the standard potential energy functional formulation. 

Now consider the evolved problem. In its current configuration, the vertical coordinates of $\mL$ and $\mR$ are $-\hL-\eta+u\ag{-\eta}$ and $-\hR+\eta-u\ag{L-\eta}$, respectively (see Fig.~\ref{fig:thoughtExperiment}(b)(ii)). Taking $x_2=0$ as the same gravitational potential energy datum, their gravitational potential energies are 
\[ 
-(\mL g) (\hL+\eta-u\ag{-\eta})
\qquad\text{and}\qquad
-(\mR g)(\hR-\eta+u\ag{L-\eta}),
\]
respectively. The corresponding potential energy functional of the evolved problem is therefore
\begin{align}
    \int_{-\eta}^{L-\eta}
    \frac{1}{2}EA\bigl(u'[x_1]\bigr)^2\,dx_1
    -\mL g\bigl(\hL+\eta-u[-\eta]\bigr)
    -\mR g\bigl(\hR-\eta+u[L-\eta]\bigr).
    \label{eq:evTotalPot}
\end{align}
Using the prescribed condition $u[-\eta]=-\eta$ and comparing with \eqref{eq:typical1DEvPE} shows that this differs from the standard potential energy functional by the terms
\begin{equation}
    -(\mL g)(\hL+2\eta)-(\mR g)(\hR-\eta)
    \label{eq:omittedEv}
\end{equation}
Again, these terms represent the contributions which are constant over $u\ag{\cdot}\in\An$ and hence have been omitted in the standard potential energy functional formulation. 

It is straightforward to verify that the difference between the equilibrium values of \eqref{eq:baseTotalPot} and \eqref{eq:evTotalPot} indeed reproduces the physical change in total potential energy, $\Delta\bar{\Pi}_A$. Having established this, we now identify which of the contributions omitted from the standard potential energy functional formulation are responsible for its failure to reproduce this change. To this end, we compare \eqref{eq:omittedBase}, the contributions omitted from the standard variational formulation of the base problem, with \eqref{eq:omittedEv}, those omitted from the standard variational formulation of the evolved problem. Their difference is
\begin{equation}
    (-2\mL+\mR)g\eta.
\end{equation}
Thus, in the present example, the contributions omitted from the standard potential energy functional, while constant over the admissible set at each fixed $\eta$, vary with $\eta$ and therefore contribute to the physical change in total potential energy across evolution states. By omitting these contributions, the standard formulation fails to reproduce this change, hence not satisfying the requirement (E2).

\vspace{1em}

\paragraph*{Evolutionarily consistent potential energy formulation}
We now construct a potential energy functional that retains the contributions identified above while preserving the equilibrium solutions obtained from the standard variational formulation. We will then verify that the variational formulation based on this functional satisfies (E1) and (E2), and hence is evolutionarily consistent. Its form will subsequently motivate modifications (M1) and (M2) in the general formulation. 

In the preceding analysis, we considered the total potential energy of the current configuration of the base problem in \eqref{eq:baseTotalPot} and that of the evolved problem in \eqref{eq:evTotalPot}. These were both expressed by taking $x_2=0$ as a common gravitational potential-energy datum for the hanging masses. Rather than introducing such an arbitrary spatial datum, we instead describe the gravitational potential energies of the hanging masses relative to their positions in the reference configuration of the base problem\footnote{This choice is also necessary to obtain a formulation whose evolutionary potential-energy change is invariant under translations of the spatial coordinate system when the applied loading itself evolves. A formulation based instead on absolute spatial position would, in general, introduce an origin-dependent contribution when the loading varies with $\eta$.}.  

To make this concrete, let us first recall the positions of the hanging masses in the reference configuration of the base problem. The left and right hanging masses are located at $x_2=-\hL$ and $x_2=-\hR$, respectively. Now let us describe their gravitational potential energies in the current configuration of the base problem relative to these positions. Since the strings are inextensible and the pulleys are fixed, in a configuration determined by $u\ag{\cdot}$, the left and right hanging masses are located at $x_2=-\hL+u\ag{0}$ and $x_2=-\hR-u[L]$, respectively (see Fig.~\ref{fig:thoughtExperiment}(a)(ii)). Thus, their changes in vertical position relative to the reference configuration are $u[0]$ and $-u[L]$. Their gravitational potential energies relative to the reference configuration are therefore
\[
    (\mL g)u[0]
    \qquad\text{and}\qquad
    -(\mR g)u[L].
\]
The corresponding potential energy functional is therefore
\begin{equation}
    \Pi^\circ[u]
    =
    \int_0^L \frac{1}{2}EA\bigl(u'[x_1]\bigr)^2\,dx_1
    +(\mL g)u[0]
    -(\mR g)u[L].
    \label{eq:consistent1DbasePE}
\end{equation}

Now consider the evolved problem. In the current configuration of the evolved problem, the left and right hanging masses are located at $x_2=-\hL-\eta+u[-\eta]$ and $x_2=-\hR+\eta-u[L-\eta]$, respectively (see Fig.~\ref{fig:thoughtExperiment}(b)(ii)). Hence, their changes in vertical position relative to their positions in the reference configuration of the base problem are 
$-\eta+u[-\eta]$ and  $\eta-u[L-\eta]$, respectively. Their gravitational potential energies relative to the reference configuration are therefore
\[
    \mL g(-\eta+u[-\eta])
    \qquad\text{and}\qquad
  -\mR g(-\eta+u[L-\eta]),
\]
respectively. The corresponding potential energy functional is therefore
\begin{align}
    \Pi^\eta[u]
    &=
    \int_{-\eta}^{L-\eta}
    \frac{1}{2}EA\bigl(u'[x_1]\bigr)^2\,dx_1
    +\mL g\bigl(-\eta+u[-\eta]\bigr)
    -\mR g\bigl(-\eta+u[L-\eta]\bigr).
    \label{eq:consistent1DEvPE}
\end{align}

We now show that this resulting formulation indeed satisfies (E1) and (E2). Since $u[-\eta]=-\eta$ for every $u\ag{\cdot}\in\An$, the difference between \eqref{eq:consistent1DEvPE} and the corresponding standard potential energy functional is constant over $\An$.  Likewise, \eqref{eq:consistent1DbasePE} differs from the standard potential energy functional for the base problem only by a term that is constant over $\AO$. Hence, the equilibrium displacement fields obtained from the standard formulation are preserved, verifying (E1).  For concreteness, recall that these equilibrium displacement fields are
\begin{equation}
\ubOroman\ag{x_1}= \frac{\mR g}{E A}x_1\quad\text{and}\quad 
\ubnroman\ag{x_1}= \frac{\mR g}{E A}(x_1+\eta)-\eta,
\end{equation}
respectively. To verify (E2), we evaluate \eqref{eq:consistent1DbasePE} and \eqref{eq:consistent1DEvPE} at these equilibrium displacement fields. This gives
\[
    \PibO
    =
    -\frac{(\mR g)^2L}{2EA} \quad \text{and}\quad  
    \Pibn
    =
    -\frac{(\mR g)^2L}{2EA}
    -2(\mL-\mR)g\eta.
\]
Consequently,
\begin{equation}
    \Pibn-\PibO
    =
    -2(\mL-\mR)g\eta
    =
    \Delta\Pib_A.
\end{equation}
Thus, the formulation preserves the equilibrium displacement fields of the standard formulation while correctly reproducing the physical change in potential energy across evolution states, satisfying both (E1) and (E2)
for the present example.

Having established the evolutionary consistency of this one-dimensional formulation, we now examine its form to motivate modifications (M1) and (M2) in the general formulation. First, consider the expressions $-\eta+u[-\eta]$ and $-\eta+u[L-\eta]$ appearing in the gravitational potential-energy terms in \eqref{eq:consistent1DEvPE}.  The quantity $-\eta$ accounts for the change in position from the reference configuration of the base problem to that of the evolved problem. In the general setting, this change in position is described by the geometric map $\xnb\ag{\cdot}$, and the corresponding geometric displacement is $\dOn\ag{\cdot}$. Thus, $-\eta$ is the one-dimensional analogue of $\dOn\ag{\cdot}$. The displacement $\u\ag{\cdot}$, in turn, accounts for the subsequent change in position from the reference configuration of the evolved problem to its current configuration. Consequently,
\[
    -\eta+u\ag{\cdot}
\]
is the one-dimensional analogue of
\[
    \dOn\ag{\cdot}+\u\ag{\cdot},
\]
which represents the \textit{total displacement} from the reference configuration of the base problem to the current configuration of the evolved problem. This motivates modification (M1) introduced above.

Next, consider the gravitational contribution associated with $\mL$ in \eqref{eq:consistent1DEvPE}, 
\[\mL g\en{-\eta+u\ag{-\eta}}.\]
The displacement of the left end of the bar, to which the mass $\mL$ is attached, is prescribed as $u[-\eta]=-\eta$. As such, the left end of the bar represents a Dirichlet boundary in the one-dimensional problem. Consequently, for each fixed $\eta$, the above contribution is constant over $\An$ as $-2\mL g\eta$ and does not affect the equilibrium displacement field. Its value, however, changes with $\eta$. Physically, this reflects the fact that the position of the mass $\mL$, which is coupled to the left end of the bar, varies across evolution states. In the present example, this change in position results from both the geometric evolution of the bar, which shifts its left end from $x_1=0$ to $x_1=-\eta$, and the $\eta$-dependent prescribed displacement $u[-\eta]$ at that end. Consequently, its gravitational potential energy can change with $\eta$.  Thus, this contribution, although immaterial to equilibrium at each fixed $\eta$, contributes to the change in total potential energy across evolution states and must be retained in an evolutionarily consistent formulation.

The same consideration applies to a Dirichlet portion of the boundary in the general setting. Although the mechanical displacement is prescribed there, its position in the current configuration may vary across evolution states. This variation may result from both the evolution of the reference configuration through the geometric map $\xnb\ag{\cdot}$ and the evolution of the prescribed displacement. External loading coupled to such a boundary may therefore contribute to the change in total potential energy, even though its potential is constant over the admissible set at each fixed $\eta$. Such contributions must consequently be retained on Dirichlet portions as well as Neumann portions of the boundary. This motivates modification (M2), under which the potential associated with the applied traction is integrated over the entire boundary.

\begin{remark} 
\small
Under modification (M2), we allow the applied traction field $\Ttn\ag{\cdot}$ to be defined over the entire boundary, including its Dirichlet portions. This may appear to imply that both displacement and traction are imposed as boundary conditions on the same portion of the boundary. This is not the intended interpretation. 
When we consider an applied traction on a Dirichlet portion of the boundary, it is not imposed as an additional boundary condition. Thus, it need not coincide with the equilibrium traction field, as will be made apparent from the equilibrium conditions we derive in \S\ref{sec:equilibrium_conditions}. We consider an applied traction on a Dirichlet portion when there exists external loading acting on it whose potential is coupled to the motion of the constrained boundary.

Physically, one may regard the applied traction on the Dirichlet boundary representing an external loading acting through the device which imposes boundary constraint. 
Our formulation is appropriate when any deformation or geometric evolution of the constraining device itself has a negligible effect on the energy comparison, so that the external loading may be regarded as coupled directly to the motion of the body's boundary. Otherwise, the device itself should be included as part of the mechanical system, with its energetic contribution incorporated into the potential energy of the enlarged system.
\end{remark}

\subsection{Common representation of the potential energy functionals}
\label{subsec:parentPE}

The physical interpretation developed in the preceding section provides the basis for comparing equilibrium potential energies associated with different evolution states. Such a comparison implicitly regards the potential energy as a single physical quantity whose value changes as the system evolves, even though the potential energy at each evolution state is represented by a distinct functional $\Pin\ag{\cdot}$ defined on a distinct admissible set $\An$. To make this viewpoint explicit, we seek to represent this physical quantity across evolution states by a single parent potential energy functional defined on a common parent displacement space.

The immediate difficulty is that the potential energy functional $\Pin\ag{\cdot}$ at each $\eta$-state is defined for displacement fields on the corresponding reference configuration $\bodyRn$, and hence the displacement fields associated with different values of $\eta$ are not defined on a common reference configuration. We therefore first place the $\eta$-family of variational problems on a common reference configuration. We take this common reference configuration to be $\bodyRO$ and pull back both the admissible displacement fields and the potential energy expressions onto it. For a fixed $p>m$, define the common displacement space
\[
\AC:=W^{1,p}(\bodyRO;\Reals[m]).
\]
The resulting pulled-back admissible sets are
\begin{equation}
\Anc=
\pr{
\u\ag{\cdot}\in\AC
~\vert~
\u\ag{\x}\cdot\nnc[\alpha]\ag{\x}
=\uhnc[\alpha]\ag{\x}
\quad\text{for a.e. }\x\in\GammagROa,
\quad \alpha=1,\ldots,m
},
\label{eq:admissblesetYcn}
\end{equation}
so that every $\Anc$ is contained in $\AC$. We refer to $\AC$ as the parent displacement space. Correspondingly, expressing the volume and boundary integrals defining $\Pin\ag{\cdot}$ on $\bodyRO$ and $\bd\bodyRO$, respectively, yields the family of pulled-back potential energy functionals
\begin{equation}
\pr{\check{\Pi}^{\eta}\ag{\cdot}:\Anc\to\Reals}_{\eta}.
\label{eq:pulledback}
\end{equation}

The pulled-back potential energy functionals in \eqref{eq:pulledback} continue to differ in their corresponding admissible sets $\Anc$. Crucially, however, these admissible sets are now all contained in the same parent displacement space $\AC$. This common setting allows us to construct a unified representation of the $\eta$-parameterized family of pulled-back potential energy functionals through a single parent potential energy functional defined on $\AC$. Such a representation arises naturally here because the expression for $\check{\Pi}^{\eta}\ag{\u}$ is well defined not only for $\u\ag{\cdot}\in\Anc$, but for any $\u\ag{\cdot}\in\AC$\footnote{This statement is understood under the general regularity and integrability requirement stated at the beginning of \S\ref{sec:VE} and the field-specific assumptions introduced in its subsequent subsections. For example, the standard $p$-growth conditions on the generalized potential-energy density recorded in Appendix B ensure that its integral is finite for every $\u\ag{\cdot}\in\AC$, while the regularity of the geometric maps, the piecewise continuity of the pulled-back applied traction, and $p>m$ ensure that the boundary integral is well defined.}.

We therefore define
\[
\Pi:\Reals\times\AC\to\Reals
\]
by taking, for each $\eta$, $\Pi\ag{\eta,\cdot}$ to be the extension of $\check{\Pi}^{\eta}\ag{\cdot}$ to $\AC$ given by the same expression. By construction,
\begin{equation}
\left.\Pi\ag{\eta,\cdot}\right|_{\Anc}
=\check{\Pi}^{\eta}\ag{\cdot}.
\label{eq:ParentPE}
\end{equation}

We now return to the equilibrium potential energies that motivated this construction. Recall that $\ubn\ag{\cdot}\in\An$ denotes the equilibrium displacement field of the $\eta$-variational problem and that $\ubnc\ag{\cdot}\in\Anc$ denotes its pullback. Since the pullback preserves the value of the potential energy, and since \eqref{eq:ParentPE} applies to $\ubnc\ag{\cdot}\in\Anc$, we have
\[
\Pin\ag{\ubn}
=\check{\Pi}^{\eta}\ag{\ubnc}
=\Pi\ag{\eta,\ubnc}.
\]
Hence, the comparison of equilibrium potential energies across the variational evolution may equivalently be viewed as the comparison of values of the single parent potential energy functional $\Pi$.

\section{Preliminaries for obtaining a closed-form expression for the \prr}
\label{sec:preliminaries}
In this section, we collect the notation and preliminary results needed to derive the closed-form expression for the \prr $F_c^\circ$ in \S\ref{sec:PERR}. 
In \S\ref{sec:Notation}, we introduce the notation and mathematical conventions used throughout the derivation; these, together with the principal symbols used to denote the quantities considered in this work, are collected in the tables in Appendix~\ref{app:notation}.
In \S\ref{sec:equilibrium_conditions}, we derive the equilibrium conditions associated with each variational problem at $\eta$. In \S\ref{sec:pullbackPinb}, we present the equilibrium potential energy at $\eta$ pulled back onto the fixed reference domains associated with the base problem, in the form that will be used to evaluate its rate of change as $\eta\to\baseval$. Finally, in \S\ref{subsec:limits_of_functions}, we collect the limits as $\eta\to\baseval$ of several $\eta$-dependent expressions and quantities that will be used in the derivation of the closed-form expression for $F_c^\circ$.

\subsection{Notation}
\label{sec:Notation}
We begin by introducing the notation and conventions used throughout the derivation, including notation for derivatives and several shorthand conventions for quantities and operations that recur in the analysis.

\subsubsection{Symbol conventions}

Throughout this work, sans-serif symbols (e.g., $\x$) denote vectors, vector-valued functions, and matrix-valued functions. Roman symbols (e.g., $x$) denote scalars and scalar-valued functions. 
For quantities denoted by sans-serif symbols, we use centered dots in subscripts to access their components with respect to a chosen basis, as follows.

Let $\x$ be an element of $\Reals[m]$, where $\Reals$ represents the set of all real numbers and $m=2$ or $m=3$. Let the canonical basis of $\Reals[m]$ be $\en{\E[1],\ldots, \E[m]}$\footnote{Each element of the canonical basis, $\E[I]$, is a tuple of length $m$ whose $I$-th entry is $1$ and whose other entries are $0$.}.
We denote the component of $\x$ with respect to the canonical basis vector\footnote{From here on, when we speak of components of elements in $\Reals[m]$, we mean components with respect to the canonical basis.} $\E[i]$ by $\x[i]$, where $i=1,\ldots,m$. 
Similarly, let $\tp{u}\ag{\cdot}: \Reals[m]\to \Reals[m]$ be a vector-valued function. We denote its component function with respect to $\E[i]$ by $\u[i]\ag{\cdot}$, where $i=1,\ldots,m$. The value of the function $\u\ag{\cdot}$ at $\x$ is denoted as $\u\ag{\x}$. 
The convention extends naturally to a matrix-valued function $\tp{H}\ag{\cdot}:\Reals[m]\to\Matrix[m\times m]$; its $iJ$-th component function at $\x$ is denoted by $\H[i][J]\ag{\x}$.

We use a similar convention when taking components with respect to the orthonormal frame field introduced in \S\ref{subsec:dispevol}. Specifically, given $\u\ag{\x}$, we denote its component with respect to $\nna\ag{\x}$ by $\u[\an]\ag{\x}$. That is,
\begin{equation*}
\u[\an]\ag{\x}:=\u\ag{\x}\cdot\nna\ag{\x}.
\end{equation*} 
\subsubsection{Derivative notation}
\label{subsec:derivNot}
We collect here the derivative notation used throughout the derivation.

\paragraph*{Fréchet derivative}
We introduce notation for the Fréchet derivative of functions that will be used in the derivation of the closed-form expression of $F_c^\circ$.  In particular, this notation will be used for derivatives of the generalized potential energy density function $\fn\ag{\cdot}$.
Consider a  mapping $f\ag{\cdot}: U\subset \mathbb{V} \to \mathbb{R}$, where $\mathbb{V}$ is a normed vector space and $U$ is an open subset of the space $\mathbb{V}$. We denote the Fr\'{e}chet derivative of $f\ag{\cdot}$ at $\tp{x}\in U$, if it exists, as the unique continuous linear operator $Df\en{\tp{x}}\ag{\cdot}:\mathbb{V} \to \mathbb{R}$ that satisfies
\begin{equation}
  f\ag{\tp{x}+\tp{h}}=f\ag{\tp{x}}+ Df\en{\tp{x}}\ag{\tp{h}}+\norm{\tp{h}}_{\mathbb{V}}\alpha\ag{\tp{h}} \quad \quad \text{with} \lim_{\norm{\tp{h}}_\mathbb{V} \to 0}\alpha\ag{\tp{h}}=0,
\end{equation}
where we assume that $\tp{h}$ is chosen such that $\tp{x}+\tp{h} \in U$. We refer to $\alpha\ag{\tp{h}}$ as the normalized remainder associated with $\f\ag{\cdot}$ at $\x$.

\paragraph*{Partial derivative of functions on $\Reals[m]$}
\label{subsec:partialderiv}
For functions whose domain is $\Reals[m]$, we introduce the following notation for their partial derivatives. We denote by $\u[i,J]\ag{\x}$ the partial derivative of $\u[i]\ag{\cdot}$ with respect to $\x[J]$, where 
\begin{equation}
\u[i,J]\ag{\x}:=\lim_{\alpha\to 0}\frac{\u[i]\ag{\x+\alpha \e[J]}-\u[i]\ag{\x}}{\alpha}, \quad i, J=1,\ldots,m.
\end{equation}
We also define 
\begin{equation}
\grad\u\ag{\cdot}:=\u[i,J]\ag{\cdot}\E[iJ],
\label{eq:nablaudef}
\end{equation}
which is the gradient of $\u\ag{\cdot}$. In \eqref{eq:nablaudef}, $\E[iJ]$ is the element of the canonical basis\footnote{The matrix representation of $\E[iJ]$ is a $m\times m$ matrix whose $(i,J)$-th entry is $1$ and all other entries are $0$.} of $\Matrix[m\times m]$ and we employ Einstein summation convention. We follow a similar convention to denote the higher-order derivatives of $\u\ag{\cdot}$. For instance,
\begin{equation}
\u[i,J,K]\ag{\cdot}=(\u[i,J])_{,K}\ag{\cdot}, \quad i,J,K=1,\ldots,m.
\end{equation}

\paragraph*{Partial derivatives of functions on $\Reals[m]\times\Reals[m]\times\Matrix[m\times m]$} We next introduce notation for partial derivatives of functions of the form $\f\ag{\cdot}:\Reals[m]\times\Reals[m]\times\Matrix[m\times m]$. Specifically, the notation will be used to denote partial derivatives of the functions related to the generalized potential energy density, namely $\fO\ag{\cdot,\cdot,\cdot}$ and $\zetaO\ag{\cdot,\cdot,\cdot}$, which we introduced in \S\ref{subsec:fnevol}.

When all the partial derivatives of $\f\ag{\cdot}$ exist at $\en{\x,\u,\H}\in \Reals[m]\times\Reals[m]\times\Matrix[m\times m]$, we describe them as follows.  
When a function $\f\ag{\cdot}$ has as a subscript an upper case letter preceded by a comma, for instance $\f[,J]$, it denotes the partial derivative of $\f\ag{\cdot}$ with respect to $\x[J]$,  $\f[,J]\ag{\cdot}:\Reals[m]\times\Reals[m]\times\Matrix[m\times m]\to\Reals$,  defined as  
\begin{align}
\f[,J]\ag{\x, \u, \H}=\lim_{\alpha\to 0}\frac{\f\ag{\x+\alpha \e[J], \u, \H}-\f\ag{\x,\u,\H}}{\alpha}, \quad J=1,\ldots,m.
\end{align}
When a function $\f\ag{\cdot}$ has as a subscript a small case letter preceded by a comma, for instance $\f[,i]$, it denotes the partial derivative of $\f$ with respect to $\u[i]$, $\f[,i]\ag{\cdot}:\Reals[m]\times\Reals[m]\times\Matrix[m\times m]\to\Reals$, defined as  
\begin{align}
\f[,i]\ag{\x,\u,\H}=\lim_{\alpha\to 0}\frac{\f\ag{\x, \u+\alpha \e[i], \H}-\f\ag{\x, \u, \H}}{\alpha}, \quad i=1,\ldots,m.
\end{align}
When a function $\f\ag{\cdot}$ has as a subscript a sequence of a small case letter and then an upper case letter preceded by a comma, for instance $\f[,iJ]$, it denotes the partial derivative of $\f$ with respect to $\u[i,J]$, $\f[,iJ]\ag{\cdot}:\Reals[m]\times\Reals[m]\times\Matrix[m\times m]\to\Reals$, defined as
\begin{align}
\f[,iJ]\ag{\x,\u,\H}=\lim_{\alpha\to 0}\frac{\f\ag{\x,\u,\H+\alpha \E[iJ]}-\f\ag{\x,\u,\H}}{\alpha}, \quad i,J=1,\ldots,m.
\end{align}
We denote the higher-order derivatives similarly. For example,
\begin{align*}
\f[,J,k]\ag{\cdot}=(\f[,J])_{,k}\ag{\cdot}, &\quad J,k=1,\ldots,m, \\
\f[,i,jK]\ag{\cdot}=(\f[,i])_{,jK}\ag{\cdot}, &\quad i,j,K=1,\ldots,m, \\
\f[,iJ,K]\ag{\cdot}=(\f[,iJ])_{,K}\ag{\cdot}, &\quad i,J,K=1,\ldots,m.
\end{align*}

\paragraph*{Total derivative of functions}
We denote the total derivative of a function with respect to $\x[J]$ by $d/d\x[J]$. In \S\ref{sec:PERR}, the total derivative $d/d\x[J]$ of $\fO\ag{\cdot,\cdot,\cdot}$ evaluated at $\en{\x,\ubO\ag{\x},\grad\ubO\ag{\x}}$ arises in the derivation of the closed-form expression of $F_c^\circ$. We expand it explicitly here for concreteness. 
\begin{equation}
\begin{split}
&\frac{d}{d\x[J]} \fO\ag{\x,\ubO\ag{\x}, \grad \ubO\ag{\x}}\\
&=\fO[,J]\ag{\x,\ubO\ag{\x}, \grad \ubO\ag{\x}}
+\fO[,i]\ag{\x,\ubO\ag{\x}, \grad \ubO\ag{\x}}\ubO[i,J]\ag{\x}
+\fO[,iK]\ag{\x,\ubO\ag{\x}, \grad \ubO\ag{\x}}\ubO[i,K,J]\ag{\x}
\end{split}
\end{equation}

\subsubsection{Shorthand conventions}
We introduce several shorthand conventions used to simplify expressions throughout the derivation.

\paragraph*{Limits}
We interchangeably write $\Bv[\eta\to\baseval]$ in place of $\lim_{\eta\to\baseval}$ where convenient for conciseness. That is, for example,
\[
\frac{\fn\ag{\x,\u,\H}-\fO\ag{\x,\u,\H}}{\eta-\baseval}\Bv[\eta\to\baseval]:=\lim_{\eta\to\baseval}\frac{\fn\ag{\x,\u,\H}-\fO\ag{\x,\u,\H}}{\eta-\baseval}.
\]
We distinguish this from the standard notation for evaluation at $\eta=\baseval$, denoted by $\Bv[\eta=\baseval]$, where the arrow is replaced by an equal sign.

\paragraph*{Evaluation and argument-suppression conventions}
To improve readability, we occasionally suppress the arguments of functions when they are clear from context. In particular, when no ambiguity can arise, we suppress the spatial argument $\ag{\x}$ in integrand expressions.
For example, we write
\[
    \int_{\bd\bodyRO}\TtO\cdot\phiO\dS
\]
we mean $\int_{\bd\bodyRO}\TtO\ag{\x}\cdot\phiO\ag{\x}\dS.$

 For the generalized potential energy density and related quantities evaluated at equilibrium, we use an overbar to suppress their arguments. Thus, for example, we write
\[
    \fbO:=\fO\ag{\x,\ubO,\grad\ubO} \qquad \fbO[,\bullet]:=\fO[,\bullet]\ag{\x,\ubO,\grad\ubO}\qquad \zetabO:=\zetaO\ag{\x,\ubO,\grad\ubO},
\]
where $\bullet$ in the subscript denotes any of $I$, $i$, or $iJ$.

\subsection{Equilibrium conditions}
\label{sec:equilibrium_conditions}

\begin{figure}[!t]
\centering
\includegraphics[draft=false,width=0.70\textwidth]{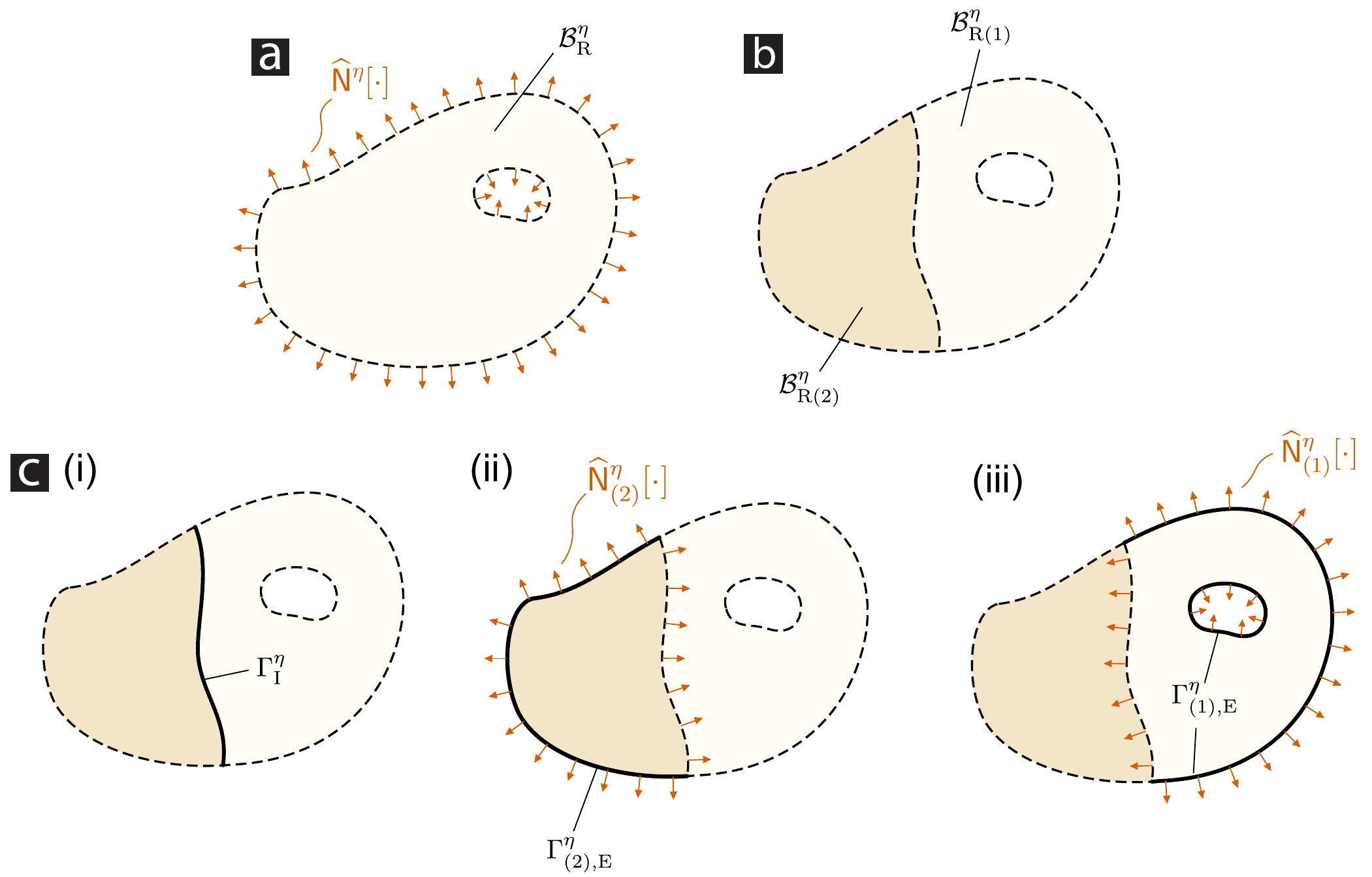}
\caption{Boundary and interface notation used in the equilibrium conditions. (a) The outward unit normal on $\bd\bodyRn$. (b) A body decomposed into two material regions, $\bodyRn[1]$ and $\bodyRn[2]$, separated by a sharp material interface. (c) The decomposition $\bd\bodyRn[r]=\GammaIn\cup\GammaErn$ of the boundary of each material region into: (i) the internal material interface $\GammaIn$, common to the boundaries of the two material regions and lying in the interior of the body; (ii) the portion $\GammaECn[2]$ of the boundary of the body that lies on the boundary of the material region indexed by $2$, together with the outward unit normal $\widehat{\tp{N}}_{(2)}^\eta$ to $\bd\bodyRn[2]$; and (iii) the portion $\GammaECn[1]$ of the boundary of the body that lies on the boundary of the material region indexed by $1$, together with the outward unit normal $\widehat{\tp{N}}_{(1)}^\eta$ to $\bd\bodyRn[1]$.}
\label{fig:boundaryNotation}
\end{figure}

In this section, we present the equilibrium conditions associated with the variational problem at each value of $\eta$. The final equilibrium conditions are summarized in \eqref{eq:equicondb}. The boundary and interface decompositions used in this section, together with the associated outward-normal fields, are illustrated in Fig.~\ref{fig:boundaryNotation}.

Specifically, we derive the equilibrium conditions from the first-order stationarity condition satisfied by $\ubn\ag{\cdot}$. Since $\ubn\ag{\cdot}$ is a local minimizer of the potential energy functional $\Pin\ag{\cdot}$ over the admissible set $\An$, the first-order stationarity condition is that

\begin{equation}
    \langle \delta\Pin\ag{\ubn}, \vu \rangle =0 \text{~for~all~} \vu \ag{\cdot}\in T\An,
    \label{eq:stationarity}
\end{equation}
where
\begin{equation*}
    \langle \delta\Pin\ag{\ubn}, \vu \rangle := \frac{d}{d\epsilon}\Pin\ag{\ubn\ag{\cdot}+\epsilon\vu\ag{\cdot}}\Bv[\epsilon=0]
\end{equation*}
is the G\^ateaux derivative of $\Pin\ag{\cdot}$ at $\ubn\ag{\cdot}$ in the direction $\vu\ag{\cdot}$ and 
\begin{align*}
T\An&=
\pr{
\vu\ag{\cdot}\in W^{1,p}\en{\bodyRn;\Reals[m]}~
\vert~\vu\ag{\z}\cdot\nna\ag{\z}=0 \quad\text{~for a.e. } \z\in\GammagRna, \quad \alpha=1,\ldots, m
}
\end{align*}
is the space of admissible variations. Substituting the definition of $\Pin\ag{\cdot}$, given in \eqref{eq:PibnDef}, into the first-order stationarity condition \eqref{eq:stationarity} and differentiating under the integral sign yields
\begin{subequations}
\begin{equation}
\langle \vPin\ag{\ubn},\vu\rangle = I^{\bodyRn}+I^{\bd\bodyRn}=0 \quad\text{~for all }\vu\ag{\cdot}\in \T\An
\label{eq:stationarityRewrite1}
\end{equation}
where
\begin{equation}
I^{\bodyRn}=\int_{\bodyRn} \frac{d}{d\epsilon}\, 
\fn\ag{\z,\ubn\ag{\z}+\epsilon\vu,\grad\ubn\ag{\z}+\epsilon\grad\vu\ag{\z}}\Bv_{\epsilon=0} \dV,
\label{eq:IbodyRn1}{}
\end{equation}
\begin{equation}
I^{\bd\bodyRn}=-\int_{\bd\bodyRn}\frac{d}{d\epsilon}\,\Ttn\ag{\z}\cdot\en{\dOn\ag{\z}+\ubn\ag{\z}+\epsilon\vu\ag{\z}}\dS\Bv[\epsilon=0].
\label{eq:GateauEpsilon}
\end{equation}
\end{subequations}
We first simplify the term $I^{\bodyRn}$. Applying the chain rule to its integrand, evaluating the resulting expression at $\epsilon=0$, and formally integrating by parts\footnote{More specifically, the regularity assumption $\ubn\ag{\cdot}\in W^{1,p}(\bodyRn;\Reals[m])$ does not by itself ensure that $\frac{d}{d\z[J]}(\fn[,iJ]\ag{\z,\ubn\ag{\z},\grad\ubn\ag{\z}})$ is an ordinary integrable field over $\bodyRn$ or that $\fn[,iJ]\ag{\z,\ubn\ag{\z},\grad\ubn\ag{\z}}\,\Nnh[J]\ag{\z}$ has a well-defined boundary value. We therefore restrict attention to equilibrium solutions possessing sufficient additional regularity for these terms to be well defined.}, we obtain (see \S\ref{sec:Notation} for relevant shorthand notations; here and throughout, Einstein summation over repeated indices is assumed), 
\begin{equation}
\begin{split}
I^{\bodyRn}&=\int_{\bodyRn} \en{\fn[,i]\ag{\z,\ubn\ag{\z},\grad\ubn\ag{\z}}-\frac{d}{d\z[J]}\fn[,iJ]\ag{\z,\ubn\ag{\z},\grad\ubn\ag{\z}}}\vu[i]\ag{\z} \dV\\
&\hspace{4cm}+\int_{\bd\bodyRn}\fn[,iJ]\ag{\z,\ubn\ag{\z},\grad\ubn\ag{\z}}\,\Nnh[J]\ag{\z}\,\,\vu[i]\ag{\z}\dS.
\label{eq:FirsttermSimp1}
\end{split}
\end{equation}
In \eqref{eq:FirsttermSimp1}, $\Nnh\ag{\cdot}:\bd\bodyRn\to\Reals[m]$ is the outward unit normal field, defined almost everywhere on the surface of the body and illustrated in Fig.~\ref{fig:boundaryNotation}(a).

We now introduce the following to further simplify \eqref{eq:FirsttermSimp1}. Specifically, we introduce the vector field $\bn\ag{\cdot}:\bodyRn\to\Reals[m]$ by  identifying its $i$-th component with
\begin{subequations}
\begin{equation}
\bn[i]\ag{\z}=\fn[,i]\ag{\z,\ubn\ag{\z},\grad\ubn\ag{\z}}.
\label{eq:bCdef}
\end{equation}
Likewise, we introduce the tensor field $\FPn\ag{\cdot}:\bodyRn\to\Matrix[m\times m]$ by identifying its $iJ$-th component with
\begin{equation}
  \FPn[i][J]\ag{\z}=\fn[,iJ]\ag{\z,\ubn\ag{\z},\grad\ubn\ag{\z}},
\label{eq:FPndef}
\end{equation}
\end{subequations}
where $\FPn\ag{\cdot}$ is identified as the first Piola--Kirchhoff stress field.
We further define the equilibrium traction field $\Tbn\ag{\cdot}:\bd\bodyRn\to\Reals[m]$, defined almost everywhere on the surface of the body, by
\begin{equation}
\Tbn\ag{\z}:=\FPn\ag{\z}\,\Nnh\ag{\z}.
\label{eq:eqTracFieldHet}
\end{equation}
Substituting \eqref{eq:bCdef}, \eqref{eq:FPndef} into \eqref{eq:FirsttermSimp1}, then subsequently using the definition \eqref{eq:eqTracFieldHet}, we obtain
\begin{equation}
I^{\bodyRn}=\int_{\bodyRn} \en{\bn[i]\ag{\z}-\frac{d}{d\z[J]}\FPn[i][J]\ag{\z}}\vu[i]\ag{\z} \dV+\int_{\bd\bodyRn}\Tbn[i]\ag{\z}\,\,\vu[i]\ag{\z}\dS.
\label{eq:FirsttermSimp3}
\end{equation}
We obtain, in a straightforward manner, that the term $I^{\bd\bodyRn}$ reduces to
\begin{equation}
I^{\bd\bodyRn}=-\int_{\bd\bodyRn}\Ttn[i]\ag{\z}\vu[i]\ag{\z}\dS
\label{eq:ThirdTermSimp1}
\end{equation}

Substituting \eqref{eq:FirsttermSimp3} and \eqref{eq:ThirdTermSimp1} into \eqref{eq:stationarity} and rearranging, we obtain 
\begin{equation}
\begin{split}
\langle \vPin\ag{\ubn},\vu\rangle &=\int_{\bodyRn} \en{\bn[i]\ag{\z}-\frac{d}{d\z[J]}\FPn[i][J]\ag{\z}}\vu[i]\ag{\z} \dV\\
&\hspace{1.5cm}+\int_{\bd\bodyRn}\en{\Tbn[i]\ag{\z}-\Ttn[i]\ag{\z}}\,\,\vu[i]\ag{\z}\dS=0 
\qquad\text{~for all }\vu\ag{\cdot}\in T\An,
\end{split}
\label{eq:StationarityRewrite2}
\end{equation}
Resolving the second integral in \eqref{eq:StationarityRewrite2} with respect to the orthonormal frame field $\en{\nna\ag{\cdot}}_{\alpha=1}^m$, and decomposing the boundary according to $\bd\bodyRn=\GammagRna\cup\GammahRna$ for each $\alpha$, we obtain
\begin{equation}
\begin{split}
\langle \vPin\ag{\ubn},\vu\rangle &=\int_{\bodyRn} \en{\bn[i]\ag{\z}-\frac{d}{d\z[J]}\FPn[i][J]\ag{\z}}\vu[i]\ag{\z} \dV\\
&\hspace{0.5cm}+\sum_{\alpha=1}^m\int_{\GammagRna\cup\GammahRna}\en{\Tbn[\an]\ag{\z}-\Ttn[\an]\ag{\z}}\,\,\vu[\an]\ag{\z}\dS
\qquad\text{~for all }\vu\ag{\cdot}\in T\An,
\end{split}
\label{eq:StationarityRewrite3}
\end{equation}
where, to recall, $\Tbn[\an]\ag{\z}=\Tbn\ag{\z}\cdot\nna\ag{\z}$ and, $\Ttn[\an]\ag{\z}$ and $\vu[\an]\ag{\z}$ are to be understood analogously.
Since $\vu\ag{\cdot}\in T\mathcal{A}$ satisfies $\vu\ag{\z}\cdot\nna\ag{\z}=0$ for $\z\in\GammagRna$, the integrand of the second integral in \eqref{eq:StationarityRewrite3} vanishes over $\GammagRna$. Hence, \eqref{eq:StationarityRewrite3} reduces to
\begin{equation}
\begin{split}
\langle \vPin\ag{\ubn},\vu\rangle &=\int_{\bodyRn} \en{\bn[i]\ag{\z}-\frac{d}{d\z[J]}\FPn[i][J]\ag{\z}}\vu[i]\ag{\z} \dV\\
&+\sum_{\alpha=1}^m\int_{\GammahRna}\en{\Tbn[\an]\ag{\z}-\Ttn[\an]\ag{\z}}\,\,\vu[\an]\ag{\z}\dS
\qquad\text{~for all }\vu\ag{\cdot}\in T\An,
\end{split}
\label{eq:StationarityRewrite4}
\end{equation}

The stationarity condition thus implies the equilibrium conditions
\begin{subequations}
\begin{align}
\bn[i]\ag{\z}-\frac{d}{d\z[J]}\FPn[i][J]\ag{\z}&=0 \quad\text{for a.e. }\z\in\bodyRn\label{eq:equiconda}\\
\Tbn_{\cdot \alpha^\eta}\ag{\z}-\Ttn_{\cdot \alpha^\eta}\ag{\z}&=0\quad\text{for a.e.~}\z\in\GammahRna, \quad \alpha=1,\ldots,m.\label{eq:equicondb}
\end{align}
\end{subequations}

\vspace{1em}
\paragraph*{Equilibrium condition when the body includes sharp material interfaces} When the body includes sharp material interfaces, such as those schematically illustrated in Fig.~\ref{fig:boundaryNotation}(b), the potential energy functional is of the form given in \eqref{eq:Pibnhet}. Substituting this definition of $\Pin\ag{\cdot}$ into the stationarity condition \eqref{eq:stationarity} and differentiating under the integral sign yields
\begin{subequations}
\begin{equation}
\langle \vPin\ag{\ubn},\vu\rangle = \sum_{r=1}^2 I^{\bodyRn[r]}-I^{\bd\bodyRn}=0 \quad\text{~for all }\vu\ag{\cdot}\in T\An,
\label{eq:stationarityHetRewrite1}
\end{equation}
where
\begin{equation}
\begin{split}
I^{\bodyRn[r]}&=\int_{\bodyRn[r]}\frac{d}{d\epsilon}\fnC[r]\ag{\z,\ubn\ag{\z}+\epsilon\vu\ag{\z},\grad\ubn\ag{\z}+\epsilon\grad\vu\ag{\z}}\Bv_{\epsilon=0} \dV, \quad r=1,2,
\label{eq:ImatRn}
\end{split}
\end{equation}
\end{subequations}
and $I^{\bd\bodyRn}$ is as given in \eqref{eq:ThirdTermSimp1}. 

Proceeding as in \eqref{eq:IbodyRn1}--\eqref{eq:FirsttermSimp3}, for each $r$ we rewrite $I^{\bodyRn[r]}$ as
\begin{subequations}
\begin{equation}
I^{\bodyRn[r]}=\int_{\bodyRn[r]} \en{\br[i]\ag{\z}-\frac{d}{d\z[J]}\FPnr[i][J]\ag{\z}}\vu[i]\ag{\z} \dV+\int_{\bd\matRn}\Tbnr[i]\ag{\z}\,\,\vu[i]\ag{\z}\dS, \quad r=1,2.
\label{eq:matRnSimp1}
\end{equation}
The fields $\br\ag{\cdot}$, $\FPnr\ag{\cdot}$, and $\Tbnr\ag{\cdot}$ appearing in \eqref{eq:matRnSimp1} are the counterparts on $\bodyRn[r]$ of the fields defined in \eqref{eq:bCdef}, \eqref{eq:FPndef}, and \eqref{eq:eqTracFieldHet}, respectively. That is, they are given by
\begin{equation}
\br[i]\ag{\z}=\fnr[,i]\ag{\z,\ubn\ag{\z},\grad\ubn\ag{\z}}, \quad \FPnr[i][J]\ag{\z}=\fnr[,iJ]\ag{\z,\ubn\ag{\z},\grad\ubn\ag{\z}}, \quad \Tbnr\ag{\z}=\FPnr\ag{\z}\Nnrh\ag{\z},
\label{eq:bMFpnMTbnM}
\end{equation}
\label{eq:matRnSimp2}
\end{subequations}
where $\Nnrh\ag{\cdot}$ is the field of outward unit normals to $\bd\bodyRn[r]$, defined almost everywhere on this boundary (see Fig.~\ref{fig:boundaryNotation}(c)(ii)--(iii)).

Substituting \eqref{eq:matRnSimp1} and \eqref{eq:ThirdTermSimp1} into \eqref{eq:stationarityHetRewrite1} and rearranging, we obtain 
\begin{equation}
\begin{split}
\langle \vPin\ag{\ubn},\vu\rangle &= 
\sum_{r=1}^2  \int_{\bodyRn[r]}\en{\br[i]\ag{\z}-\frac{d}{d\z[J]}\FPnr[i][J]\ag{\z}}\vu[i]\ag{\z} \dV\\
&\hspace{1cm}+\sum_{r=1}^2 \int_{\bd\bodyRn[r]}\Tbnr[i]\ag{\z}\,\,\vu[i]\ag{\z}\dS-\int_{\bd\bodyRn}\Ttn[i]\ag{\z}\,\,\vu[i]\ag{\z}\dS=0 \qquad \text{~for all }\vu\ag{\cdot}\in T\An.
\end{split}
\label{eq:stationarityHetRewrite2}
\end{equation}

To further simplify \eqref{eq:stationarityHetRewrite2}, we decompose the boundary of each material region into the portion lying in the interior of the body and the portion lying on the boundary of the body, as illustrated in Fig.~\ref{fig:boundaryNotation}(c)(i)--(iii). Specifically, for each $r=1,2$, we write
\[
    \bd\bodyRn[r]=\GammaIn\cup\GammaErn,
\]
where this decomposition and those below are understood up to sets of surface measure zero, and where
\[
    \GammaIn:=\bd\bodyRn[r]\cap\bodyRn \quad\text{and}\quad \GammaErn:=\bd\bodyRn[r]\cap\bd\bodyRn.
\]
Here, $\GammaIn$ is the internal material interface, which is common to the boundaries of the two material regions and lies in the interior of the body, while $\GammaErn$ is the portion of the boundary of the body that lies on the boundary of the material region indexed by $r$.  Consequently, $\bd\bodyRn=\GammaECn[1]\cup\GammaECn[2]$.

Using these decompositions in \eqref{eq:stationarityHetRewrite2} and collecting integrals over common boundary portions gives
\begin{equation}
\begin{split}
\langle \vPin\ag{\ubn},\vu\rangle &= 
\sum_{r=1}^2  \int_{\bodyRn[r]}\en{\br[i]\ag{\z}-\frac{d}{d\z[J]}\FPnr[i][J]\ag{\z}}\vu[i]\ag{\z} \dV\\
&+\int_{\GammaIn}\en{\TbnC[1][i]\ag{\z}+\TbnC[2][i]\ag{\z}}\,\,\vu[i]\ag{\z}\dS+\sum_{r=1}^2\int_{\GammaErn}\en{\Tbnr[i]\ag{\z}-\Ttn[i]\ag{\z}}\,\,\vu[i]\ag{\z}\dS =0 \\
&\hspace{10cm}\text{~for all }\vu\ag{\cdot}\in T\An.
\label{eq:stationarityHetRewrite2p5}
\end{split}
\end{equation}
Proceeding analogously to \eqref{eq:StationarityRewrite2}--\eqref{eq:StationarityRewrite4}, we resolve the integrands of the third integral with respect to $\en{\nna\ag{\cdot}}_{\alpha=1}^m$, and for each $r$ and $\alpha$, decompose $\GammaErn$ according to their intersections with $\GammagRna$ and $\GammahRna$. The contributions over $\GammagRna$ vanish by the definition of $T\An$. Hence,
we obtain 
\begin{equation}
\begin{split}
\langle \vPin\ag{\ubn},\vu\rangle 
&=\sum_{r=1}^2  \int_{\bodyRn[r]}\en{\br[i]\ag{\z}-\frac{d}{d\z[J]}\FPnr[i][J]\ag{\z}}\vu[i]\ag{\z} \dV+\int_{\GammaIn}\en{\TbnC[1][i]\ag{\z}+\TbnC[2][i]\ag{\z}}\,\,\vu[i]\ag{\z}\dS\\
&+\sum_{r=1}^2\sum_{\alpha=1}^m\int_{\GammaErn\cap\GammahRna}\en{\Tbnr[\an]\ag{\z}-\Ttn[\an]\ag{\z}}\,\,\vu[\an]\ag{\z}\dS= 0 \qquad \text{~for all }\vu\ag{\cdot}\in T\An,
\label{eq:stationarityHetRewrite3}
\end{split}
\end{equation}
where, to recall, $\Tbnr[\an]\ag{\z}$ denotes $\Tbnr\ag{\z}\cdot\nna\ag{\z}$.

The stationarity condition \eqref{eq:stationarityHetRewrite2} implies the equilibrium conditions

\begin{subequations}
\begin{align}
\br[i]\ag{\z}-\frac{d}{d\z[J]}\FPnr[i][J]\ag{\z}&=0 \quad\text{for~a.e.~}\z\in\bodyRn[r], \quad r=1,2 \label{eq:bodyRnpEq}\\
\TbnC[1]\ag{\z}+\TbnC[2]\ag{\z}&=0 \quad\text{for~a.e.~}\z\in\GammaIn\label{eq:GammaInEq}\\
\Tbnr[\an]\ag{\z}-\Ttn[\an]\ag{\z}&=0\quad\text{for~a.e.~}\z\in\GammaErn\cap\GammahRna \quad \alpha=1,\ldots,m, \quad r=1,2. \label{eq:GammaPExtEq}
\end{align}
\end{subequations}

\subsection{Pullback of the equilibrium potential energy}
\label{sec:pullbackPinb}

In this subsection, we derive a fixed-domain representation of the equilibrium potential energy that will be used to evaluate its rate of change as $\eta\to\baseval$. 
 Recall that the equilibrium potential energy at $\eta$ is obtained by evaluating the potential energy functional \eqref{eq:PibnDef} at the equilibrium displacement $\ubn\ag{\cdot}$. We denote this quantity by $\Pibn:=\Pin\ag{\ubn}$. Thus,

\begin{equation}
\Pibn=\int_{\bodyRn} \fn\ag{\z,\ubn\ag{\z},\grad\ubn\ag{\z}} \dV -\int_{\bd\bodyRn}\Ttn\ag{\z}\cdot\en{\dOn\ag{\z}+\ubn\ag{\z}}\dS.
\label{eq:Pibn2}
\end{equation}
In \eqref{eq:Pibn2}, the integrals depend on $\eta$ both through the domains of integration and the integrands. We therefore rewrite them over the fixed domains of the $\baseval$ problem so that their $\eta$-dependence is contained entirely in the integrands. Specifically, we pull back the first and second integrals onto $\bodyRO$ and $\bd\bodyRO$, respectively, via the geometric map $\xnb\ag{\cdot}$, thereby obtaining the explicit form of the pulled-back potential energy functional $\check{\Pi}^\eta\ag{\cdot}$ introduced in \S\ref{subsec:parentPE}. This gives

\begin{equation}
\begin{split}
    \Pibn&=
    \int_{\bodyRO} 
        \fn\ag{
             \xnb\ag{\x},
             \en{\ubn\comp\xnb}\ag{\x},
              \en{\grad\ubn\comp\xnb}\ag{\x}
    } \Gn\ag{\x}\dV
    -\int_{\bd\bodyRO}
            \en{
            \en{\Ttn\comp\xnb}\ag{\x} 
            \cdot\en{\en{\dOn\comp\xnb}\ag{\x}+
             \en{\ubn\comp\xnb}\ag{\x}
             }}
            \Gs\ag{\x}\dS.
    \label{eq:Pibn3}
    \end{split}
\end{equation}
In \eqref{eq:Pibn3}, the map $\Gn\ag{\cdot}: \bodyRO \to \Reals$ is the Jacobian determinant function associated with the geometric map $\xnb\ag{\cdot}$, given by 
\begin{equation}
    \Gn\ag{\x}
        := \Det[\grad\xnb\ag{\x}].
    \label{eq:Gdef}
\end{equation}
The map $\Gs\ag{\cdot}:\bd\bodyRO\to\Reals$ appearing in the second integral of \eqref{eq:Pibn3} is the surface-area scaling factor associated with the geometric map $\xnb\ag{\cdot}$, which is also sometimes referred to as the surface Jacobian. That is, it measures how much the surface area around the point $\x\in\bd\bodyRO$ is stretched or compressed by the map $\xnb\ag{\cdot}$. It is given by
\begin{equation}
    \Gs\ag{\x}
        := \norm{
                \cof[\grad\xnb\ag{\x}]
                \,\Nh\ag{\x}
            }_{\Reals[m]},
    \label{eq:Gsdef}
\end{equation}
where $\Nh\ag{\x}$ is the outer surface normal to $\partial \bodyRO$ at $\x$ and $\cof\ag{\grad\xnb\ag{\x}}$ is the cofactor matrix of $\grad\xnb\ag{\x}$. 
To write \eqref{eq:Pibn3} in terms of the pullback fields introduced in \S\ref{sec:VE}, we recall
\[\en{\ubn\comp\xnb}\ag{\x}=\ubnc\ag{\x}, \quad \en{\Ttn\comp\xnb}\ag{\x}=   \Ttnc\ag{\x}, \quad \en{\dOn\comp\xnb}\ag{\x}=   \dOnc\ag{\x},\quad \en{\grad\ubn\comp\xnb}\ag{\x}=\grad\ubnc\ag{\x}\en{\grad\xnb}^{-1}\ag{\x}.\]
The last identity follows by differentiating both sides of $\en{\ubn\comp\xnb}\ag{\x}=\ubnc\ag{\x}$ with respect to $\x$ and applying the chain rule. Substituting these identities into \eqref{eq:Pibn3} yields
\begin{equation}
\begin{split}
    \Pibn=
    \int_{\bodyRO} 
        \fn\ag{
             \xnb\ag{\x},
             \ubnc\ag{\x},
              \grad\ubnc\ag{\x}\en{\grad\xnb}^{-1}\ag{\x}
    } \Gn\ag{\x}\dV-\int_{\bd\bodyRO}
            \en{
            \Ttnc\ag{\x} 
            \cdot
             \en{\dOn\ag{\x}+\ubnc\ag{\x}}
             }
            \Gs\ag{\x}\dS.
    \label{eq:PibnDef4}
    \end{split}
\end{equation}
The representation \eqref{eq:PibnDef4} will be used in the derivation of the closed-form expression of $F_c^\circ$ in \S\ref{sec:PERR}.

\paragraph*{Pullback in the presence of sharp material interfaces} When the body contains sharp material interfaces, the equilibrium potential energy is obtained instead by evaluating \eqref{eq:Pibnhet} at $\ubn\ag{\cdot}$:
\begin{equation}
\Pibn=\sum_{r=1}^2\int_{\bodyRn[r]} \fnr\ag{\z,\ubn\ag{\z},\grad\ubn\ag{\z}} \dV-\int_{\bd\bodyRn}\Ttn\ag{\z}\cdot\en{\dOn\ag{\z}+\ubn\ag{\z}}\dS.
\label{eq:Pibnhet1}
\end{equation}

Applying the same pullback procedure used to pass from \eqref{eq:Pibn2} to \eqref{eq:PibnDef4} to the volume integrals over each material region gives

\begin{equation}
\begin{split}
    \Pibn&=
    \sum_{r=1}^2\int_{\bodyRO[r]} 
        \fnr\ag{
             \xnb\ag{\x},
             \ubnc\ag{\x},
              \grad\ubnc\ag{\x}\en{\grad\xnb}^{-1}\ag{\x}
    } \Gn\ag{\x}\dV\\
    &\hspace{5cm}-\int_{\bd\bodyRO}
            \en{
            \Ttnc\ag{\x} 
            \cdot
             \en{\dOnc\ag{\x}+\ubnc\ag{\x}}
             }
            \Gs\ag{\x}\dS.
    \label{eq:Pibnhet2}
    \end{split}
\end{equation}

\subsection{Limits of functions and expressions as $\eta\to\baseval$}
\label{subsec:limits_of_functions}
In the derivation of closed-form expression of $F_c^\circ$ in \S\ref{sec:PERR}, we will require the limits, as $\eta\to\baseval$, of several functions and expressions associated with the variational evolution. We collect these limits here for later use.

\subsubsection{Those related to the geometric map $\xnb\ag{\cdot}$}
Recall that in \S\ref{subsec:domainevol}, we introduced the evolution map $\xnb\ag{\cdot}$ at $\eta$ and assumed the asymptotic expansions
\[
    \xnb\ag{\x}=\x+\en{\eta-\baseval}\phiO\ag{\x}+\oC[\eta-\baseval], \qquad \grad\xnb\ag{\x}=\Id+\en{\eta-\baseval}\grad\phiO\ag{\x}+\oC[\eta-\baseval]
\]
as $\eta\to\baseval$, uniformly in $\x\in\clos[\bodyRO]$. 
It immediately follows from these expansions that 
\begin{subequations}
\begin{align}
\frac{\xnb\ag{\x}-\x}{\eta-\baseval}\Bv[\eta\to\baseval]&=\phiO\ag{\x},\label{eq:xnbminusxEval}\\
\grad\xnb\ag{\x}\Bv[\eta\to\baseval]&=\Id.
\end{align}
Further, recalling the definition $\dOnc\ag{\x}:=\xnb\ag{\x}-\x$ from \eqref{eq:dOncDef}, we may equivalently write \eqref{eq:xnbminusxEval} as
\begin{align}
\frac{\dOnc\ag{\x}}{\eta-\baseval}\Bv[\eta\to\baseval]&=\phiO\ag{\x}.\label{eq:dOncEval}
\end{align}
\label{eq:geometricLimits}
\end{subequations}   
Using the standard first-order perturbation expansion for the matrix inverse~\cite{higham2008functions}, we obtain
\begin{align}
\en{\grad\xnb}^{-1}\ag{\x}=\Id-\en{\eta-\baseval}\grad\phiO\ag{\x}+\oC[\eta-\baseval].
\label{eq:gradxnbinvAsymp}
\end{align}
Taking the transpose of \eqref{eq:gradxnbinvAsymp} and letting $\eta\to\baseval$ gives
\begin{align}
\en{\grad\xnb}^{-T}\ag{\x}\Bv[\eta\to\baseval]=\Id.
\label{eq:gradxnbinvTev}
\end{align}
In the above, the limits in \eqref{eq:geometricLimits}--\eqref{eq:gradxnbinvTev}, as well as the remainder term in \eqref{eq:gradxnbinvAsymp}, are understood uniformly in $\x\in\overline{\bodyRO}$.

\subsubsection{Those related to the generalized potential energy density $\fn\ag{\cdot}$}
\label{subsec:fnLimits}
Recall that in \S\ref{subsec:fnevol}, we introduced the generalized potential energy density $\fn\ag{\cdot}$ at $\eta$ as one that admits an asymptotic expansion of the form
\[
    \fn\ag{\x,\u,\H}=\fO\ag{\x,\u,\H}+\en{\eta-\baseval}\zetaO\ag{\x,\u,\H}+\oC[\eta-\baseval]
\]
as $\eta\to\baseval$, for each fixed $(\x,\u,\H)$. 
 From this pointwise expansion, it immediately follows that
\begin{equation}
\frac{\fn\ag{\x,\u,\H}-\fO\ag{\x,\u,\H}}{\eta-\baseval}\Bv[\eta\to\baseval]=\zetaO\ag{\x,\u,\H}\label{eq:evtozetaO}.
\end{equation}    
Further, for the derivation in \S\ref{sec:volumeterm}, we assume the additional regularity and integrability condition that, upon evaluation at the equilibrium arguments $\en{\x,\ubO\ag{\x},\grad\ubO\ag{\x}}$, the convergence in \eqref{eq:evtozetaO} holds in $L^1\en{\bodyRO}$. Specifically, we assume that
\begin{equation}
\norm{
\frac{\fn\ag{\x,\ubO\ag{\x},\grad\ubO\ag{\x}}-\fO\ag{\x,\ubO\ag{\x},\grad\ubO\ag{\x}}}{\eta-\baseval}-\zetaO\ag{\x,\ubO\ag{\x},\grad\ubO\ag{\x}}}_{L^1\en{\bodyRO}}
\longrightarrow0
\qquad\text{as }\eta\to\baseval.
\end{equation}

\subsubsection{Those related to the applied traction field $\Ttn\ag{\cdot}$}

Recall that in \S\ref{subsec:appliedtrac}, we assumed the asymptotic expansion of the pullback of the applied traction field:
\[
    \Ttnc\ag{\x}=\TtO\ag{\x}+\en{\eta-\baseval}\chiO\ag{\x}+\oC[\eta-\baseval]
\]
as $\eta\to\baseval$, uniformly in $\x\in\bd\bodyRO$ in the sense specified in \S\ref{subsec:appliedtrac}. It immediately follows that
\begin{subequations}
\begin{align}
    \Ttnc\ag{\x}\Bv[\eta\to\baseval]&=\TtO\ag{\x},\label{eq:evtoTtO}\\
    \frac{\Ttnc\ag{\x}-\TtO\ag{\x}}{\eta-\baseval}\Bv[\eta\to\baseval]&=\chi^\circ\ag{\x}\label{eq:evtoChiO}.
\end{align}
\end{subequations}

\subsubsection{Those related to the equilibrium displacement field $\ubn\ag{\cdot}$}

Recall that in \S\ref{subsec:equilDispevol}, we assumed the following asymptotic expansion of the pullback of the equilibrium displacement field:
\[
    \ubnc\ag{\x}=\ubO\ag{\x}+\en{\eta-\baseval}\psiO\ag{\x}+\oC[\eta-\baseval]
\]
as $\eta\to\baseval$, uniformly in $\x\in\clos[\bodyRO]$.
From this expansion, it immediately follows that
\begin{subequations}
\begin{align}
\ubnc\ag{\x}\Bv[\eta\to\baseval]&=\ubO\ag{\x},\label{eq:ubncev}\\
\frac{\ubnc\ag{\x}-\ubO\ag{\x}}{\eta-\baseval}\Bv[\eta\to\baseval]&=\psiO\ag{\x}\label{eq:evtopsiO}.
\end{align}
\end{subequations}

In the derivation of the closed-form expression of $F_c^\circ$ in \S\ref{sec:PERR}, we will also require the limit
\begin{equation}
\frac{\grad\ubnc\ag{\cdot}\en{\grad\xnb}^{-1}\ag{\cdot}-\grad\ubO\ag{\cdot}}{\eta-\baseval}\Bv[\eta\to\baseval]=\grad\psiO\ag{\cdot}-\grad\ubO\ag{\cdot} \grad\phiO\ag{\cdot}\quad \text{in~} L^p(\bodyRO;\Matrix[m\times m]).
\label{eq:gradubnpullback}
\end{equation}
To see this, recall that $\grad\ubnc\ag{\cdot}$ admits the asymptotic expansion in $L^p$:
\[
    \grad\ubnc\ag{\cdot}=\grad\ubO\ag{\cdot}+\en{\eta-\baseval}\grad\psiO\ag{\cdot}+\oC[\eta-\baseval].
\]
Multiplying this $L^p$ expansion by the uniform expansion of $\en{\grad\xnb}^{-1}\ag{\cdot}$ in \eqref{eq:gradxnbinvAsymp} 
and collecting terms through first order in $\eta-\baseval$, we obtain, in $L^p$,
\[
    \grad\ubnc\ag{\cdot}\en{\grad\xnb}^{-1}\ag{\cdot}=\grad\ubO\ag{\cdot}+\en{\eta-\baseval}(\grad\psiO\ag{\cdot}-\grad\ubO\ag{\cdot}\grad\phiO\ag{\cdot})+\oC[\eta-\baseval].
\]
Dividing by $\eta-\baseval$ and passing to the limit in $L^p$ gives \eqref{eq:gradubnpullback}.

\subsubsection{Those related to the Jacobian of geometric map $\Gn\ag{\cdot}$}
\label{sec:AsymJD}
Substituting \eqref{eq:gradxnbAsymp} into the definition of $\Gn\ag{\x}$ in \eqref{eq:Gdef} and using the first-order expansion of the determinant, we obtain the following uniform expansion:
\begin{equation}
\Gn\ag{\x}=1+\en{\eta-\baseval}\,Tr\ag{\grad\phiO\ag{\x}}+\oC[\eta-\baseval].
\label{eq:Gexpand}
\end{equation}
as $\eta\to\baseval$, uniformly in $\x\in\bodyRO$. Consequently,
\begin{subequations}
\begin{align}
\Gn\ag{\x}\Bv[\eta\to\baseval]&=1,\label{eq:Gev}\\
\frac{\Gn\ag{\x}-1}{\eta-\baseval}\Bv[\eta\to\baseval]&=Tr\ag{\grad\phiO\ag{\x}}.
\label{eq:dGdnev}
\end{align}
\label{eq:GnLimits}
\end{subequations}

\subsubsection{Those related to the surface Jacobian of geometric map $\Gs\ag{\cdot}$}
\label{sec:dGsdnEval}
For convenience, we recall the definition of $\Gs\ag{\cdot}: \partial \bodyRO \to \mathbb{R}_{\ge 0}$ given in \eqref{eq:Gsdef}:
\begin{equation}
\Gs\ag{\x}=\norm{\cof[\grad\xnb\ag{\x}]\,\Nh\ag{\x}}_{\Reals[m]},
\label{eq:Gns}
\end{equation}
where $\Nh\ag{\x}$ is the outer surface normal to $\partial \bodyRO$ at $\x$, and $\cof\ag{\grad\xnb\ag{\x}}$ is the co-factor matrix of $\grad\xnb\ag{\x}$.
Using the identity $\cof\ag{\H}=\det[\H]\en{\H}^{-T}$ for $\H\in\Matrix[m\times m]$~\cite{sokolowski1992introduction},  together with the first-order expansion of $\det[\grad\xnb\ag{\x}]=:\Gn\ag{\x}$ given in \eqref{eq:Gexpand} and that of $\en{\grad\xnb}^{-T}\ag{\x}$, which follows directly by transposing \eqref{eq:gradxnbinvAsymp},  we obtain
\[
    \cof\ag{\grad\xnb\ag{\x}}=\Id+\en{\eta-\baseval}\en{\,Tr\ag{\grad\phiO\ag{\x}}\,\Id-\en{\grad\phiO}^{T}\ag{\x}}+\oC[\eta-\baseval].
\]
Substituting this expansion into the definition of $\Gs\ag{\x}$ in \eqref{eq:Gns}  and expanding the Euclidean norm to first order\footnote{For $\tp{v},\tp{a}\in\Reals[m]$, 
\[
    \norm{\tp{v}+\eta\,\tp{a}+\oC[\eta]}_{\Reals[m]}=\norm{\tp{v}}_{\Reals[m]}+\eta \frac{\tp{v}\cdot\tp{a}}{\hspace{0.34cm}\norm{\tp{v}}_{\Reals[m]}}+\oC[\eta], \quad \tp{v}\neq\zerovec
\]
} yields
\[
    \Gs\ag{\x}=\norm{\Nh\ag{\x}}_{\Reals[m]}+\en{\eta-\baseval}\en{Tr\ag{\grad\phiO\ag{\x}}-\,\Nh\ag{\x}\cdot\en{\grad\phiO}^{T}\ag{\x}\Nh\ag{\x}}+\oC[\eta-\baseval].
\]
The quantity in parentheses is the surface divergence~\cite{sokolowski1992introduction} of $\phiO\ag{\cdot}$ on $\bd\bodyRO$, which we denote by $\Divs[\phiO]\ag{\x}$. That is,
\begin{equation}
    \Divs[\phiO]\ag{\x}:=Tr\ag{\grad\phiO\ag{\x}}-\,\Nh\ag{\x}\cdot\en{\grad\phiO}^{T}\ag{\x}\Nh\ag{\x}.
\label{eq:SurfaceDivergence}
\end{equation}
Noting also that $\norm{\Nh\ag{\x}}_{\Reals[m]}=1$, the expansion of $\Gs\ag{\x}$ simplifies to
\[
    \Gs\ag{\x}=1+\en{\eta-\baseval}\,\Divs[\phiO]\ag{\x}+\oC[\eta-\baseval].
\]
as $\eta\to\baseval$, uniformly in $\x\in\bd\bodyRO$.
Consequently,
\begin{subequations}
\begin{align}
\Gs\ag{\x}\Bv[\eta\to\baseval]&=1\label{eq:Gsev}\\
\frac{\Gs\ag{\x}-1}{\eta-\baseval}\Bv[\eta\to\baseval]&=\Divs[\phiO]\ag{\x}.
\label{eq:Gsm1ovetaEv}
\end{align}
\end{subequations}

\section{The potential energy release rate formula}
\label{sec:PERR}
\noindent\textit{For reference, the principal symbols and mathematical conventions used throughout this section are collected in Appendix \ref{app:notation}.}

We now wish to derive the \prr $F_c^\circ$ at $\eta=\baseval$. Denoting this quantity by $F_c^\circ$, we have
\begin{equation}
    F_c^\circ
        := -\lim_{\eta\to\baseval}\frac{\Pibn-\PibO}{\eta-\baseval},
    \label{eq:Fc}
\end{equation}
where, to recall, $\Pibn$ and $\PibO$ denote the equilibrium potential energy at a given $\eta>\baseval$ and $\eta=\baseval$, respectively. 

For $\Pibn$, we use the representation \eqref{eq:PibnDef4}, while $\PibO$ is obtained by setting $\eta=\baseval$ in \eqref{eq:Pibn2}. Substituting these expressions into \eqref{eq:Fc}, we obtain
\begin{subequations}
\label{eq:Pibnmanip3}
    \begin{equation}
        F_c^\circ
            = F_c^{\bodyRO}
            + F_c^{\partial \bodyRO},
        \label{eq:FcSplit}
    \end{equation}
    where
    \begin{align}
    \begin{split}
        F_c^{\bodyRO}
        &:=-\lim_{\eta\to\baseval}\frac{1}{\eta-\baseval}\,\Biggl(\, 
        \int_{\bodyRO} \fn\ag{
                            \xnb\ag{\x},
                            \ubnc\ag{\x},
                            \grad\ubnc\ag{\x}
                            \en{\grad\xnb}^{-1}\ag{\x}
                        }\Gn\ag{\x}\dV-\int_{\bodyRO}\fO\ag{\x,\ubO\ag{\x},\grad\ubO\ag{\x}}\dV\Biggr) 
        \label{eq:FcBody}
        \end{split}
        \\
        \begin{split}
        F_c^{\partial \bodyRO}
            &:=-\lim_{\eta\to\baseval}\frac{1}{\eta-\baseval}\Biggl(
        -\int_{\bd\bodyRO}\en{\Ttnc\ag{\x}\cdot\en{\dOnc\ag{\x}+\ubnc\ag{\x}}}\Gs\ag{\x}\dS+\int_{\bd\bodyRO}\TtO\ag{\x}\cdot\ubO\ag{\x}\dS\Biggr)
        \label{eq:FcBoundary}
        \end{split}
    \end{align}
\end{subequations}
Here, we have used $\dOO\ag{\cdot}=\zerovec$, which follows directly from the definition of the geometric displacement. As you will note in \eqref{eq:Pibnmanip3} we have separated out
the volume and surface contributions to $F_c^\circ$
as $F_c^{\bodyRO}$ and $F_c^{\partial\bodyRO}$, respectively.
We refer to $F_c^{\bodyRO}$ as the body potential energy release rate,
and to $F_c^{\partial \bodyRO}$ as the surface potential energy release rate.

We show in \S\ref{sec:volumeterm} that $F_c^{\bodyRO}$ evaluates to the following expression, written in the shorthand notation introduced in \S\ref{sec:Notation}:
\begin{subequations}
\begin{equation}
\label{eq:FcVolume}
    F_c^{\bodyRO}
       = -\int_{\bodyRO}
            \zetabO
        \dV
        -
        \int_{\bodyRO}
            \frac{d}{d\x[J]}
            \en{\SigmaO[p][J]\phiO[p]}
        \dV
        -\int_{\bd\bodyRO}
            \TbO\cdot\psiO\dS, \,\,
    \end{equation}
where $\SigmaO\ag{\cdot}:\bodyRO\to\Matrix[m\times m]$  is the celebrated Eshelby energy momentum tensor of the $\baseval$ problem in the presence of body forces. Its components are
\begin{equation}
\SigmaO[p][J]  := \fbO\,\delta_{pJ}-\FPO[i][J]\ubO[i,p].
\label{eq:SigmaODef}
\end{equation}
\label{eq:FcVolumeandSigmaO}
\end{subequations}

We show in \S\ref{sec:surafecterm} that 
\begin{align}
\label{eq:FcSurface}
    F_c^{\bd\bodyRO}
        &= \int_{\bd\bodyRO}
                \chiO\cdot\ubO
            \dS
        +
           \int_{\bd\bodyRO}
                \TtO\cdot\en{\phiO+\psiO}
            \dS
        +
           \int_{\bd\bodyRO}
              \en{\TtO\cdot\ubO}
                \Divs[\phiO]
            \dS.
\end{align}
Here, the expressions in  \eqref{eq:FcSurface} are written according to the shorthand convention introduced in \S\ref{sec:Notation}. To recall, $\Divs[\phiO]$ has been defined in \eqref{eq:SurfaceDivergence}.

In \S\ref{sec:PRRinTermspfEshelby}, we show that it follows from \eqref{eq:FcVolume} and \eqref{eq:FcSurface} that 
\begin{align}
\begin{split}
\label{eq:configForce5}
    F_c^\circ=
        &-\int_{\bodyRO}
            \zetabO
        \dV
        -
        \int_{\bodyRO}
            \frac{d}{d\x[J]}
            \en{\SigmaO[p][J]\phiO[p]}
        \dV
        -
        \sum_{\alpha=1}^m\int_{\GammagROa}
            \en{\TbO[\aO]-\TtO[\aO]}\en{\psihOa-\ubO\cdot\OmegaO\nOa}
        \dS\\
        &+
        \int_{\bd\bodyRO}
            \chiO\cdot\ubO
        \dS
        +
        \int_{\bd\bodyRO}
            \TtO\cdot\phiO
        \dS
        +
        \int_{\bd\bodyRO}
            \en{\TtO\cdot\ubO}
            \Divs[\phiO]
        \dS,
\end{split}
\end{align}
Importantly, all quantities appearing in \eqref{eq:configForce5} are determined by the equilibrium solution at $\eta=\baseval$ and the quantities that prescribe the variational evolution. Thus, evaluation of $F_c^\circ$ does not require solving the variational problem at neighboring values of $\eta$.

Finally, in \S\ref{subsec:configurational_het}, we derive the corresponding \prr expression for a body containing sharp material interfaces.

In \S\ref{sec:Demos}, we demonstrate the utility of the framework presented in \S\ref{sec:VE} and the closed-form expression for $F_c^\circ$ presented in this section.  

\subsection{The body potential energy release rate}
\label{sec:volumeterm}

{
In this section, we derive \eqref{eq:FcVolume} starting from \eqref{eq:FcBody}.
} %
To increase readability, we first rewrite \eqref{eq:FcBody} as
\begin{equation}
F_c^{\bodyRO}:=-\lim_{\eta\to\baseval}\frac{1}{\eta-\baseval}\en{\int_{\bodyRO}\fn\ag{\Xn}\Gn\ag{\x}\dV-\int_{\bodyRO}\fO\ag{\XO}\dV}. 
\label{eq:FcBodysimp}
\end{equation}
In \eqref{eq:FcBodysimp}, we have replaced the arguments of $\fn\ag{\cdot,\cdot,\cdot}$ and $\fO\ag{\cdot,\cdot,\cdot}$ in the first and second integrals of \eqref{eq:FcBody} by $\Xn:=\en{\xnb\ag{\x}, \ubnc\ag{\x}, \grad\ubnc\ag{\x}\en{\grad\xnb}^{-1}\ag{\x}}$ and $\XO:=\en{\x, \ubO\ag{\x}, \grad\ubO\ag{\x}}$, respectively. 

Rearranging the integrand \eqref{eq:FcBodysimp} and separating the resulting contributions, we obtain

\begin{align}
\begin{split}
F_c^{\bodyRO}=&-\lim_{\eta\to\baseval}\int_{\bodyRO}\fO\ag{\XO}\, \frac{\Gn\ag{\x}-1}{\eta-\baseval} \dV-\lim_{\eta\to\baseval}\int_{\bodyRO}\frac{\fn\ag{\XO}-\fO\ag{\XO}}{\eta-\baseval}\Gn\ag{\x}\dV\\
&-\lim_{\eta\to\baseval}\int_{\bodyRO}\frac{\fn\ag{\Xn}-\fn\ag{\XO}}{\eta-\baseval}\,\Gn\ag{\x} \dV.
\label{eq:FcBodysimp2}
\end{split}
\end{align}

Under the additional regularity conditions stated in Appendix~\ref{app:fnRegularity}, the third integrated limit in \eqref{eq:FcBodysimp2} may be replaced by its first-order Fréchet linearization with respect to the arguments of $\fn\ag{\cdot}$. Thus, we obtain

\begin{align}
\begin{split}
F_c^{\bodyRO}=&-\lim_{\eta\to\baseval}\int_{\bodyRO}\fO\ag{\XO}\, \frac{\Gn\ag{\x}-1}{\eta-\baseval} \dV-\lim_{\eta\to\baseval}\int_{\bodyRO}\frac{\fn\ag{\XO}-\fO\ag{\XO}}{\eta-\baseval}\Gn\ag{\x}\dV\\
&-\lim_{\eta\to\baseval}\int_{\bodyRO}\frac{D\fO\en{\XO}[\Xn-\XO]}{\eta-\baseval} \dV.
\label{eq:FcBodysimp3}
\end{split}
\end{align}
We next evaluate the third integrated limit by decomposing $\Xn-\XO$ componentwise with respect to the canonical bases of its three argument spaces:
\begin{align}
\begin{split}
\Xn-\XO
&=\sum_{I=1}^m \en{\xnb[I]\ag{\x}-\x[I]}\en{\E[I],\zerovec,\zeroMat}+\sum_{i=1}^m \en{\ubnc[i]\ag{\x}-\ubO[i]\ag{\x}}\en{\zerovec,\E[i],\zeroMat}\\
&+\sum_{i,k,J=1}^m\en{\ubnc[i,k]\ag{\x}\en{\xnb[k,J]}^{-1}\ag{\x}-\ubO[i,J]\ag{\x}}\en{\zerovec,\zerovec,\E[iJ]}.
\end{split}
\label{eq:Xn-XO}
\end{align}
In \eqref{eq:Xn-XO}, to recall, $\E[I]$ (similarly, $\E[i]$) and $\E[iJ]$ are the canonical basis of $\Reals[m]$ and $\Matrix[m\times m]$, and  $\zerovec$ and $\zeroMat$ are their respective null elements.

Substituting \eqref{eq:Xn-XO} into \eqref{eq:FcBodysimp3}, using the linearity of $D\fO\en{\cdot}\ag{\cdot}$, and adopting the Einstein summation convention, we obtain
\begin{align}
\begin{split}
F_c^{\bodyRO}=&-\lim_{\eta\to\baseval}\int_{\bodyRO}\fO\ag{\XO}\, \frac{\Gn\ag{\x}-1}{\eta-\baseval} \dV-\lim_{\eta\to\baseval}\int_{\bodyRO}\frac{\fn\ag{\XO}-\fO\ag{\XO}}{\eta-\baseval}\Gn\ag{\x}\dV\\
&-\lim_{\eta\to\baseval}\int_{\bodyRO}\frac{\xnb[I]\ag{\x}-\x[I]}{\eta-\baseval}\, D\fO\en{\XO}\ag{(\E[I],\zerovec,\zeroMat)} \dV\\
&-\lim_{\eta\to\baseval}\int_{\bodyRO}\frac{\ubnc[i]\ag{\x}-\ubO[i]\ag{\x}}{\eta-\baseval}\, D\fO\en{\XO}\ag{(\zerovec,\E[i],\zeroMat)} \dV\\
&-\lim_{\eta\to\baseval}\int_{\bodyRO}\frac{\ubnc[i,k]\ag{\x}\en{\xnb[k,J]}^{-1}\ag{\x}-\ubO[i,J]\ag{\x}}{\eta-\baseval}\, D\fO\en{\XO}\ag{(\zerovec,\zerovec,\E[iJ])}  \dV
\label{eq:FcBodysimp6}
\end{split}
\end{align}

In \eqref{eq:FcBodysimp6}, the Fr\'echet derivatives $D\fO\en{\XO}\ag{\en{\E[I],\zerovec,\zeroMat}}$, $D\fO\en{\XO}\ag{\en{\zerovec,\E[i],\zeroMat}}$, and $D\fO\en{\XO}\ag{\en{\zerovec,\zerovec,\E[iJ]}}$ are, respectively,  nothing but the partial derivatives $\fO[,I]\ag{\cdot}$, $\fO[,i]\ag{\cdot}$, and $\fO[,iJ]\ag{\cdot}$ evaluated at $\XO$ (see \S\ref{subsec:partialderiv}). In view of the convergence results established in \S\ref{subsec:limits_of_functions}, and assuming sufficient additional regularity and integrability (see, e.g., Appendix~\ref{app:fnRegularity}) to pass the limits through the corresponding integrals, we obtain
\begin{align}
\begin{split}
F_c^{\bodyRO}=&-\int_{\bodyRO}\fO\ag{\XO}\, \frac{\Gn\ag{\x}-1}{\eta-\baseval}\Bv[\eta\to\baseval] \dV
-\int_{\bodyRO}\frac{\fn\ag{\XO}-\fO\ag{\XO}}{\eta-\baseval}\Bv[\eta\to\baseval]\Gn\ag{\x}\Bv[\eta\to\baseval]\dV\\
&-\int_{\bodyRO}\frac{\xnb[I]\ag{\x}-\x[I]}{\eta-\baseval}\Bv[\eta\to\baseval]\, \fO[,I]\ag{\XO} \dV-\int_{\bodyRO}\frac{\ubnc[i]\ag{\x}-\ubO[i]\ag{\x}}{\eta-\baseval}\Bv[\eta\to\baseval]\, \fO[,i]\ag{\XO} \dV\\
&-\int_{\bodyRO}\frac{\ubnc[i,k]\ag{\x}\en{\xnb[k,J]}^{-1}\ag{\x}-\ubO[i,J]\ag{\x}}{\eta-\baseval}\Bv[\eta\to\baseval]\, \fO[,iJ]\ag{\XO}  \dV
\label{eq:FcBodysimp7}
\end{split}
\end{align}
Several different limits appear in \eqref{eq:FcBodysimp7}. The limits involving $\Gn\ag{\cdot}$ have been given in \S\ref{sec:AsymJD}. Specifically, as $\eta\to\baseval$, we established that $\Gn\ag{\x}\to1$ and $\en{\Gn\ag{\x}-1}/\en{\eta-\baseval}\to\phiO[p,p]\ag{\x}$ (see \eqref{eq:Gev} and \eqref{eq:dGdnev}, respectively). Here, $\phiO[p,p]\ag{\x}$ is the divergence of $\phiO\ag{\cdot}$ at $\x$, written in component form. Making these substitutions, 
\begin{align}
\begin{split}
F_c^{\bodyRO}=&-\int_{\bodyRO}\fO\ag{\XO}\, \phiO[p,p]\ag{\x} \dV
-\int_{\bodyRO}\frac{\fn\ag{\XO}-\fO\ag{\XO}}{\eta-\baseval}\Bv[\eta\to\baseval]\dV\\
&-\int_{\bodyRO}\frac{\xnb[I]\ag{\x}-\x[I]}{\eta-\baseval}\Bv[\eta\to\baseval]\, \fO[,I]\ag{\XO} \dV-\int_{\bodyRO}\frac{\ubnc[i]\ag{\x}-\ubO[i]\ag{\x}}{\eta-\baseval}\Bv[\eta\to\baseval]\, \fO[,i]\ag{\XO} \dV\\
&-\int_{\bodyRO}\frac{\ubnc[i,k]\ag{\x}\en{\xnb[k,J]}^{-1}\ag{\x}-\ubO[i,J]\ag{\x}}{\eta-\baseval}\Bv[\eta\to\baseval]\, \fO[,iJ]\ag{\XO}  \dV
\label{eq:FcBodysimp8}
\end{split}
\end{align}

The remaining limits are
\[
\frac{\fn\ag{\XO}-\fO\ag{\XO}}{\eta-\baseval}\Bv[\eta\to\baseval],\quad \frac{\xnb[I]\ag{\x}-\x[I]}{\eta-\baseval}\Bv[\eta\to\baseval], \quad \frac{\ubnc[i]\ag{\x}-\ubO[i]\ag{\x}}{\eta-\baseval}\Bv[\eta\to\baseval], \quad\text{and}\quad \frac{\ubnc[i,k]\ag{\x}\en{\xnb[k,J]}^{-1}\ag{\x}-\ubO[i,J]\ag{\x}}{\eta-\baseval}\Bv[\eta\to\baseval] ,
\]
which respectively appear in the second, third, fourth, and fifth integrals in \eqref{eq:FcBodysimp8}. We showed in \eqref{eq:evtozetaO}, \eqref{eq:xnbminusxEval}, \eqref{eq:evtopsiO}, and \eqref{eq:gradubnpullback} that these evaluate to $\zetaO\ag{\XO}$, $\phiO[I]\ag{\x}$, $\psiO[i]\ag{\x}$, and $\psiO[i,J]\ag{\x}-\ubO[i,p]\ag{\x}\phiO[p,J]\ag{\x}$, respectively. 
With these simplifications, we obtain
\begin{align}
\begin{split}
F_c^{\bodyRO}=&-\int_{\bodyRO}\fO\ag{\XO}\phiO[p,p]\ag{\x} \dV-\int_{\bodyRO}\zetaO\ag{\XO} \dV-\int_{\bodyRO}\phiO[I]\ag{\x}\,\fO[,I]\ag{\XO}\dV\\
&-\int_{\bodyRO}\psiO[i]\ag{\x}\,\fO[,i]\ag{\XO} \dV-\int_{\bodyRO}\en{\psiO[i,J]\ag{\x}-\ubO[i,p]\ag{\x}\phiO[p,J]\ag{\x}}\,\fO[,iJ]\ag{\XO} \dV
\label{eq:FcBodysimp9}
\end{split}
\end{align}

Recalling the identifications $\fO[,i]\ag{\XO}=\bO[i]\ag{\x}$ and $\fO[,iJ]\ag{\XO}=\FPO[i][J]\ag{\x}$ (see \eqref{eq:bCdef} and \eqref{eq:FPndef}, respectively) and adopting the shorthand convention introduced in \S\ref{sec:Notation}, we rewrite \eqref{eq:FcBodysimp8} as
\begin{align}
\begin{split}
F_c^{\bodyRO}=&-\int_{\bodyRO}\fbO\phiO[p,p] \dV-\int_{\bodyRO}\zetabO\dV\\
&-\int_{\bodyRO}\fbO[,I]\phiO[I]\,\dV
-\int_{\bodyRO}\bO[i]\psiO[i]\, \dV
-\int_{\bodyRO}\FPO[i][J]\en{\psiO[i,J]-\ubO[i,p]\phiO[p,J]}\, \dV.
\label{eq:FcBodysimp10}
\end{split}
\end{align}

Through standard, albeit tedious, algebraic manipulations of \eqref{eq:FcBodysimp10}, we obtain
\begin{align}
\begin{split}
F_c^{\bodyRO} = -\int_{\bodyRO}\zetabO \dV-\int_{\bodyRO}\frac{d}{d\x[J]}\biggl(\en{\fbO\delta_{pJ}-\FPO[i][J]\ubO[i,p]}\phiO[p]\biggr)\dV-\int_{\bodyRO}\frac{d}{d\x[J]}\en{\FPO[i][J]\psiO[i]}\,&\dV \\
-\int_{\bodyRO}\en{\bO[i]-\frac{d}{d\x[J]}\FPO[i][J]}\en{\psiO[i]-\ubO[i,p]\phiO[p]}\, &\dV.
\label{eq:FcBodysimp11}
\end{split}
\end{align}

We finally arrive at \eqref{eq:FcVolume} by applying divergence theorem to the third integral in \eqref{eq:FcBodysimp10}, and noticing that 
by the equilibrium condition \eqref{eq:equiconda}  given  in \S\ref{sec:equilibrium_conditions}, the fourth integral of \eqref{eq:FcBodysimp10} vanishes identically.

\subsection{The surface potential energy release rate}
\label{sec:surafecterm}

In this section, we derive \eqref{eq:FcSurface} starting from \eqref{eq:FcBoundary}.

{
Rearranging the integrand in \eqref{eq:FcBoundary} and interchanging the limit and integration,\footnote{This interchange is justified by the convergence results established in \S\ref{subsec:limits_of_functions} and the finiteness of the surface measure of $\bd\bodyRO$.}, we obtain  
\begin{align}
\begin{split}
 F_c^{\bd \bodyRO}
            :=
             &\int_{\bd\bodyRO}\en{\frac{\Ttnc\ag{\x}-\TtO\ag{\x}}{\eta-\baseval}\Bv[\eta\to\baseval]\cdot\ubO\ag{\x}}\,\, \Gs\ag{\x}\Bv[\eta\to\baseval]\dS\\
        &+\int_{\bd\bodyRO}\Ttnc\ag{\x}\Bv[\eta\to\baseval]\cdot\en{\frac{\dOnc\ag{\x}}{\eta-\baseval}\Bv[\eta\to\baseval]+\frac{\ubnc\ag{\x}-\ubO\ag{\x}}{\eta-\baseval}\Bv[\eta\to\baseval]}\Gs\ag{\x}\Bv[\eta\to\baseval]\dS\\  
    &+\int_{\bd\bodyRO}\en{\TtO\ag{\x}\cdot\ubO\ag{\x}}\,\frac{\Gs\ag{\x}-1}{\eta-\baseval}\Bv[\eta\to\baseval]\dS
    \end{split}
    \label{eq:surfacetermmanip2}
\end{align}
}

Several limits appear in \eqref{eq:surfacetermmanip2}. 
Conceptually, it is straightforward to see that $\Gs\ag{\x}\to1$ as $\eta\to\baseval$, as established in \eqref{eq:Gsev}. As we further established in \eqref{eq:Gsm1ovetaEv},  $\en{\Gs\ag{\x}-1}/\en{\eta-\baseval} \to \Divs[\phiO]\ag{\x}$ as $\eta\to\baseval$. Substituting these results into \eqref{eq:surfacetermmanip2}, we obtain 
\begin{align}
\begin{split}
 F_c^{\bd \bodyRO}
            :=
             &\int_{\bd\bodyRO}\frac{\Ttnc\ag{\x}-\TtO\ag{\x}}{\eta-\baseval}\Bv[\eta\to\baseval]\cdot\ubO\ag{\x}\dS\\
        &+\int_{\bd\bodyRO}\Ttnc\ag{\x}\Bv[\eta\to\baseval]\cdot\en{\frac{\dOnc\ag{\x}}{\eta-\baseval}\Bv[\eta\to\baseval]+\frac{\ubnc\ag{\x}-\ubO\ag{\x}}{\eta-\baseval}\Bv[\eta\to\baseval]}\dS\\  
    &+\int_{\bd\bodyRO}\en{\TtO\ag{\x}\cdot\ubO\ag{\x}}\,\Divs\ag{\phiO}\ag{\x}\dS
    \end{split}
    \label{eq:surfacetermmanip3}
\end{align}

The remaining limits which appear in \eqref{eq:surfacetermmanip3} are
\[
\frac{\Ttnc\ag{\x}-\TtO\ag{\x}}{\eta-\baseval}\Bv[\eta\to\baseval],\quad
\Ttnc\ag{\x}\Bv[\eta\to\baseval], \quad
\frac{\dOnc\ag{\x}}{\eta-\baseval}\Bv[\eta\to\baseval], \quad 
\text{and}\ \ 
\frac{\ubnc\ag{\x}-\ubO\ag{\x}}{\eta-\baseval}\Bv[\eta\to\baseval].
\]
These limits evaluate to $\chiO\ag{\x}$, $\TtO\ag{\x}$, $\phiO\ag{\x}$, and $\psiO\ag{\x}$, as we established in \eqref{eq:evtoChiO}, \eqref{eq:evtoTtO}, \eqref{eq:dOncEval}, and \eqref{eq:evtopsiO}, respectively.

We complete the derivation of \eqref{eq:FcSurface} upon replacing the limits in \eqref{eq:surfacetermmanip2} with their respective evaluated values.

\subsection{The potential energy release rate in terms of the Eshelby energy momentum tensor}
\label{sec:PRRinTermspfEshelby}
In this section, we derive \eqref{eq:configForce5} from \eqref{eq:FcVolume} and \eqref{eq:FcSurface}.

Adding \eqref{eq:FcVolume} and \eqref{eq:FcSurface}, we obtain that

\begin{align}
\begin{split}
\label{eq:configForce6}
    F_c^\circ=
        &-\int_{\bodyRO}
            \zetabO
        \dV
        -
        \int_{\bodyRO}
            \frac{d}{d\x[J]}
            \en{\SigmaO[p][J]\phiO[p]}
        \dV
         -
        \int_{\bd\bodyRO}
            \en{\TbO-\TtO}\cdot\psiO
        \dS\\
        &+
        \int_{\bd\bodyRO}
            \chiO\cdot\ubO
        \dS
        +
        \int_{\bd\bodyRO}
            \TtO\cdot\phiO
        \dS
        +
        \int_{\bd\bodyRO}
            \en{\TtO\cdot\ubO}
            \Divs[\phiO]
        \dS.
\end{split}
\end{align}
We resolve the third integral in \eqref{eq:configForce6} with respect to the orthonormal frame field $\en{\nOa\ag{\cdot}}_{\alpha=1}^m$ and decompose the boundary according to $\bd\bodyRO=\GammagROa\cup\GammahROa$ for each $\alpha$. The contribution over $\GammahROa$ then vanishes by the equilibrium condition \eqref{eq:equicondb} at $\eta=\baseval$, yielding
\begin{align}
\begin{split}
\label{eq:configForce7}
    F_c^\circ=
        &-\int_{\bodyRO}
            \zetabO
        \dV
        -
        \int_{\bodyRO}
            \frac{d}{d\x[J]}
            \en{\SigmaO[p][J]\phiO[p]}
        \dV
        -
        \sum_{\alpha=1}^m\int_{\GammagROa}
            \en{\TbO[\aO]-\TtO[\aO]}\psiO[\aO] \dS\\
        &+
        \int_{\bd\bodyRO}
            \chiO\cdot\ubO
        \dS
        +
        \int_{\bd\bodyRO}
            \TtO\cdot\phiO
        \dS
        +
        \int_{\bd\bodyRO}
            \en{\TtO\cdot\ubO}
            \Divs[\phiO]
        \dS.
\end{split}
\end{align}

Although $\psiO[\aO]$ appearing in \eqref{eq:configForce7} is not a prescribed quantity, its value on $\GammagROa$can be expressed in terms of prescribed quantities and the equilibrium solution of the $\baseval$ problem. To see this, recall that on each $\GammagRna$, the equilibrium displacement field satisfies the condition  $\ubn\ag{\z}\cdot\nna\ag{\z}=\uhn[\alpha]\ag{\z}$. We pull this condition back onto $\GammagROa$ via the geometric map $\xnb\ag{\cdot}$ and write
\begin{equation}
\ubnc\ag{\x}\cdot\nnca\ag{\x}=\uhnc[\alpha]\ag{\x} \quad \text{for a.e. }\x\in\GammagROa
\end{equation}
We expand each of $\ubnc\ag{\cdot}$, $\nnca\ag{\cdot}$, and $\uhnc[\alpha]\ag{\cdot}$ about $\eta=\baseval$ according to the expansions \eqref{eq:ubncAsymp}, \eqref{eq:nncaAsymp}, and \eqref{eq:uhncAsymp}. Comparison of the first-order terms yields
 \begin{equation}
\ubO\ag{\x}\cdot\nOa\ag{\x}+\en{\eta-\baseval}\en{\psiO[\aO]\ag{\x}+\ubO\ag{\x}\cdot\OmegaO\ag{\x}\nOa\ag{\x}}=\uhO[\alpha]\ag{\x}+\en{\eta-\baseval}\,\psihOa\ag{\x} \quad \text{for a.e. }\x\in\GammagROa
\end{equation}
Since the equilibrium displacement field at $\eta=\baseval$ satisfies $\ubO\ag{\x}\cdot\nOa\ag{\x}=\uhO[\alpha]\ag{\x}$, we obtain
\begin{equation}
\psiO[\aO]\ag{\x}=\psihOa\ag{\x}-\ubO\ag{\x}\cdot\OmegaO\ag{\x}\nOa\ag{\x} \quad \text{for a.e. }\x\in\GammagROa
\label{eq:psiOaOResult}
\end{equation}
Substituting this relation into \eqref{eq:configForce7} yields the final potential energy release-rate expression given in \eqref{eq:configForce5}.

\subsection{Potential-energy release rate in the presence of sharp material interfaces}
\label{subsec:configurational_het}

We now derive the potential energy release rate in the presence of a sharp material heterogeneity. To \eqref{eq:Fc}, we now substitute the representation \eqref{eq:Pibnhet2} for $\Pibn$, while for $\PibO$, we substitute \eqref{eq:Pibnhet1} evaluated at $\eta=\baseval$. The potential energy release rate can then be decomposed as
\begin{subequations}
    \begin{equation}
        F_c^\circ
            = \sum_{r=1}^2F_c^{\bodyRO[r]}
            + F_c^{\partial \bodyRO},
        \label{eq:FcHetSplit}
    \end{equation}
    where 
    \begin{align}
    \begin{split}
        F_c^{\bodyRO[r]}
        &:=-\lim_{\eta\to\baseval}\frac{1}{\eta-\baseval}\,\Biggl(\, 
        \int_{\bodyRO[r]} \fnr\ag{
                            \xnb\ag{\x},
                            \ubnc\ag{\x},
                            \grad\ubnc\ag{\x}
                            \en{\grad\xnbi\comp\xnb}\ag{\x}
                        }\Gn\ag{\x}\dV\\
                        &\hspace{6cm}-\int_{\bodyRO[r]}\fOr\ag{\x,\ubO\ag{\x},\grad\ubO\ag{\x}}\dV\Biggr), \quad r=1,2,
        \label{eq:FcHet}
        \end{split}
    \end{align}
\end{subequations}
and $F_c^{\bd\bodyRO}$ is given by \eqref{eq:FcBoundary}. 
The simplification of $F_c^{\bodyRO[r]}$ proceeds analogously to that of $F_c^{\bodyRO}$ in \S\ref{sec:volumeterm}, giving

\begin{equation}
    F_c^{\bodyRO[r]}
       = -\int_{\bodyRO[r]}
            \zetabOr
        \dV
        -
        \int_{\bodyRO[r]}
            \frac{d}{d\x[J]}
            \en{\SigmaOr[p][J]\phiO[p]}
        \dV
        -\int_{\bd\bodyRO[r]}
           \TbOr\cdot\psiO\dS.\,\,
    \label{eq:FcHetVolume}
        \end{equation}
Substituting \eqref{eq:FcHetVolume} and the expression for $F_c^{\bd\bodyRO}$ from \eqref{eq:FcSurface} into \eqref{eq:FcHetSplit}, we obtain
\begin{equation}
\begin{split}
F_c^\circ= &-\sum_{r=1}^2\int_{\bodyRO[r]}
            \zetabOr
        \dV
        -\sum_{r=1}^2\int_{\bodyRO[r]}
            \frac{d}{d\x[J]}
            \en{\SigmaOr[p][J]\phiO[p]}
        \dV\\
        &-\sum_{r=1}^2\int_{\bd\bodyRO[r]}
           \TbOr\cdot\psiO\dS\,\,
        +\int_{\bd\bodyRO}
                \TtO\cdot\psiO
            \dS\\
        &+\int_{\bd\bodyRO}
                \TtO\cdot\phiO \dS
        +\int_{\bd\bodyRO}
                \chiO\cdot\ubO\dS
        +\int_{\bd\bodyRO}
                \en{\TtO\cdot\ubO}
                \Divs[\phiO]
            \dS.
\end{split}
\label{eq:FcHetintermed}
\end{equation}
The boundary integrals involving $\psiO$ in \eqref{eq:FcHetintermed} can be treated exactly as the boundary integrals involving $\vu$ in the passage from \eqref{eq:stationarityHetRewrite2} to \eqref{eq:stationarityHetRewrite2p5}, with $\vu$ replaced by $\psiO$. 

\begin{equation}
\begin{split}
F_c^\circ= &-\sum_{r=1}^2\int_{\bodyRO[r]}
            \zetabOr
        \dV
        -\sum_{r=1}^2\int_{\bodyRO[r]}
            \frac{d}{d\x[J]}
            \en{\SigmaOr[p][J]\phiO[p]}
        \dV\\
        &-\int_{\GammaIO}\en{\TbOC[1]+\TbOC[2]}\cdot\psiO \dS
        -\sum_{r=1}^2\int_{\GammaErO}
           \en{\TbOr-\TtO}\cdot\psiO\dS\,\,\\
        &+\int_{\bd\bodyRO}
                \TtO\cdot\phiO \dS
        +\int_{\bd\bodyRO}
                \chiO\cdot\ubO\dS
        +\int_{\bd\bodyRO}
                \en{\TtO\cdot\ubO}
                \Divs[\phiO]
            \dS.
\end{split}
\label{eq:FcHetintermed2}
\end{equation}

The contribution over the internal interface $\GammaIO$ vanishes by the interface equilibrium condition \eqref{eq:GammaInEq}. For each integral over $\GammaErn$, we resolve its integrand with respect to $\en{\nOa\ag{\cdot}}_{\alpha=1}^m$ and, for each $r$ and $\alpha$, decompose $\GammaErO$ according to its intersections with $\GammagROa$ and $\GammahROa$. The contributions over $\GammaErO\cap\GammahROa$ vanish by the equilibrium condition \eqref{eq:GammaPExtEq}. Hence, we obtain
\begin{equation}
\begin{split}
F_c^\circ= &-\sum_{r=1}^2\int_{\bodyRO[r]}
            \zetabOr
        \dV
        -\sum_{r=1}^2\int_{\bodyRO[r]}
            \frac{d}{d\x[J]}
            \en{\SigmaOr[p][J]\phiO[p]}
        \dV\\
        -&\sum_{r=1}^2\sum_{\alpha=1}^m\int_{\GammaErO\cap\GammagROa}\en{\TbOr[\aO]-\TtO[\aO]}\psiO[\aO]\dS\\
        &+\int_{\bd\bodyRO}
                \TtO\cdot\phiO \dS
        +\int_{\bd\bodyRO}
                \chiO\cdot\ubO\dS
        +\int_{\bd\bodyRO}
                \en{\TtO\cdot\ubO}
                \Divs[\phiO]
            \dS.
\end{split}
\label{eq:FcHetintermed3}
\end{equation}
Finally, substituting \eqref{eq:psiOaOResult} for $\psiO[\aO]$ into \eqref{eq:FcHetintermed3} yields the expression for \prr in the presence of sharp material interfaces as
\begin{equation}
\begin{split}
F_c^\circ= &-\sum_{r=1}^2\int_{\bodyRO[r]}
            \zetabOr
        \dV
        -\sum_{r=1}^2\int_{\bodyRO[r]}
            \frac{d}{d\x[J]}
            \en{\SigmaOr[p][J]\phiO[p]}
        \dV\\
        &-\sum_{r=1}^2\sum_{\alpha=1}^m\int_{\GammaErO\cap\GammagROa}\en{\TbOr[\aO]-\TtO[\aO]}\en{\psihOa
        -\ubO\cdot\OmegaO\nOa} \dS\,\,\\
        &+\int_{\bd\bodyRO}
                \TtO\cdot\phiO \dS
        +\int_{\bd\bodyRO}
                \chiO\cdot\ubO\dS
        +\int_{\bd\bodyRO}
                \en{\TtO\cdot\ubO}
                \Divs[\phiO]
            \dS.
\end{split}
\end{equation}

\section{Applications/Demonstrations}
\label{sec:Demos}
\begin{figure}[htbp]
\centering
\includegraphics[draft=false,width=0.8\textwidth]{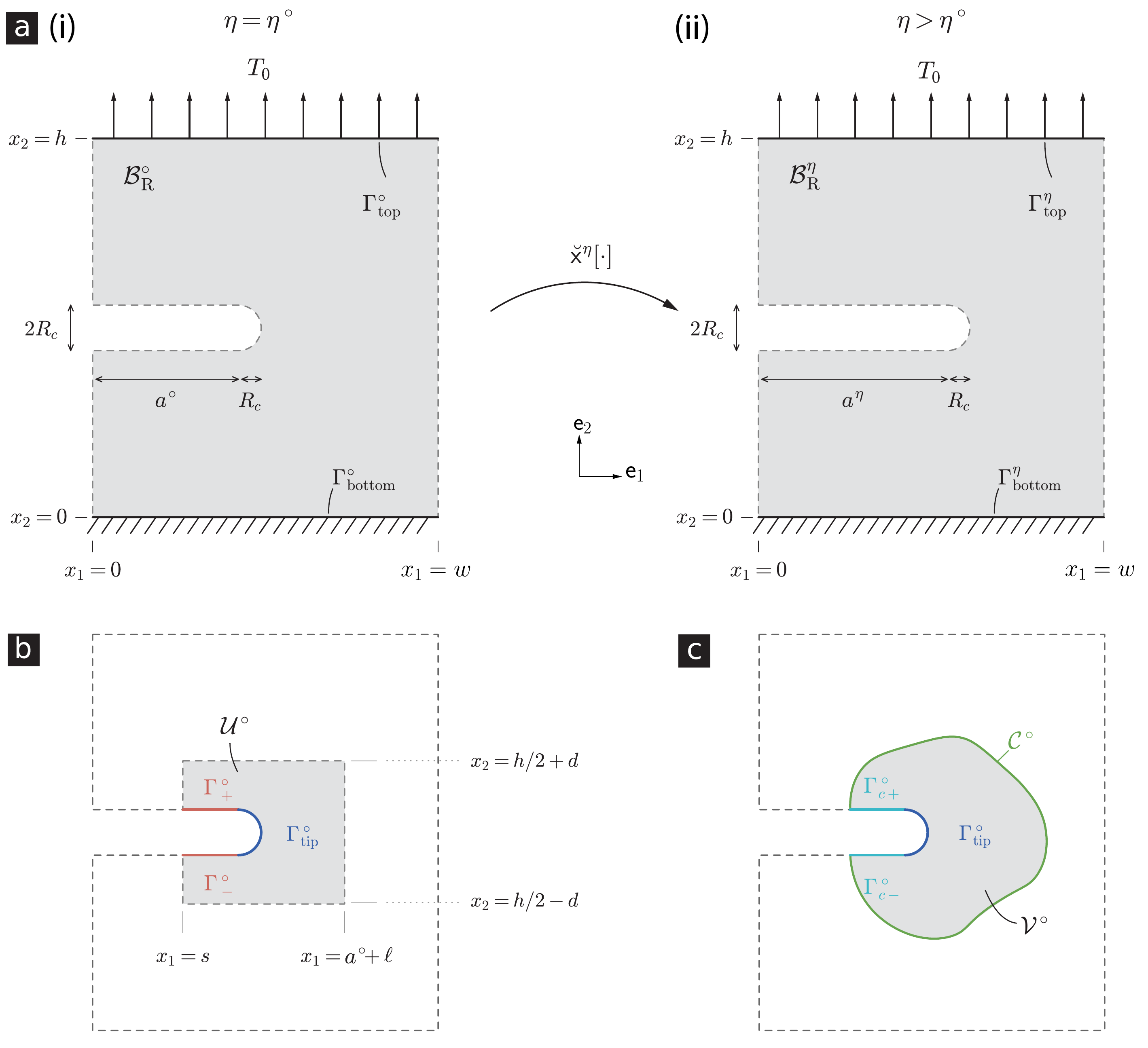}
\caption{Notch-extension problem used to recover the classical $J$-integral. (a)(i) The base state and (a)(ii) an evolved state obtained by extending the notch while holding the material properties and applied boundary conditions fixed. (b) The region $\UO$ used to localize the geometric evolution near the notch tip. (c) An arbitrary contour $\mathcal{C}^\circ$ and the enclosed region $\mathcal{V}^\circ$ used in the path-independence argument.}
\label{fig:JintegralProblem}
\end{figure}

We now demonstrate the utility of the expression for the rate of change of potential energy derived in \eqref{eq:configForce5} by considering the extension of a two-dimensional, notched elastic body. In this example, only the geometry evolves; the material properties and applied boundary conditions remain fixed throughout the evolution. We focus on the state $\eta=\baseval$ and determine the rate of change of potential energy associated with further extension of the notch. We show that the resulting potential-energy release rate reduces to the classical $J$-integral, thereby demonstrating how the general formulation developed above recovers a classical result of configurational mechanics. The problem and the geometric constructions used in the derivation are summarized in Fig.~\ref{fig:JintegralProblem}. We begin by specifying the corresponding variational evolution below.

\subsection{Problem description}

\noindent{\itshape For the present two-dimensional example, we simplify the notation by using $\x=x_1\e[1]+x_2\e[2]$, where $x_i=\x[i]$ for $i=1,2$, to denote a generic point of the ambient space $\Reals[2]$. Thus, within this section, $\x$ may denote a point in either $\bodyRO$ or $\bodyRn$, or in any region outside these bodies. We therefore do not use the separate symbol $\z$ for the image $\xnb\ag{\x}$; the relevant configuration is instead indicated by the domain or superscript.\par}

\vspace{1em}

Consider the state $\eta=\baseval$ at which we wish to evaluate the rate of change of potential energy. At this state, the elastic body occupies the rectangular notched domain shown in Fig.~\ref{fig:JintegralProblem}(a)(i). We represent this domain as
\begin{equation}
\bodyRO:=\en{(0,w)\times(0,h)}\setminus\mathscr{N}^\circ,
\end{equation}
where $w$ and $h$ are the width and height of the rectangular outer boundary, respectively, and $\mathscr{N}^\circ$ denotes the region corresponding to the notch. We specify this region by
\begin{equation}
\mathscr{N}^\circ:=
\pr{(x_1,x_2)\in\Reals[2]~\vert~0\leq x_1\leq a^\circ,\quad \abs{\,x_2-\frac{h}{2}\,}<R_c\,}
\,\,\cup\,\,
\pr{(x_1,x_2)\in\Reals[2]~|~(x_1-a^\circ)^2+\left(x_2-\frac{h}{2}\right)^2<R_c^2,\quad x_1>a^\circ\,},
\end{equation}
where $a^\circ$ is the length of the straight portion of the notch at $\eta=\baseval$, and $R_c$ is the radius of its semicircular tip.

We take the body to be homogeneous and hyperelastic, with strain-energy density $W^\circ\ag{\cdot}$, and assume that body forces are absent. The generalized potential-energy density is therefore
\begin{equation}
\fO\ag{\x,\u,\H}=W^\circ\ag{\H}.
\end{equation}  

The body is subjected to the following boundary conditions. To facilitate their description, we identify the top and bottom edges of the elastic body being considered as
\begin{subequations}
\begin{align}
\Gamma_{\rm top}^\circ&=\pr{(x_1,x_2)\in\Reals[2]~|~ 0\leq x_1\leq w,\quad x_2=h},\\
\Gamma_{\rm bottom}^\circ&=\pr{(x_1,x_2)\in\Reals[2]~|~ 0\leq x_1\leq w,\quad x_2=0},
\end{align}
\end{subequations}
respectively.

The body is subjected to a uniform applied traction $T_0\,\e[2]$ on $\Gamma_{\rm top}^\circ$. The applied traction is zero on the remainder of the boundary. Thus, we take
\begin{equation}
\TtO\ag{\x}=
\begin{cases}
T_0\,\e[2]&\x\in\Gamma_{\rm{top}}^\circ,\\[2pt]
\zerovec& \x\in\bd\bodyRO\setminus\Gamma_{\rm{top}}^\circ.
\end{cases}
\end{equation}

The body's displacement is fixed at zero on its bottom edge, $\Gamma_{\rm{bottom}}^\circ$. Thus, recalling that prescribed displacement is specified by its magnitude $\uhO[\alpha]\ag{\cdot}$ along the directions $\nO[\alpha]\ag{\cdot}$, we take
\begin{equation}
\uhO[1]\ag{\x}=\uhO[2]\ag{\x}=0, \quad \x\in\Gamma_{\rm bottom}^\circ.
\end{equation}
For concreteness, we take $\nO[1]\ag{\cdot}=\e[1]$ and $\nO[2]\ag{\cdot}=\e[2]$, although any orthonormal frame field may be chosen in the present setting.

At equilibrium, the equilibrium traction field satisfies
\begin{subequations}
\begin{align}
\TbO\ag{\x}&=T_0\,\e[2],\qquad \x\in\Gamma_{\rm top}^\circ,\\
\TbO\ag{\x}&=\mathrlap{\zerovec,}\phantom{T_0\,\e[2],}\qquad \x\in\bd\bodyRO\setminus\en{\Gamma_{\rm top}^\circ\cup\Gamma_{\rm bottom}^\circ}.
\end{align}
\end{subequations}

\vspace{1em}
\paragraph*{Geometric evolution (notch extension)}
We consider an extension of the notch in the $\e[1]$ direction. This evolution is modeled through the geometric map $\xnb\ag{\cdot}$. For concreteness, we present one construction of such a map.

Since the notch extension constitutes a local change in geometry, we seek a geometric map that advances the notch tip while leaving the body unchanged sufficiently far from it. To localize the geometric change in this manner, we introduce the region $\UO$ surrounding the notch tip shown in Fig.~\ref{fig:JintegralProblem}(b), outside of which the geometric map will remain the identity. We take $\UO$ to be the portion of a rectangular neighborhood of the notch tip lying within the elastic body:
 \begin{equation}
 \UO:=\pr{(x_1,x_2)\in\bodyRO~|~s <x_1< a^\circ+\mathcal{l},\quad\abs{x_2-h/2}< d},
 \end{equation}
 where we take $0<s<a^\circ$ and $\mathcal{l},d>R_c$.
We now define the geometric map by
\begin{equation}
    \xnb\ag{\x}=\x+(\eta-\baseval)\phiO\ag{\x} \qquad \x\in\bodyRO,
    \label{eq:J-geometric-evolution}
\end{equation}
and choose the smooth vector field $\phiO\ag{\cdot}$ to localize the geometric change to $\UO$. Specifically, for $c>0$, we take it to equal $c\,\e[1]$ in a neighborhood of the notch tip, transition smoothly to zero within $\UO$, and vanish outside $\UO$. Thus, $\xnb\ag{\cdot}$ differs from the identity only within $\UO$, while translating the notch-tip neighborhood in the $\e[1]$ direction at the rate $c$. Accordingly, the straight portion of the notch in the evolved configuration has length
\[
a^\eta:=a^\circ+c\en{\eta-\baseval}.
\]
In particular, to evaluate the rate of change of potential energy for this problem, we will need the value of $\phiO\ag{\cdot}$ on the boundary of $\UO$. To this end, we decompose 
\begin{equation}
\bd\UO=\en{\bd\UO\cap\bd\bodyRO}\cup\en{\bd\UO\setminus\bd\bodyRO}
\end{equation}
where
\[
\bd\UO\cap\bd\bodyRO=\Gamma_{+}^\circ\cup\Gamma^\circ_{\rm tip}\cup\Gamma_{-}^\circ.
\]
Here, $\Gamma_{\rm tip}^\circ$ denotes the semicircular notch tip boundary, and $\Gamma_{+}^\circ$ and $\Gamma_{-}^\circ$ denote the upper and lower straight faces of the notch, respectively, belonging to $\bd\UO$. That is,
\begin{subequations}
\begin{align}
\Gamma_{\rm tip}^\circ
&=\pr{(x_1,x_2)\in\Reals[2]~|~(x_1-a^\circ)^2+(x_2-h/2)^2=R_c^2,\quad x_1>a^\circ},\\
\Gamma_{+}^\circ
&=\pr{(x_1,x_2)\in\Reals[2]~|~s < x_1 < a^\circ, \quad x_2=h/2+R_c},\\
\Gamma_{-}^\circ
&=\pr{(x_1,x_2)\in\Reals[2]~|~s < x_1 < a^\circ, \quad x_2=h/2-R_c}.
\end{align}
\end{subequations}
 
With this decomposition, the values of $\phiO\ag{\cdot}$ on $\bd\UO$ are
\begin{equation}
\phiO\ag{\x}=
\begin{cases}
c\,\e[1], \quad &\x\in\Gamma_{\rm tip}^\circ\\
c\,q\ag{\x}\e[1], \quad &\x\in\Gamma_{+}^\circ\cup\Gamma_{-}^\circ\\
\zerovec, \quad &\x\in\bd\UO\setminus\bd\bodyRO,
\end{cases}
\label{eq:phiOeqs}
\end{equation}
where $q\ag{\cdot}$ is a smooth scalar function which transitions from zero to unity along $\Gamma_\pm^\circ$, from $x_1=s$ to $x_1=a^\circ$. The complete, explicit construction of $\phiO\ag{\cdot}$ is deferred to Appendix \ref{app:phiOConstruction}, where we verify that it satisfies the regularity conditions introduced in \S\ref{subsec:domainevol} and that, for $\eta$ sufficiently close to $\baseval$, the corresponding geometric map $\xnb\ag{\cdot}$ is a diffeomorphism onto its image.

\vspace{1em}
\paragraph*{Description of the evolved problem}

Through the geometric evolution described above, we obtain a family of elasticity problems that differ only in the geometry of the notch. That is, as the notch extends, the material properties of the body and the applied boundary conditions remain fixed. We now make this description precise.

At a nearby $\eta>\baseval$, the body occupies the evolved domain illustrated in Fig.~\ref{fig:JintegralProblem}(a)(ii):
\[\bodyRn=\xnb\ag{\bodyRO},\]
and the straight portion of its notch has length $a^\eta=a^\circ+c\en{\eta-\baseval}$, with its boundary portions obtained by the same geometric map. In particular, the top and bottom edges are 
\begin{equation}
\Gamma_{\rm top}^\eta:=\xnb\ag{\Gamma_{\rm top}^\circ} \quad \text{and}\quad
\Gamma_{\rm bottom}^\eta:=\xnb\ag{\Gamma_{\rm bottom}^\circ},
\end{equation}
respectively. As the geometric evolution is localized to the neighborhood $\UO$ near the notch tip, the geometric map is simply the identity on the top and bottom edges. Thus, $\Gamma_{\rm top}^\eta=\Gamma_{\rm top}^\circ$ and $\Gamma_{\rm bottom}^\eta=\Gamma_{\rm bottom}^\circ$.

Since we consider the material properties of the body to remain fixed, the body at the evolved state consists of the same homogeneous hyperelastic material as at $\eta=\baseval$. Its strain-energy density therefore remains $W^\circ\ag{\cdot}$, so
\begin{equation}
\fn\ag{\cdot}=\fO\ag{\cdot}.
\label{eq:fnExample}
\end{equation}
Likewise, we consider the evolved body to be subjected to the same physical boundary conditions as those prescribed at $\eta=\baseval$: the uniform applied traction $T_0\,\e[2]$ remains prescribed on the top edge, with zero applied traction on the remainder of the boundary, while the bottom edge remains fixed at zero displacement. Thus, 
\begin{equation}
\Ttn\ag{\x}=
\begin{cases}
T_0\,\e[2],&\x\in\Gamma_{\rm{top}}^\eta,\\[2pt]
\zerovec,& \x\in\bd\bodyRn\setminus\Gamma_{\rm{top}}^\eta,
\end{cases}\qquad
\text{and}
\qquad \uhn[1]\ag{\x}=\uhn[2]\ag{\x}=0, \quad \x\in\Gamma_{\rm bottom}^\eta,
\label{eq:etaBC}
\end{equation}
where we take the associated directions $\nn[1]\ag{\cdot}$ and $\nn[2]\ag{\cdot}$ to be the same as that chosen at $\eta=\baseval$.
Expressing the boundary conditions in \eqref{eq:etaBC} in the pulled-back form introduced in \S\ref{subsec:appliedtrac}--\S\ref{subsec:dispevol}, we obtain 
\begin{equation}
\Ttnc\ag{\cdot}=\TtO\ag{\cdot},\quad\uhnc[\alpha]\ag{\cdot}=\uhO[\alpha]\ag{\cdot},\quad\nnca\ag{\cdot}=\nOa\ag{\cdot}. 
\label{eq:BCexamples}
\end{equation}
Recall that this is precisely how non-evolving applied traction and prescribed displacement fields are described in our framework (cf. \S\ref{subsec:appliedtrac} and \S\ref{subsec:dispevol}).
Comparison of \eqref{eq:fnExample} with \eqref{eq:fAsymptoticForm} and \eqref{eq:BCexamples} with \eqref{eq:TtncAsymp}, \eqref{eq:uhncAsymp} and \eqref{eq:nncaAsymp} shows that the associated first-order terms in the asymptotic expansions vanish, i.e.,
\begin{equation}
\zetaO\ag{\cdot}=0, \quad \chiO\ag{\cdot}=\zerovec, \quad \psihOa\ag{\cdot}=0,\quad \OmegaO\ag{\cdot}=\zeroMat.
\label{eq:firstorderterms}
\end{equation}

\vspace{1em}
\subsection{Specialization of the \prr}

We now evaluate \prr of this evolution by applying \eqref{eq:configForce5}.

Substituting \eqref{eq:firstorderterms} into \eqref{eq:configForce5}, the expression for $F_c$ reduces to
\begin{equation}
F_c^\circ=
        -\int_{\bodyRO}
            \frac{d}{d\x[J]}
            \en{\SigmaO[p][J]\phiO[p]}
        \dV+\int_{\bd\bodyRO}
            \TtO\cdot\phiO
        \dS+\int_{\bd\bodyRO}
            \en{\TtO\cdot\ubO}
            \Divs[\phiO]
        \dS,
        \label{eq:FcSpecialized1}
\end{equation}
Because $\phiO$ is localized to the region $\UO$ shown in Fig.~\ref{fig:JintegralProblem}(b), the first integral in \eqref{eq:FcSpecialized1} may be restricted to $\UO$. We also have that $\phiO=\zerovec$ on $\bd\bodyRO\setminus\bd\UO$, and consequently $\Divs\ag{\phiO}=0$ there. Hence, the second and third integrals may be restricted to $\bd\bodyRO\cap\bd\UO$. However, the applied traction field vanishes on $\bd\bodyRO\cap\bd\UO$, i.e., $\TtO=\zerovec$ (see Fig.~\ref{fig:JintegralProblem}(b)). The second and third integrals therefore vanish entirely, and we obtain
\begin{equation}
    F_c^\circ
    =
    -\int_{\UO}
    \frac{d}{d\x[J]}
    \left(
        \SigmaO[p][J]
        \phiO[p]
    \right)\dV
\end{equation}

Applying the divergence theorem gives
\begin{equation}
    F_c^\circ
    =
    -\int_{\partial\UO}
    \SigmaO[p][J]
        \phiO[p]\,\NUh[J] \,dS
        \label{eq:intermedFcCirc}
\end{equation}
where $\NUh\ag{\cdot}:\bd\UO\to\Reals[m]$ in the equation above denotes the field of outward unit normal to the region enclosed by $\UO$.
Identifying $\NUh\ag{\x}=\Nh\ag{\x}$ on $\x\in\bd\bodyRO\cap\bd\UO$, then substituting \eqref{eq:phiOeqs} into \eqref{eq:intermedFcCirc}, we obtain
\begin{equation}
    F_c^\circ
    =
    -c \int_{\Gamma_{\rm tip}^\circ}
    \SigmaO[1][J]\,
        \Nh[J] \,dS
    -c \int_{\Gamma_{+}^\circ\cup\Gamma_{-}^\circ}
    q\,\SigmaO[1][J]
        \Nh[J] \,dS.
    \label{eq:Fc-tip}
\end{equation}
To show that the second integral vanishes entirely, we recall the definition of the Eshelby energy-momentum tensor, $\SigmaO[1][J]=\fbO\delta_{1J}-\FPO[i][J]\ubO[i,1]$, and that $\TbO:=\FPO\Nh$, and rewrite the second integral in \eqref{eq:Fc-tip} as
\begin{equation}
    F_c^\circ
    =
    -c
    \int_{\Gamma_{\rm tip}^\circ}\SigmaO[1][J]\,
        \Nh[J]\dS
    -c \int_{\Gamma_{+}^\circ\cup\Gamma_{-}^\circ}
    q\,\en{\fbO\Nh[1]-\TbO[i]\,\ubO[i,1]} \,dS
    \label{eq:Fc-tip-expanded}
\end{equation}
On $\Gamma_{+}^\circ\cup\Gamma_{-}^\circ$, we have that $\Nh[1]=0$ and, by the equilibrium condition, $\TbO\ag{\x}=\zerovec$. Thus, \eqref{eq:Fc-tip-expanded} reduces to
\begin{equation}
    F_c^\circ
    =
    -c
    \int_{\Gamma_{\rm tip}^\circ}\SigmaO[1][J]\,
        \Nh[J]\dS
    \label{eq:Fc-tip-expanded2}
\end{equation}

We now show that, for an arbitrary contour $\mathcal{C}^\circ$ enclosing the notch tip and terminating on the straight notch faces, as illustrated in Fig.~\ref{fig:JintegralProblem}(c), we may rewrite \eqref{eq:Fc-tip-expanded2} as
\begin{equation}
    F_c^\circ
    =
    c
    \int_{\mathcal{C}^\circ}\SigmaO[1][J]\,\NCh[J]\dS,
    \label{eq:Fc-tip-expanded3}
\end{equation}
where $\NCh\ag{\cdot}$ denotes the field of outward unit normal to the region enclosed by $\mathcal{C}^\circ$. 

To establish \eqref{eq:Fc-tip-expanded3}, we first recall that, for the homogeneous hyperelastic body considered here, the Eshelby energy-momentum tensor satisfies\footnote{Indeed, substituting the definition of the Eshelby energy-momentum tensor, we have
\begin{align}
\frac{d}{d\x[J]}\SigmaO[1][J]&=\frac{d\fbO}{d\x[1]}-\FPO[i][J,J]\ubO[i,1]-\FPO[i][J]\ubO[i,1,J].
    \label{eq:J-div-free2}
\end{align}
Since the body is homogeneous and we have specified $\fO\ag{\x,\u,\H}=W^\circ\ag{\H}$, we may write $d\fbO/d\x[1] =\fbO[,iJ]\ubO[i,J,1]$.  Further recalling that $\fbO[,iJ]=\FPO[i][J]$ and rearranging, we obtain
\begin{align}
\frac{d}{d\x[J]}\SigmaO[1][J]&=-\FPO[i][J,J]\ubO[i,1]+\FPO[i][J]\en{\ubO[i,J,1]-\ubO[i,1,J]}.
\label{eq:J-div-free3}
\end{align}
By equilibrium (in the absence of body force), we have $\FPO[i][J,J]=0$. Assuming sufficient regularity of $\ubO$ for the present problem, the mixed derivatives commute, hence we also have $\ubO[i,J,1]=\ubO[i,1,J]$. The right-hand side of \eqref{eq:J-div-free3} thereby vanishes, establishing \eqref{eq:J-div-free1}.}
\begin{equation}
\frac{d}{d\x[J]}\SigmaO[1][J]=0.
\label{eq:J-div-free1}
\end{equation}
We now consider the region $\mathcal{V}^\circ\subset\bodyRO$ bounded by $\Gamma_{\rm tip}^\circ$, $\mathcal{C}^\circ$, and the portions $\Gamma_{\mathcal{C}+}^\circ$ and $\Gamma_{\mathcal{C}-}^\circ$ of the upper and lower straight notch faces, respectively, that extend between them. That is,
\begin{equation}
\bd\mathcal{V}^\circ=\Gamma_{\mathcal{C}+}^\circ\cup\Gamma_{\rm tip}^\circ\cup\Gamma_{\mathcal{C}-}^\circ\cup\mathcal{C}^\circ.
\label{eq:bdVdecomp}
\end{equation}
Integrating \eqref{eq:J-div-free1}  over $\mathcal{V}^\circ$, applying the divergence theorem,  and decomposing the resulting boundary integral according to  \eqref{eq:bdVdecomp} gives
\begin{equation}
    \int_{\Gamma_{\rm tip}^\circ}\SigmaO[1][J]\NVh[J]\dS +\int_{\Gamma_{\mathcal{C}+}^\circ\cup\Gamma_{\mathcal{C}-}^\circ}\SigmaO[1][J]\NVh[J]\dS+\int_{\mathcal{C}^\circ}\SigmaO[1][J]\NVh[J]\dS= 0.
    \label{eq:decompOvbdV}
\end{equation}
In \eqref{eq:decompOvbdV}, $\NVh\ag{\cdot}:\bd\mathcal{V}^\circ\to\Reals[m]$ denotes the field of outward unit normal to $\bd\mathcal{V}^\circ$. Analogously to the passage from \eqref{eq:Fc-tip} to \eqref{eq:Fc-tip-expanded2}, the contributions from the straight notch faces, $\Gamma_{\mathcal{C}+}^\circ\cup\Gamma_{\mathcal{C}-}^\circ$, vanish. 
Upon identifying $\NVh=\Nh$ on $\Gamma_{\rm tip}^\circ$ and $\NVh=\NCh$ on $\mathcal{C}^\circ$, we obtain
\begin{equation}
    -\int_{\Gamma_{\rm tip}^\circ}
    \SigmaO[1][J]\Nh[J] \,dS
    =\int_{\mathcal{C}^\circ}
   \SigmaO[1][J]\NCh[J]\,dS.
    \label{eq:J-path-independent}
\end{equation}
Comparing \eqref{eq:Fc-tip-expanded2} and \eqref{eq:J-path-independent}, we arrive at \eqref{eq:Fc-tip-expanded3}. 

The expression in \eqref{eq:Fc-tip-expanded3} may already be identified as the standard $J$-integral. To make this identification explicit, we recall the definition of the Eshelby energy-momentum tensor. Noting that $\fbO=W^\circ$ for the problem considered here, and defining the traction associated with $\mathcal{C}^\circ$ by $\tp{t}^\circ=\FPO\NCh$, \eqref{eq:Fc-tip-expanded3} becomes
\begin{equation}
    F_c^\circ
    =
    c
    \int_{\mathcal{C}^\circ}
    \en{W^\circ\NCh[1]-\tp{t}^\circ_{\cdot i}\ubO[i,1]}\dS.
\end{equation}
We have thus recovered the familiar expression for the $J$-integral associated with notch extension in the $\e[1]$ direction. Hence,
\begin{equation}
    \boxed{F_c^\circ=cJ}.
\end{equation}

\appendix

\section{Notation}
\label{app:notation}

\subsection*{Principal notation}
The principal notation used throughout the paper is collected below. The subscript $\mathrm{R}$ identifies sets associated with the reference configuration of the corresponding variational problem. A superscript $\circ$ denotes evaluation at the arbitrary state $\eta=\baseval$ about which the local evolution is considered. A check accent denotes a quantity pulled back to the fixed domain (or boundary) associated with that state. For bodies with sharp material interfaces, the subscript $(r)$, $r=1,2$, identifies the material region associated with a quantity. Symbols are listed alphabetically by their principal symbol; within a symbol family, $\eta$-dependent quantities precede their $\circ$ counterparts.

\renewcommand{\arraystretch}{1.16}
\begin{longtable}{@{}p{0.22\textwidth}p{0.74\textwidth}@{}}
\toprule
\textbf{Symbol} & \textbf{Meaning}\\
\midrule
\endfirsthead
\toprule
\textbf{Symbol} & \textbf{Meaning}\\
\midrule
\endhead
\bottomrule
\endfoot

$\An$ & Admissible set for the $\eta$-variational problem.\\
$\Anc$ & Pulled-back admissible set on $\bodyRO$.\\
$\AC$ & Parent displacement space containing every pulled-back admissible set $\Anc$.\\

$\bn\ag{\cdot}$ & Vector field with components $\fn[,i]\ag{\cdot}$ evaluated at equilibrium; identified with the body-force term in the equilibrium equations.\\
$\bO\ag{\cdot}$ & $\bn\ag{\cdot}$ evaluated at $\eta=\baseval$.\\
$\br\ag{\cdot}$ & Counterpart of $\bn\ag{\cdot}$ in material region $r$.\\

$\bodyRn$ & Domain of the elastic body in $\Reals[m]$ in the reference configuration of the $\eta$-problem.\\
$\bodyRO$ & Domain of the elastic body in $\Reals[m]$ in the reference configuration at $\eta=\baseval$.\\
$\bodyRn[r]$ & Domain of material region $r$ of the elastic body at the $\eta$-state.\\
$\bodyRO[r]$ & Domain of material region $r$ of the elastic body at $\eta=\baseval$.\\
$\bd\bodyRn$ & Boundary of $\bodyRn$.\\
$\bd\bodyRO$ & Boundary of $\bodyRO$.\\
$\bd\bodyRn[r]$ & Boundary of $\bodyRn[r]$.\\
$\bd\bodyRO[r]$ & Boundary of $\bodyRO[r]$.\\

$\chiO\ag{\cdot}$ & First-order term in the asymptotic expansion of the pulled-back applied traction field about $\eta=\baseval$.\\
$\dOn\ag{\cdot}$ & Geometric displacement field of the $\eta$-problem, defined relative to the $\eta=\baseval$ state.\\
$\dOnc\ag{\cdot}$ & Pullback of $\dOn\ag{\cdot}$ to $\bodyRO$; equivalently, $\dOnc\ag{\x}=\xnb\ag{\x}-\x$.\\

$\eta$ & Evolution parameter indexing the one-parameter family of variational problems.\\
$\baseval$ & Arbitrary state of the evolution about which the local evolution is characterized.\\

$\fn\ag{\cdot}$ & Generalized potential energy density of the $\eta$-problem.\\
$\fO\ag{\cdot}$ & Generalized potential energy density at $\eta=\baseval$.\\
$\fnr\ag{\cdot}$ & Generalized potential energy density associated with material region $r$.\\
$\fOr\ag{\cdot}$ & Generalized potential energy density of material region $r$ at $\eta=\baseval$.\\
$\fbO$ & Shorthand for the generalized potential energy density evaluated at equilibrium at $\eta=\baseval$, $\fbO:=\fO\ag{\x,\ubO,\grad\ubO}$.\\
$\bar f^\circ_{,I},\ \bar f^\circ_{,i},\ \bar f^\circ_{,iJ}$&Shorthand for partial derivatives of $\fO\ag{\cdot}$ evaluated at equilibrium at $\eta=\eta^\circ$, e.g.,  $\fbO[,I]:=\fO[,I]\ag{\x,\ubO,\grad\ubO}$.\\

$F_c^{\eta}$ & Negative of the rate of change of equilibrium potential energy with respect to the evolution parameter.\\
$F_c^{\circ}$ & $F_c^{\eta}$ evaluated at $\eta=\baseval$.\\
$F_c^{\bodyRO}$ & Volume contribution to $F_c^{\circ}$.\\
$F_c^{\bd\bodyRO}$ & Surface contribution to $F_c^{\circ}$.\\

$\Gn\ag{\cdot}$ & Jacobian determinant associated with $\xnb\ag{\cdot}$.\\
$\Gs\ag{\cdot}$ & Surface Jacobian (surface-area scaling factor) associated with $\xnb\ag{\cdot}$.\\

$\GammaErn$ & Portion of the external boundary occupied by material region $r$ at the $\eta$-state.\\
$\GammaErO$ & Portion of the external boundary occupied by material region $r$ at $\eta=\baseval$.\\
$\GammagRna$ & $\alpha$-Dirichlet boundary at the $\eta$-state.\\
$\GammagROa$ & $\alpha$-Dirichlet boundary at $\eta=\baseval$.\\
$\GammahRna$ & Complement of the $\alpha$-Dirichlet boundary $\GammagRna$ in $\bd\bodyRn$.\\
$\GammahROa$ & Complement of the $\alpha$-Dirichlet boundary $\GammagROa$ in $\bd\bodyRO$.\\
$\GammaIn$ & Internal material interface at the $\eta$-state.\\
$\GammaIO$ & Internal material interface at $\eta=\baseval$.\\

$\Id$ & Identity tensor.\\

$\nna\ag{\cdot}$ & $\alpha$-th member of the auxiliary orthonormal frame field at the $\eta$-state.\\
$\nnca\ag{\cdot}$ & Pullback of $\nna\ag{\cdot}$ to $\bd\bodyRO$.\\
$\nOa\ag{\cdot}$ & $\alpha$-th member of the auxiliary orthonormal frame field at $\eta=\baseval$.\\

$\Nnh\ag{\cdot}$ & Outward unit normal field on $\bd\bodyRn$.\\
$\Nh\ag{\cdot}$ & Outward unit normal field on $\bd\bodyRO$.\\
$\Nnrh\ag{\cdot}$ & Outward unit normal field on $\bd\bodyRn[r]$.\\

$\OmegaO\ag{\cdot}$ & Infinitesimal rotation field governing the first-order evolution of the auxiliary orthonormal frame.\\

$\Phin\ag{\cdot}$ & Body-force potential of the $\eta$-problem.\\
$\phiO\ag{\cdot}$ & First-order field characterizing the local geometric evolution of $\xnb\ag{\cdot}$ about $\eta=\baseval$.\\

$\Pin\ag{\cdot}$ & Evolutionarily consistent potential energy functional of the $\eta$-problem.\\
$\PiO\ag{\cdot}$ & Evolutionarily consistent potential energy functional at $\eta=\baseval$.\\
$\Pin_{\rm S}\ag{\cdot}$ & Standard potential energy functional of the $\eta$-problem.\\
$\PiO_{\rm S}\ag{\cdot}$ & Standard potential energy functional at $\eta=\baseval$.\\
$\check{\Pi}^{\eta}\ag{\cdot}$ & Potential energy functional pulled back to the fixed domain $\bodyRO$.\\
$\Pi\ag{\eta,\cdot}$ & Parent potential energy functional on the parent admissible space $\AC$.\\
$\Pibn$ & Equilibrium potential energy at the $\eta$-state, $\Pibn:=\Pin\ag{\ubn}$.\\
$\PibO$ & Equilibrium potential energy at $\eta=\baseval$.\\

$\psihOa\ag{\cdot}$ & First-order term in the asymptotic expansion of the pulled-back prescribed displacement value.\\
$\psiO\ag{\cdot}$ & First-order term in the asymptotic expansion of $\ubnc\ag{\cdot}$ about $\eta=\baseval$.\\

$\SigmaO\ag{\cdot}$ & Eshelby energy--momentum tensor at $\eta=\baseval$.\\
$\SigmaOr\ag{\cdot}$ & Eshelby energy--momentum tensor associated with material region $r$ at $\eta=\baseval$.\\

$T\An$ & Space of admissible variations at the $\eta$-state.\\
$\Ttn\ag{\cdot}$ & Applied traction field on $\bd\bodyRn$.\\
$\Ttnc\ag{\cdot}$ & Pullback of the applied traction field to $\bd\bodyRO$ via $\xnb\ag{\cdot}$.\\
$\TtO\ag{\cdot}$ & Applied traction field at $\eta=\baseval$.\\
$\Tbn\ag{\cdot}$ & Equilibrium traction field on $\bd\bodyRn$, $\Tbn:=\FPn\Nnh$.\\
$\TbO\ag{\cdot}$ & Equilibrium traction field evaluated at $\eta=\baseval$.\\
$\Tbnr\ag{\cdot}$ & Equilibrium traction associated with material region $r$, $\Tbnr:=\FPnr\Nnrh$.\\

$\u\ag{\cdot}$ & Generic admissible displacement field.\\
$\ubn\ag{\cdot}$ & Equilibrium displacement field of the $\eta$-problem.\\
$\ubnc\ag{\cdot}$ & Pullback of $\ubn\ag{\cdot}$ to $\bodyRO$.\\
$\ubO\ag{\cdot}$ & Equilibrium displacement field at $\eta=\baseval$.\\
$\uhn[\alpha]\ag{\cdot}$ & Prescribed displacement value along $\nna\ag{\cdot}$ on $\GammagRna$.\\
$\uhnc[\alpha]\ag{\cdot}$ & Pullback of the prescribed displacement value $\uhn[\alpha]\ag{\cdot}$ to $\GammagROa$.\\
$\uhO[\alpha]\ag{\cdot}$ & Prescribed displacement value at $\eta=\baseval$.\\
$\vu\ag{\cdot}$ & Admissible displacement variation.\\

$\Wn\ag{\cdot}$ & Strain energy density of the $\eta$-problem.\\

$\Xn$ & Shorthand for the argument triple $\en{\xnb,\ubnc,\grad\ubnc\en{\grad\xnb}^{-1}}$.\\
$\XO$ & Shorthand for the argument triple $\en{\x,\ubO,\grad\ubO}$.\\
$\xnb\ag{\cdot}$ & Geometric map taking $\bodyRO$ to $\bodyRn$.\\

$\zetaO\ag{\cdot}$ & First-order term in the asymptotic expansion of $\fn\ag{\cdot}$ about $\eta=\baseval$.\\
$\zetaOr\ag{\cdot}$ & First-order term in the asymptotic expansion of $\fnr\ag{\cdot}$ about $\eta=\baseval$.\\
$\zetabO$ & Shorthand for $\zetaO\ag{\cdot}$ evaluated at equilibrium at $\eta=\baseval$, $\zetabO:=\zetaO\ag{\x,\ubO,\grad\ubO}$.\\
$\zetabOr$ & Shorthand for $\zetaOr\ag{\cdot}$ evaluated at equilibrium at $\eta=\baseval$, $\zetabOr:=\zetaOr\ag{\x,\ubO,\grad\ubO}$.\\

$\FPn\ag{\cdot}$ & First Piola--Kirchhoff stress field of the $\eta$-problem.\\
$\FPO\ag{\cdot}$ & First Piola--Kirchhoff stress field evaluated at $\eta=\baseval$.\\
$\FPnr\ag{\cdot}$ & First Piola--Kirchhoff stress field of the $\eta$-problem in material region $r$.\\

\end{longtable}

\subsection*{Mathematical conventions and operators}

The mathematical conventions and operators used throughout the paper are collected below.

\renewcommand{\arraystretch}{1.16}
\begin{longtable}{@{}p{0.22\textwidth}p{0.74\textwidth}@{}}
\toprule
\textbf{Symbol / convention} & \textbf{Meaning}\\
\midrule
\endfirsthead
\toprule
\textbf{Symbol / convention} & \textbf{Meaning}\\
\midrule
\endhead
\bottomrule
\endfoot

Sans-serif symbols &
Vectors, vector-valued functions, and matrix-valued functions. 
\\
Roman symbols &
Scalars and scalar-valued functions.
\\

$\Reals[m]$, $\Matrix[m\times m]$ &
$m$-dimensional real Euclidean space and the space of real $m\times m$ matrices, respectively, with $m=2$ or $3$ in the present work.
\\

$L^p(\Omega;E)$ &
Lebesgue space of measurable $E$-valued fields $v$ on $\Omega$ satisfying $\int_\Omega\lVert v\rVert_E^p<\infty$; $L^p(\Omega)$ denotes the scalar-valued case.
\\

$W^{1,p}(\Omega;\Reals[m])$ &
Sobolev space of $\Reals[m]$-valued fields on $\Omega$ whose components and first weak derivatives belong to $L^p(\Omega)$.
\\

$\E[I]$, $\E[iJ]$ &
Canonical basis elements of $\Reals[m]$ and $\Matrix[m\times m]$, respectively.
\\

$\x[J]$, $\u[i]\ag{\cdot}$, $\H[i][J]\ag{\cdot}$ &
$J$-th component of $\x$, $i$-th component function of $\u\ag{\cdot}$, and $iJ$-th component function of $\H\ag{\cdot}$, respectively, with respect to the canonical bases.
\\

$\u[\an]\ag{\cdot}$,\,$\u[\aO]\ag{\cdot}$ &Components of $\u\ag{\cdot}$ with respect to $\nna\ag{\cdot}$ and $\nOa\ag{\cdot}$, respectively; that is, $\u[\an]\ag{\x}:=\u\ag{\x}\cdot\nna\ag{\x}$ and $\u[\aO]\ag{\x}:=\u\ag{\x}\cdot\nOa\ag{\x}$.\\

$\delta_{ij}$ &
Kronecker delta.
\\

$\zerovec$, $\zeroMat$ &
Null elements of $\Reals[m]$ and $\Matrix[m\times m]$, respectively.
\\

$Df(\cdot)\ag{\cdot}$ &
Fr\'echet derivative of $f\ag{\cdot}$; $Df(\x)\ag{\tp{h}}$ denotes the derivative evaluated at $\x$ and applied to the direction $\tp{h}$.
\\

$\u[i,J]\ag{\cdot}$, $\u[i,J,K]\ag{\cdot}$ &
First and successive spatial partial derivatives of the $i$-th component function of $\u\ag{\cdot}$; in particular, $\u[i,J,K]\ag{\cdot}:=(\u[i,J])_{,K}\ag{\cdot}$.
\\

$\grad\u\ag{\cdot}$ &
Gradient of $\u\ag{\cdot}$, given by $\grad\u\ag{\cdot}:=\u[i,J]\ag{\cdot}\E[iJ]$.
\\

$\dfrac{d}{d\x[J]}$ &
Total derivative with respect to the $J$-th component of the $\x$.
\\

$\Divs[\cdot]$ &
Surface divergence operator.
\\

$\det\ag{\cdot}$, $\cof\ag{\cdot}$, $\rm{Tr}\ag{\cdot}$ &
Determinant, cofactor matrix, and trace operators, respectively.
\\

$\norm{\cdot}_{\Reals[m]}$, $\norm{\cdot}_{\mathbb{V}}$ &
Euclidean norm on $\Reals[m]$ and, more generally, the norm on a normed vector space $\mathbb{V}$, respectively.
\\

$\Bv[\eta\to\baseval]$ &
Shorthand indicating the limit as $\eta\to\baseval$ when appended to an expression.
\\

\end{longtable}

\section{Regularity of the generalized potential-energy density}
\label{app:fnRegularity}

Here, we record sufficient conditions for the limit relation used in \S\ref{sec:volumeterm}. We introduce the product space
\begin{equation}
\mathbb{V}
:=
C^0\en{\overline{\bodyRO};\mathbb{R}^m}
\times
C^0\en{\overline{\bodyRO};\mathbb{R}^m}
\times
L^p\en{\bodyRO;\Matrix[m\times m]},
\label{eq:VDef}
\end{equation}
equipped, for $\X\in\mathbb{V}$, with the norm
\begin{equation}
\norm{\X}_{\mathbb{V}}
:=
\norm{\X[1]}_{C^0}
+
\norm{\X[2]}_{C^0}
+
\norm{\X[3]}_{L^p}.
\label{eq:VNorm}
\end{equation}
The three components of $\X$ correspond, respectively, to the spatial, displacement, and displacement-gradient arguments of the generalized potential-energy density. For each $\eta$, define the weighted integrated functional $\mathcal{F}^\eta\ag{\cdot}:\mathbb{V}\to\Reals$ by
\begin{equation}
\mathcal{F}^\eta[\X]
:=
\int_{\bodyRO}
\Gn\ag{\x}\,\fn\ag{\X\ag{\x}}\,dV,
\label{eq:integratedFDef}
\end{equation}
where $G^\circ[\cdot]\equiv1$.

Suppose that
\begin{equation}
\norm{\Xn-\XO}_{\mathbb{V}}
=
O\en{\abs{\eta-\baseval}}.
\label{eq:XnBigO}
\end{equation}
We assume that $\mathcal{F}^\eta\ag{\cdot}$ is Fr\'echet differentiable at $\XO$ and that
\begin{equation}
\norm{
D\mathcal{F}^\eta(\XO)
-
D\mathcal{F}^\circ(\XO)
}_{\mathrm{op}}
\longrightarrow0
\qquad\text{as }\eta\to\baseval.
\label{eq:DFnConv}
\end{equation}
We further assume that
\begin{equation}
\mathcal{F}^\eta[\Xn]
-
\mathcal{F}^\eta[\XO]
=
D\mathcal{F}^\eta(\XO)\ag{\Xn-\XO}
+
\alpha^\eta\norm{\Xn-\XO}_{\mathbb{V}},
\label{eq:FExpansion}
\end{equation}
where
\begin{equation}
\alpha^\eta\longrightarrow0
\qquad\text{as }\eta\to\baseval.
\label{eq:alphaConv}
\end{equation}
When $\Xn=\XO$, we take $\alpha^\eta=0$.

It follows from \eqref{eq:FExpansion} and the triangle inequality that
\begin{equation}
\begin{split}
&
\abs{
\frac{
\mathcal{F}^\eta[\Xn]
-
\mathcal{F}^\eta[\XO]
}{
\eta-\baseval
}
-
D\mathcal{F}^\circ(\XO)
\left[
\frac{\Xn-\XO}{\eta-\baseval}
\right]
}
\\
&\qquad\leq
\norm{
D\mathcal{F}^\eta(\XO)
-
D\mathcal{F}^\circ(\XO)
}_{\mathrm{op}}
\frac{
\norm{\Xn-\XO}_{\mathbb{V}}
}{
\abs{\eta-\baseval}
}
+
\abs{\alpha^\eta}
\frac{
\norm{\Xn-\XO}_{\mathbb{V}}
}{
\abs{\eta-\baseval}
}.
\end{split}
\label{eq:FEstimate}
\end{equation}
By \eqref{eq:XnBigO}, the quotients involving $\norm{\Xn-\XO}_{\mathbb{V}}$ on the right-hand side are bounded. The first factor in the first term tends to zero by \eqref{eq:DFnConv}, while $\abs{\alpha^\eta}\to0$ by \eqref{eq:alphaConv}. Hence,
\begin{equation}
\lim_{\eta\to\baseval}
\frac{
\mathcal{F}^\eta[\Xn]
-
\mathcal{F}^\eta[\XO]
}{
\eta-\baseval
}
=
\lim_{\eta\to\baseval}
D\mathcal{F}^\circ(\XO)
\left[
\frac{\Xn-\XO}{\eta-\baseval}
\right],
\label{eq:FLinearizationLimit}
\end{equation}
whenever the limit on the right-hand side exists.

We further assume that differentiation of $\mathcal{F}^\circ\ag{\cdot}$ with respect to $\X$ may be performed under the integral. Because $G^\circ[\cdot]\equiv1$, this gives
\begin{equation}
D\mathcal{F}^\circ(\XO)[\tp{Y}]
=
\int_{\bodyRO}
D\fO\en{\XO\ag{\x}}
\ag{\tp{Y}\ag{\x}}\,dV,
\qquad
\tp{Y}\in\mathbb{V}.
\label{eq:DFIntegralRepresentation}
\end{equation}
Moreover, by \eqref{eq:integratedFDef},
\begin{equation}
\mathcal{F}^\eta[\Xn]
-
\mathcal{F}^\eta[\XO]
=
\int_{\bodyRO}
G^\eta[\x]
\en{
\fn\ag{\Xn}
-
\fn\ag{\XO}
}\,dV.
\label{eq:LHS}
\end{equation}
Substituting \eqref{eq:LHS} into the left-hand side of \eqref{eq:FLinearizationLimit} and applying the representation \eqref{eq:DFIntegralRepresentation} to its right-hand side, we obtain
\begin{equation}
\begin{split}
&
\lim_{\eta\to\baseval}
\int_{\bodyRO}
G^\eta[\x]
\frac{
\fn\ag{\Xn}
-
\fn\ag{\XO}
}{
\eta-\baseval
}\,dV=
\lim_{\eta\to\baseval}
\int_{\bodyRO}
D\fO\en{\XO}
\left[
\frac{
\Xn-\XO
}{
\eta-\baseval
}
\right]dV,
\end{split}
\label{eq:integratedLinearizationLimit}
\end{equation}
provided that the limit of $(\Xn-\XO)/\en{\eta-\baseval}$ exists in $\mathbb{V}$.

\subsection{A standard sufficient setting}

We record one familiar set of conditions ensuring the following properties used in the preceding argument: the weighted functionals $\mathcal{F}^\eta\ag{\cdot}$ are well defined and Fr\'echet differentiable at $\XO$; their derivatives converge in operator norm as in \eqref{eq:DFnConv}; the linearization remainder in \eqref{eq:FExpansion} satisfies $\alpha^\eta\to0$; and differentiation with respect to $\X$ may be performed under the integral as in \eqref{eq:DFIntegralRepresentation}. These conditions also ensure the integrability of the expressions appearing under the volume integrals in \eqref{eq:FcBodysimp6}, thereby justifying the passage to \eqref{eq:FcBodysimp7}.

Assume that the map
\[
\en{\eta,\x,\u,\H}
\longmapsto
\fn\ag{\x,\u,\H}
\]
is jointly $C^1$. Suppose further that, uniformly for $\eta$ sufficiently close to $\baseval$ and for $\en{\x,\u}$ in bounded sets,
\begin{equation}
\abs{\fn\ag{\x,\u,\H}}
+
\abs{\fn_{,I}\ag{\x,\u,\H}}
+
\abs{\fn_{,i}\ag{\x,\u,\H}}
\leq
C\en{1+\abs{\H}^{p}},
\end{equation}
and
\begin{equation}
\abs{\fn_{,iJ}\ag{\x,\u,\H}}
\leq
C\en{1+\abs{\H}^{p-1}},
\end{equation}
for all $I,i,J=1,\ldots,m$.

Because the displacement-gradient component of $\XO$ belongs to $L^p\en{\bodyRO;\Matrix[m\times m]}$, these conditions imply that
\begin{equation}
\fO\ag{\XO},\quad
\fO_{,I}[\XO],\quad
\fO_{,i}[\XO]
\quad\in L^1\en{\bodyRO},
\end{equation}
and
\begin{equation}
\fO_{,iJ}[\XO]
\in L^{p'}\en{\bodyRO},
\qquad
\frac{1}{p}+\frac{1}{p'}=1.
\end{equation}
Consequently, Hölder's inequality, together with the convergence results established in \S\ref{subsec:limits_of_functions}, justifies passing the limits through the first, third, fourth, and fifth integrals in \eqref{eq:FcBodysimp6}. (The passage of the limit through the second integral is justified separately by the \(L^1\) convergence assumption stated in \S\ref{subsec:fnLimits}.)

\section{Explicit construction of $\phiO\ag{\cdot}$}
\label{app:phiOConstruction}

We provide an explicit construction of the vector field $\phiO\ag{\cdot}$ used in \S\ref{sec:Demos}. Recall that
\begin{equation}
 \UO:=\pr{(x_1,x_2)\in\bodyRO~|~s <x_1< a^\circ+\mathcal{l},\quad\abs{x_2-h/2}< d},
\end{equation}
where $0<s<a^\circ$ and $\ell,d>R_c$. The field $\phiO\ag{\cdot}$ is required to equal $c\,\e[1]$ in a neighborhood of the notch tip, transition smoothly to zero within $\UO$, and vanish outside $\UO$.

We may define such a $\phiO\ag{\cdot}$ as follows. Our construction is based on smooth cutoff functions in the $\e[1]$ and $\e[2]$ directions. By a smooth cutoff function, we mean a smooth scalar function which is identically unity over a region where geometric evolution of interest is being considered (in the current example, the translation of semicircular notch tip in the $\e[1]$ direction), identically zero beyond the neighborhood of interest, and transitions smoothly between these values. To construct such cutoff functions, we first introduce the smooth step function $S\ag{\cdot}:\Reals\to\Reals$, defined by
\begin{equation}
    S\ag{\xi}
    :=
    \begin{cases}
        0,
        & \xi\leq 0,
        \\[2mm]
        \displaystyle
        \frac{\exp(-1/\xi)}
        {\exp(-1/\xi)+\exp[-1/(1-\xi)]},
        & 0<\xi<1,
        \\[4mm]
        1,
        & \xi\geq 1.
    \end{cases}
    \label{eq:smoothStepPhi}
\end{equation}
The function $S\ag{\cdot}\in C^\infty(\Reals)$ transitions smoothly from zero to unity as its argument increases from zero to one. By appropriately shifting and scaling its argument, we can therefore use $S\ag{\cdot}$ to prescribe the locations over which the cutoff functions transition between zero and unity.

We first construct a cutoff function $q_1\ag{\cdot}$ in the $\e[1]$ direction. We require $q_1\ag{\cdot}$ to transition from zero to unity as $\x[1]$ increases from $s$ to $a^\circ$, remain unity over the range of $\x[1]$ occupied by the notch tip, and transition back to zero as $\x[1]$ increases from $a^\circ+R_c$ to $a^\circ+\ell$. Accordingly, we define
\begin{equation}
    q_1\ag{\x}
    :=
    S\ag{\frac{\x[1]-s}{a^\circ-s}}
    \en{
        1-
        S\ag{
            \frac{\x[1]-(a^\circ+R_c)}{\ell-R_c}
        }
    }.
    \label{eq:q1Phi}
\end{equation}
Similarly, we construct a cutoff function $q_2\ag{\cdot}$ in the $\e[2]$ direction. We require $q_2\ag{\cdot}$ to remain unity over the range of $\x[2]$ occupied by the notch tip and to transition smoothly to zero between the notch tip and the upper and lower boundaries of $\UO$, defined by $x_2=h/2\pm d$. We therefore define
\begin{equation}
    q_2\ag{\x}
    :=
    S\ag{
        \frac{\x[2]-(h/2-d)}{d-R_c}
    }
    \en{
        1-
        S\ag{
            \frac{\x[2]-(h/2+R_c)}{d-R_c}
        }
    }.
    \label{eq:q2Phi}
\end{equation}

We may now define 
\begin{equation}
    \phiO\ag{\x}
    :=
    c\,q_1\ag{\x}q_2\ag{\x}\e[1].
    \label{eq:explicitPhiO}
\end{equation}
The product $q_1\ag{\x}q_2\ag{\x}$ combines the two directional cutoffs to localize the geometric evolution within $\UO$. In particular, it equals unity for
\begin{equation}
    a^\circ\leq\x[1]\leq a^\circ+R_c,
    \qquad
    \abs{\x[2]-h/2}\leq R_c,
\end{equation}
which contains $\Gamma_{\rm tip}^\circ$. Thus,
\begin{equation}
    \phiO\ag{\x}=c\,\e[1],
    \qquad
    \x\in\Gamma_{\rm tip}^\circ.
\end{equation}
If $\x\notin\UO$, at least one of $q_1\ag{\x}$ and $q_2\ag{\x}$ vanishes. It immediately follows that
\begin{equation}
    \phiO\ag{\x}=\zerovec,
    \qquad
    \x\in\bd\bodyRO\setminus\bd\UO.
\end{equation}
On the straight notch faces $\Gamma_+^\circ\cup\Gamma_-^\circ$, we have $\x[2]=h/2\pm R_c$, and therefore $q_2\ag{\x}=1$. Moreover, since $s<\x[1]<a^\circ$ on those faces, 
\begin{equation}
    q_1\ag{\x}=q\ag{\x}:=
    S\ag{\frac{\x[1]-s}{a^\circ-s}}.
\end{equation}
Thus, we have
\begin{equation}
    \phiO\ag{\x}=c\,q\ag{\x}\e[1],
    \qquad
    \x\in\Gamma_+^\circ\cup\Gamma_-^\circ,  
\end{equation}
where $q\ag{\cdot}$ transitions smoothly from zero to unity as $\x[1]$ increases from $s$ to $a^\circ$.
This completes the explicit construction of $\phiO\ag{\cdot}$ used in the example problem in \S\ref{sec:Demos}.

\paragraph*{Verification of admissibility}
The formula above defines a smooth, compactly supported extension of $\phiO\ag{\cdot}$ to $\Reals[2]$, and hence $\phiO\ag{\cdot}$ is Lipschitz on $\bodyRO$ with some constant $L_{\phiO}<\infty$. Therefore,
\begin{equation}
    \norm{\xnb\ag{\x}-\xnb\ag{\y}}_{\Reals[2]}
    \geq
    \en{1-\abs{\eta-\baseval}L_{\phiO}}
    \norm{\x-\y}_{\Reals[2]},
    \qquad \x,\y\in\bodyRO.
\end{equation}
Thus, $\xnb\ag{\cdot}$ is injective whenever $\abs{\eta-\baseval}L_{\phiO}<1$. Under the same condition,
\begin{equation}
    \grad\xnb
    =
    \tp{I}+\en{\eta-\baseval}\grad\phiO
\end{equation}
is nonsingular throughout $\bodyRO$. It follows from the inverse function theorem that $\xnb\ag{\cdot}$ is a smooth diffeomorphism onto its image for $\eta$ sufficiently close to $\baseval$.

\bibliographystyle{elsarticle-num}
\bibliography{references}

\end{document}